\documentclass[twocolumn,aps,pra,notitlepage,amsmath,amssymb,superscriptaddress]{revtex4-1}

\usepackage{graphicx}
\usepackage{bm}
\usepackage{amsmath}
\usepackage{verbatim}     
\usepackage[dvipsnames]{xcolor}       
\usepackage{bbold}
\newcommand{\qo}[1]{``#1''}                               		
\newcommand{\braket}[2]{\langle#1|#2\rangle}  		
\newcommand{\ket}[1]{|#1\rangle}                      		
\newcommand{\bra}[1]{\langle #1|}                     		
\newcommand{\me}[1]{\langle #1\rangle}    		
\newcommand{\abs}[1]{\left| #1 \right|}			
\newcommand{\sref}[1]{\ref{#1}}	
\newcommand{\beq}{\begin{equation}}
\newcommand{\eeq}{\end{equation}}
\newcommand{\bei}{\begin{itemize}}			
	\newcommand{\eei}{\end{itemize}}			

\newcommand{\bk}{\mathbf{k}}
\newcommand{\bmxy}{\mathbf{m}}

\newcommand{\bq}{\mathbf{q}}
\newcommand{\br}{\mathbf{r}}
\newcommand{\bR}{\mathbf{R}}

\usepackage{hyperref}
\hypersetup{%
   pdfpagemode=None, 
   pdfstartpage=1,
   pdfmenubar=true,
   pdftoolbar=true,
   colorlinks = true,
   linkcolor=blue,
   citecolor=blue,
   urlcolor=blue,
   bookmarksopen=false
 } 

\newcommand{\edit}[1]{{\color{
black}#1}}
\usepackage[normalem]{ulem}   
\begin{document}
\title{Two-dimensional topological quantum walks 
in the momentum space of structured~light}
\author{Alessio D'Errico}\thanks{AD'E and FC contributed equally to this work}
\affiliation{Dipartimento di Fisica \qo{Ettore Pancini}, Universit\`{a} di Napoli Federico II, Complesso Universitario di Monte Sant'Angelo, Via Cintia, 80126 Napoli, Italy}
\author{Filippo Cardano}\email{filippo.cardano2@unina.it}
\affiliation{Dipartimento di Fisica \qo{Ettore Pancini}, Universit\`{a} di Napoli Federico II, Complesso Universitario di Monte Sant'Angelo, Via Cintia, 80126 Napoli, Italy}
\author {Maria Maffei}
\affiliation{Dipartimento di Fisica \qo{Ettore Pancini}, Universit\`{a} di Napoli Federico II, Complesso Universitario di Monte Sant'Angelo, Via Cintia, 80126 Napoli, Italy}
\affiliation{ICFO -- Institut de Ciencies Fotoniques, The Barcelona Institute of Science and Technology, 08860 Castelldefels (Barcelona), Spain}
\author {Alexandre Dauphin}\email{alexandre.dauphin@icfo.eu}
\affiliation{ICFO -- Institut de Ciencies Fotoniques, The Barcelona Institute of Science and Technology, 08860 Castelldefels (Barcelona), Spain}
\author{Raouf Barboza}
\affiliation{Dipartimento di Fisica \qo{Ettore Pancini}, Universit\`{a} di Napoli Federico II, Complesso Universitario di Monte Sant'Angelo, Via Cintia, 80126 Napoli, Italy}
\author{Chiara Esposito}
\affiliation{Dipartimento di Fisica \qo{Ettore Pancini}, Universit\`{a} di Napoli Federico II, Complesso Universitario di Monte Sant'Angelo, Via Cintia, 80126 Napoli, Italy}
\author{Bruno Piccirillo}
\affiliation{Dipartimento di Fisica \qo{Ettore Pancini}, Universit\`{a} di Napoli Federico II, Complesso Universitario di Monte Sant'Angelo, Via Cintia, 80126 Napoli, Italy}
\author {Maciej Lewenstein}
\affiliation{ICFO -- Institut de Ciencies Fotoniques, The Barcelona Institute of Science and Technology, 08860 Castelldefels (Barcelona), Spain}
\affiliation{ICREA -- Instituci{\'o} Catalana de Recerca i Estudis Avan\c{c}ats, Pg.\ Lluis Companys 23, 08010 Barcelona, Spain}
\author {Pietro Massignan}\email{pietro.massignan@upc.edu}
\affiliation{ICFO -- Institut de Ciencies Fotoniques, The Barcelona Institute of Science and Technology, 08860 Castelldefels (Barcelona), Spain}
\affiliation{Departament de F\'isica, Universitat Polit\`ecnica de Catalunya, Campus Nord B4-B5, 08034 Barcelona, Spain}
\author{Lorenzo Marrucci}
\affiliation{Dipartimento di Fisica \qo{Ettore Pancini}, Universit\`{a} di Napoli Federico II, Complesso Universitario di Monte Sant'Angelo, Via Cintia, 80126 Napoli, Italy}
\affiliation{CNR-ISASI, Institute of Applied Science and Intelligent Systems, Via Campi Flegrei 34, 80078 Pozzuoli (NA), Italy}

\begin{abstract}
Quantum walks are powerful tools for quantum applications and for designing topological systems. Although they are simulated in a variety of platforms, genuine two-dimensional realizations are still challenging. Here we present an innovative approach to the photonic simulation of a quantum walk in two dimensions, where walker positions are encoded in the transverse wavevector components of a single light beam.
The desired dynamics is obtained by means of a sequence of liquid-crystal devices, which apply polarization-dependent transverse \qo{kicks} to the photons in the beam. 
We engineer our quantum walk so that it realizes a periodically-driven Chern insulator, and we probe its topological features by detecting the anomalous displacement of the photonic wavepacket under the effect of a constant force. Our compact, versatile platform offers exciting prospects for the photonic simulation of two-dimensional quantum dynamics and topological systems.
\end{abstract}



\maketitle
\section{Introduction}

Quantum walks (QWs) are the \edit{deterministic quantum analogues} of classical random walks, and describe particles (walkers) whose discrete dynamics is conditioned by the instantaneous configuration of their spin-like degree of freedom (the coin)~\cite{Venegas-Andraca2012}. 
QWs \edit{were originally introduced as versatile} candidates for implementing quantum search algorithms and universal quantum computation~\cite{Shenvi2003,Childs2009}, and have been used to model energy transport in photosynthetic processes~\cite{Mohseni2008}. 
These systems bear close analogies with electrons in periodic potentials, and it was shown that QWs can host all possible symmetry-protected topological phases displayed by non-interacting fermions in one or two spatial dimensions (1D or 2D)~\cite{Kitagawa2010a}. 
Practical implementations of quantum walks have been demonstrated, for instance, with ultracold atoms in optical lattices~\cite{Karski2009,Genske2013,Preiss2015,Dadras2018}, superconducting circuits~\cite{Flurin2017} and photonic systems~\cite{Broome2010,Peruzzo2010,Sansoni2012,Schreiber2012,Cardano2015,Defienne2016}.

In optical architectures, the lattice coordinates have been encoded in different degrees of freedom of light, such as the arrival time of a pulse at a detector~\cite{Schreiber2011,Schreiber2012,Jeong2013}, the optical path of the beam~\cite{Broome2010,Peruzzo2010,Sansoni2012,Kitagawa2012,Poulios2014}, or the orbital angular momentum~\cite{Cardano2015}\edit{, while} the coin is typically encoded in the polarization degree of freedom or in the entrance port of beam splitters. 
In a remarkable series of experiments, QWs proved instrumental in studying the evolution of correlated photons~\cite{Peruzzo2010,Sansoni2012}, the effects of decoherence~\cite{Broome2010,Schreiber2011} and interactions~\cite{Schreiber2012}, Anderson localization~\cite{Crespi2013a}, quantum transport in presence of disorder~\cite{Harris2017}, and topological phenomena in Floquet systems~\cite{Kitagawa2012,Zeuner2015,Cardano2016,Cardano2017,Xiao2017,Flurin2017,Zhan2017}. An excellent review of the state of the art on topological photonics and artificial gauge fields in quantum simulators is given in Refs.~\cite{Khanikaev2017,Ozawa2018} and \cite{Aidelsburger_2018}, respectively.
Despite being so fruitful, experimental research on QWs has been almost entirely focused on 1D systems. Few exceptions are the studies presented in Refs.~\cite{Schreiber2012,Jeong2013,Chen2018,Chalabi2019}, where a 2D walk was cleverly simulated by folding a 2D lattice in a 1D chain, and in Ref.~\cite{Wang2018b}, where path and OAM encoding were combined. Very recently, a continuous-time walk has been realized in a 2D array of coupled-waveguides~\cite{Tang2018,Tang2018b}.
\\
Here, we report a novel approach to the photonic simulation of quantum dynamics on 2D discrete lattices, based on the encoding of the individual sites in the transverse wavevector (or photon momentum), which is an inherently 2D degree of freedom. In our specific case, we simulate a quantum walk process. Unlike the common approach of using distinct optical paths (as in parallel waveguides), in our system the photonic evolution takes place within a single light beam that acquires a complex internal structure as it propagates. The core of our photonic QW simulator is a stack of closely-spaced liquid-crystal devices, conceptually similar to standard $q$-plates~
\cite{Marrucci2006,Rubano2019}. We present a proof-of-principle demonstration of our platform by generating up to 5 steps of a 2D QW, with both localized and extended initial inputs~\cite{Jeong2013,DiFranco2011,DiFranco2011a
}. 

We design the unitary evolution of our QW so that it realizes a periodically-driven (Floquet) Chern insulator. To characterize this system, we first analyze the energy dispersion of one of the bands of the effective Hamiltonian by tracking the free displacement of a wavepacket. Then we probe the Berry curvature of the band by repeating the tracking under the action of a constant force, that is simulated by means of simple translations of specific plates. Upon sampling uniformly across the whole band, the average transverse displacement provides us a straightforward and accurate measurement of the Chern number of that band.
\begin{figure*}
\centering
\vskip 0 pt
\includegraphics[width=\linewidth]{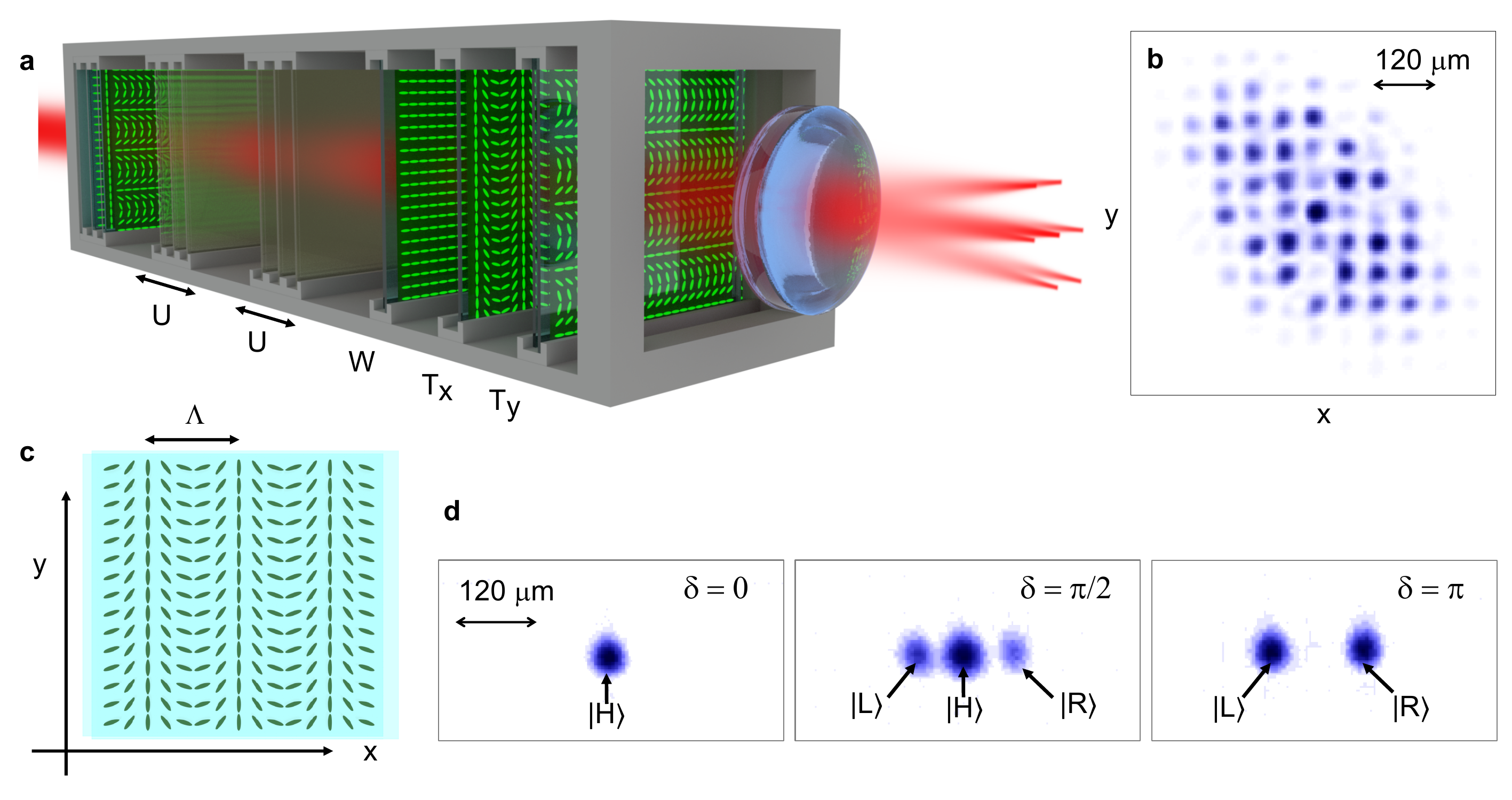}
\caption{
{\bf Experimental concept and apparatus.}
{\bf a}, A collimated beam crosses a sequence of liquid crystal (LC) devices. Different LC patterns implement coin rotations ($W$) and spin-dependent walker discrete translations ($T_x$ and $T_y$). Each evolution step $U=T_y T_x W$ is realized with three LC devices. The walker position is encoded in the transverse momentum of photons, so that walker steps physically correspond to transverse kicks which tilt slightly the photon propagation direction. 
The transverse diffraction of light remains negligible across the whole set-up, and the entire evolution effectively occurs in a single beam. At the exit of the walk, a lens (with focal distance equal to 50 cm) Fourier-transforms transverse momentum into position, allowing us to resolve and measure individual modes.
{\bf b}, The recorded intensity pattern is a regular grid of small Gaussian spots, whose intensities are proportional to the walker's spatial probability distribution. We set the modes beam radius to $w_0=5$ mm, that corresponds to a spot size of $\simeq 20 \,\mu$m (radius) on the camera plane.
{\bf c}, LC optic-axis pattern for a $g$-plate which realizes a $T_x$ operator. The spatial period $\Lambda$ fixes the transverse momentum lattice spacing $\Delta k_\perp = 2\pi/\Lambda$. We use $\Lambda= 5$ mm, so that $\Delta k_\perp=1.26\,$ mm$^{-1}$, corresponding to a spacing between spots of $\simeq 63\,\mu$m on the camera. 
{\bf d}, Action of a single $g$-plate  $T_x$  on a linearly polarized beam $\ket{\Psi_0}=\ket{0,0,H}$, where $\ket{H}=(\ket{L}+\ket{R})/\sqrt{2}$, for three different values of $\delta$.
}
\label{fig:setup}
\end{figure*}

\section{Results and discussion}
\subsection{Quantum walks in the transverse wavevector of light}
A discrete-time QW on a square lattice in 2D results from the repeated action of a unitary operator $U$ on a quantum system, the walker, and its internal spin-like degree of freedom, the coin \cite{Aharanov1993}
. 
After $t$ discrete steps, a given initial state $\ket{\Psi_0}$ evolves according to $\ket{\Psi(t)}=U^{t}\ket{\Psi_0}$.
The step operator $U$ typically includes a spin rotation $W$, and discrete displacements of the walker along the directions $x$ and $y$, generated by spin-dependent translation operators $T_x$ and $T_y$. In the simplest case, the Hilbert space of the coin has dimension two~\cite{Kitagawa2010a,Jeong2013}.
In our photonic QW implementation, we encode the coin into light polarization. For definiteness, we use left and right circular polarizations ($\ket L,\ket R$) as the basis states (with handedness defined from the point of view of the receiver), like in our earlier realization of a QW with twisted light~\cite{Cardano2015}.

The main novelty of the setup considered here lies in encoding the discrete dimensionless coordinate of the walker $\bmxy = (m_x,m_y)$ on a 2D square lattice in the transverse momentum of light. 
In particular, we use Gaussian modes whose mean transverse wavevector assumes the discrete values $\bk_\perp=\Delta k_\perp\bmxy$.
The lattice constant $\Delta k_\perp$ is taken to be much smaller than the longitudinal wavevector component $k_z\approx 2\pi/\lambda$ (where $\lambda$ is the light's wavelength), so that these modes propagate along a direction that is only slightly tilted with respect to the $z$ axis. More explicitly, a generic light mode in our setup reads:
\beq
\label{eq:gaussianbeamtilted}
\ket{\bmxy,\phi}=A(x,y,z) e^{i[\Delta k_\perp (m_x x+m_y y) +k_z z]}\otimes \ket{\phi},
\eeq
where $A(x,y,z)$ is a Gaussian envelope with large beam radius $w_0$ in the transverse $xy$ plane and $\ket{\phi}$ denotes the polarization state (see Sec.\ S1 in the Supplemental Materials (SM) for more details). Accordingly, arbitrary superpositions of these modes still form a single optical beam, travelling approximately along the $z$ axis. Only in the far field, or equivalently in the focal plane of a lens, these modes become spatially separated and their relative distribution can be easily read-out (see Figs.~\ref{fig:setup}{\bf a,b}). 
The parameters $w_0$ and $\Delta k_\perp$ are chosen so that these modes almost perfectly overlap spatially while propagating in the whole QW apparatus (as long as $|m_x|$ and $|m_y|$ are not too large), and have negligible crosstalk in the lens focal plane.

The QW dynamics is implemented with the apparatus depicted schematically in Fig.~\ref{fig:setup}{\bf a}, and described in greater detail in Sec.\ S2 in the SM. A collimated Gaussian laser beam passes through a sequence of closely-spaced liquid crystal (LC) plates which realize both walker-translation and coin-rotation operators. At the exit of the walk, a camera placed in the focal plane of a lens reads out the field intensity, providing the coordinates distribution of the walker (as in Fig.~\ref{fig:setup}{\bf b}; see also Sec.\ S3 in the SM). If needed, also the polarization components may be straightforwardly read out (see 
Fig.\ S2 in the SM).

The elements yielding the QW dynamics are optical devices consisting of thin layers of LC sandwiched between glass plates. The local orientation $\alpha(x,y)$ of the LC optic axis in the plane of the plate can follow arbitrary patterns, imprinted during the fabrication by a photo-alignment technique. The birefringent optical retardation $\delta$ of the LC may instead be controlled dynamically through an external electric field~
\cite{Piccirillo2010,Rubano2019}. \edit{In the basis of circular polarizations $\ket{L}=(1,0)^T$ and $\ket{R}=(0,1)^T$, these plates act as follows:
\beq\label{eq:LCplates}
L_\delta(x,y)\equiv
\left( {\begin{array}{cc}
   \cos(\delta/2) & i \sin(\delta/2) e^{-2i\alpha(x,y)}  \\
   i \sin(\delta/2) e^{2i\alpha(x,y)} & \cos(\delta/2) \\
  \end{array} } \right).
\eeq
}
Such plates give rise to coin rotations or walker translations, depending on the optic axis pattern. For example, a spin-dependent translation operator in the $x$ direction is obtained when the local orientation $\alpha$ increases linearly along $x$:
\begin{align}\label{eq:Lambda}
\alpha(x,y)=\pi x/\Lambda+\alpha_0,
\end{align}
where $\Lambda$ is the spatial periodicity of the angular pattern and $\alpha_0$ is a constant (see Fig.\ \ref{fig:setup}{\bf c}). This patterned birefringent structure is also known as a ``polarization grating'', and hence we refer here to these devices as ``$g$-plates'' (as opposed to the $q$-plates used in our previous works, which have azimuthally-varying patterns~\cite{Rubano2019}). \edit{By inserting Eq.\ \ref{eq:Lambda} in Eq.\ \ref{eq:LCplates}, one gets the action of a $g$-plate :}
\beq\label{eq:gplates}
T_x\equiv
\left( {\begin{array}{cc}
   \cos(\delta/2) & i \sin(\delta/2) e^{-2i\alpha_0}\hat t_x \\
   i \sin(\delta/2) e^{2i\alpha_0}\hat t^\dagger_x & \cos(\delta/2) \\
  \end{array} } \right),
\eeq
where $\hat t_x$ and $\hat t^\dagger_x$ are the (spin-independent) left and right translation operators along $x$, acting respectively as $\hat t_x\ket{m_x,m_y,\phi}=\ket{m_x-1,m_y,\phi}$ and $\hat t_x^\dagger \ket{m_x,m_y,\phi}=\ket{m_x+1,m_y,\phi}$. The spatial periodicity $\Lambda$ of the LC pattern controls the momentum lattice spacing $\Delta k_\perp=2\pi/\Lambda$. It is sufficient to set $\Lambda \sim w_0$ to avoid mode crosstalk. \edit{The action of a single $g$-plate  $T_x$ is shown in Fig.~\ref{fig:setup}{\bf d}.}
The $T_y$ operator is implemented analogously, imposing a gradient of the LC angle $\alpha$ along $y$. Finally, the spin rotations $W$ are realized with uniform LC plates acting as standard quarter-wave plates, that is with constant $\alpha=0$ and $\delta=\pi/2$. \edit{By using these values in Eq.\ \ref{eq:LCplates}, we get that in the basis of circular polarizations this operator acts as:}
\beq\label{eq:W}
W=
\frac{1}{\sqrt 2}\left( {\begin{array}{cc}
   1 & i \\
   i & 1 \\
  \end{array} } \right).
\eeq

\begin{figure}[t!]
\centering
\includegraphics[width= \columnwidth]{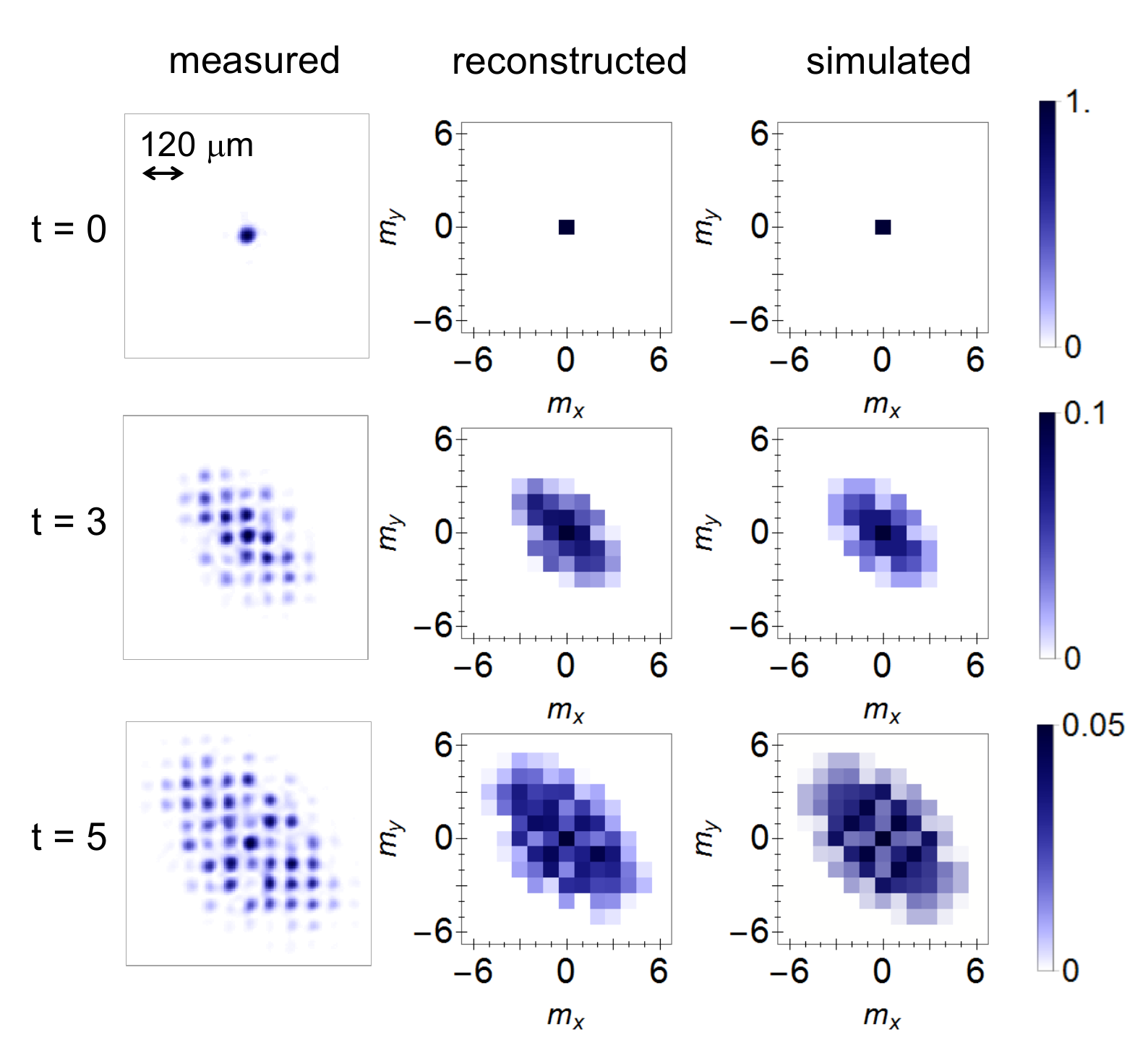}
\caption{\label{fig:2Dexperiment}
{\bf 2D Quantum Walk on a square lattice.}
Spatial probability distributions for a quantum walk with initial condition $\ket{0,0,H}$ and optical retardation
 $\delta=\pi/2$. From top to bottom, we display results after 0, 3, and 5 evolution steps. 
Datapoints are averages of four independent measures. 
 }
\end{figure}

\subsection{Engineering a 2D topological quantum walk} 
Among possible protocols obtained by combining our plates, we considered the QW generated by the unit step operator
\beq\label{eq:U}
U=T_yT_x W,
\eeq
with $T_x$ and $T_y$ tuned at the same value of $\delta$. We implemented five complete steps of this QW, that represents a generalization of the alternated protocol described in Ref.\ 
\cite{DiFranco2011,Jeong2013}. In particular, it realizes a periodically-driven Chern insulator, exhibiting different topological phases according to the value of the parameter $\delta$, as we discuss in detail below. 

We start with a localized walker state $\ket{\bmxy=(0,0),\phi}$, which physically corresponds to a wide input Gaussian beam with radius $w_0=5$ mm, propagating along the $z$ direction.
In Fig.~\ref{fig:2Dexperiment} we show representative data for $\delta = \pi/2$ and a linearly polarized input. The walker distribution remains concentrated along the diagonal $m_x=-m_y$ during the whole evolution, as a consequence of the absence of coin rotation operations between every action of $T_x$ and $T_y$. 
All data show an excellent agreement with numerical simulations. A quantitative comparison is provided by computing the similarity 
$S=\bigr(\sum_\bmxy \sqrt{P_e P_s}\bigr)^2/\left(\sum_\bmxy P_e\sum_\bmxy P_s\right)$ between simulated ($P_s$) and experimental ($P_e$) distributions.
For the data shown in Fig.~\ref{fig:2Dexperiment}, we have $S=(98.2\pm0.5)\%,\,(98.0\pm0.3)\%,\,(98.0\pm0.2)\%$ for the distributions at $t = 0,3,5$, respectively. The uncertainties on these values are the standard errors of the mean, obtained by repeating each experiment four times.
Distributions obtained for other choices of the coin input state are reported in 
Figs.~S5, S6 in the SM.

\subsection{Quasi-momentum, quasi-energy bands and group velocity}
\begin{figure*}[t!]
\centering
\vskip 0 pt
\includegraphics[width=\linewidth]{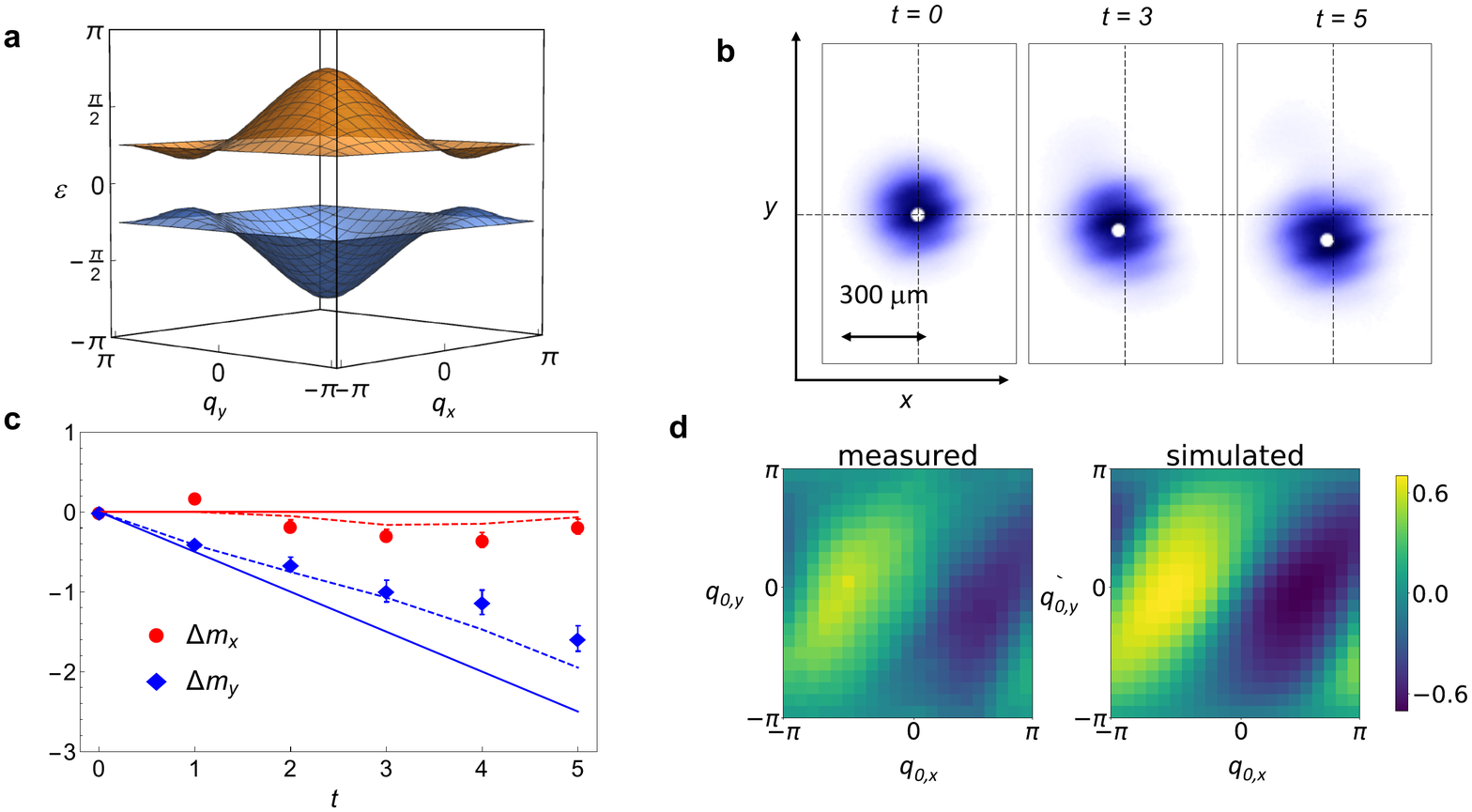}
\caption{
\label{fig:groupvelocity}
{\bf Detection of the group velocity at $\delta=\pi/2$.} 
{\bf a}, Quasi-energy spectrum of the effective Hamiltonian $H_{\rm eff}$.
{\bf b}, Light intensity distribution measured for a wavepacket with ${\bq_0}=(\pi/2, \pi)$ in the upper band, where the expected group velocity is $\mathbf{v}^{(+)}=(0,-0.5)$. The white marker indicates the center of mass of the wavepacket. 
The radius $w_g$ of the input beam is $(0.62\pm0.02)$ mm. In the camera plane, we measure a beam diameter of $(0.32\pm0.01)$ mm, corresponding to $\approx 5$ lattice sites.
{\bf c}, Displacement of the wavepacket center of mass, extracted from images as in panel {\bf b}. Experimental results (datapoints) are compared to semiclassical predictions of uniform motion (straight continuous lines), and to complete numerical simulations (dashed lines). Statistical uncertainties include estimated misalignment effects, as discussed in \edit{the main text}. 
{\bf d}, Experimental mapping of the upper band's group velocity $\mathbf{v}^{(+)}$ along $x$ across the whole Brillouin zone, compared to a complete numerical simulation. 
Each datapoint is obtained from a linear fit of the center-of-mass displacement of a Gaussian wavepacket in 5 steps.
}
\end{figure*}

A quantum walk can be regarded as the stroboscopic evolution generated by the (dimensionless) effective Floquet Hamiltonian $H_{\rm eff} \equiv i \ln U $. 
The eigenvalues of $H_{\rm eff}$ are therefore defined only up to integer multiples of $2\pi$, and are termed quasi-energies. 
Both $U$ and $H_{\rm eff}$ admit a convenient representation in the reciprocal space $\bq$ associated with the coordinate $\bmxy$ of the walker.
As discussed above, the dimensionless coordinate $\bmxy=\bk_\perp/\Delta k_\perp$ is encoded  in our setup in the transverse momentum $\mathbf{k}_\perp$ of the propagating beam. 
As such, its conjugate variable corresponds physically to the position vector $\br_\perp$ in the $xy$ transverse plane. 
We introduce therefore the dimensionless quasi-momentum $\bq=-2\pi\br_\perp/\Lambda$, belonging to the square Brillouin zone $[-\pi,\pi]^2$, as the conjugate variable to the walker position $\bmxy$. The negative sign in the definition of $\bq$ provides the standard representation for plane waves $\braket{\bmxy}{\bq}\propto e^{i\bmxy\cdot\bq}$.
In the space of quasi-momenta the effective Hamiltonian assumes the diagonal form 
$H_{\rm eff}(\bq)=\varepsilon (\bq){\bf n}(\bq)\cdot \bm{\sigma}$. 
Here ${\bf n}(\bq)$ is a unit vector, $\bm{ \sigma}=(\sigma_{x},\sigma_{y},\sigma_{z})$ represents the three Pauli matrices, 
and \edit{$\pm$}$\varepsilon(\bq)$ yields the quasi-energies of two bands (as shown in Fig.~\ref{fig:groupvelocity}{\bf a}). In the following, we will denote the complete eigenstates of the system by $\ket{\bq,\phi_\pm(\bq)}$, where $\pm$ refers to the upper/lower band. 
\edit{Let us note here that, although the operator $U$ is obtained by cascading independent displacements along the $x$ and $y$ axes, the overall evolution is non separable, that is the effective Hamiltonian cannot be expressed as the sum of two contributions depending on a single spatial coordinate. This can be seen clearly in the expression of the quasi-energy dispersion, that is given by the relation:
\begin{align}\label{eq:qenergy}
&\cos\varepsilon= \frac{1}{\sqrt 2}(\cos^2(\delta/2)-\cos(\delta/2) \sin(\delta/2)\times\\ 
&\times\left(\cos(q_x)+\cos(q_y)\right)-\sin^2(\delta/2)\cos(q_x-q_y)).
\end{align}
The complete expression of the Hamiltonian is provided in Sec.\ S4 of the SM.}\\

\begin{figure*}[t!]
\centering
\includegraphics[width=\linewidth]{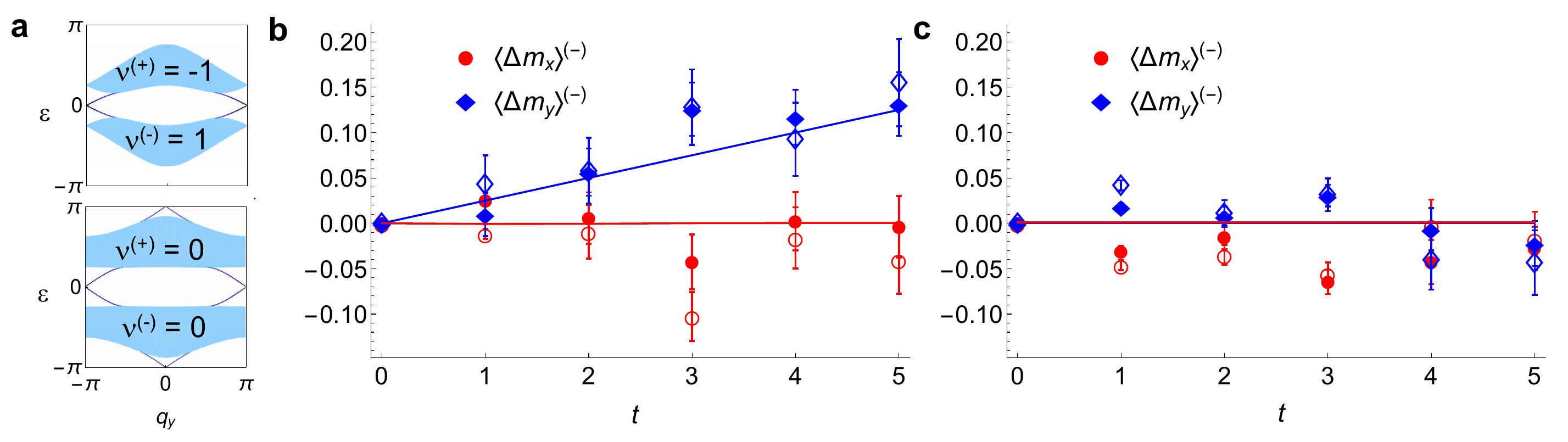}
\caption{{\bf Anomalous displacement for trivial and non-trivial Chern bands.}
{\bf a}, Quasi-energy spectra computed on a cylinder open along $x$ for $\delta=\pi/2$ (top) and $\delta=7\pi/8$ (bottom), showing the Chern numbers $\nu$ of the various bands.
In our Floquet system, edge states (shown as darker lines inside the gaps) may be present even if all bands have vanishing Chern numbers.
{\bf b}, Center-of-mass average displacement $\langle \Delta m_{x}\rangle^{(-)}(t)$ and $\langle \Delta m_{y}\rangle^{(-)}(t)$ measured for $\delta=\pi/2$ in the lower band.
Empty markers show results from the simple protocol $U$, while solid symbols show the improved results obtained by combining protocol $U$ with its inverse  $U^{-1}$.
Straight lines correspond to the theoretical results dictated by the semi-classical equations of motion, predicting an anomalous displacement proportional to the band Chern number. 
{\bf c}, Center of mass displacements measured with $\delta=7\pi/8$. Meaning of all symbols and lines as in panel {\bf b}. Statistical uncertainties include estimated misalignment effects, as discussed in the \edit{main text}. }
\label{fig:2Danomalous}
\end{figure*}

In our experiment, we can directly explore the band structure of the system by observing the propagation of walker wavepackets $\ket{\Psi_g(\bq_0,\pm)}$ that are sharply peaked around a given quasi-momentum $\bq_0$ and belong to the upper/lower band. These wavepackets are physically generated as narrow Gaussian light beams (with beam radius $w_g\ll \Lambda$) propagating along the direction $z$, centered around a specific transverse position $\br_{\perp0}=-\bq_0 \Lambda/(2\pi)$ at the input port of the QW, and with polarization $\ket{\phi_\pm(\bq_0)}$ (see Sec.\ S1 in the SM for \edit{further instructions for preparing these states}). 
In the experiment, the choice of transverse position $\br_{\perp0}$ is easily controlled by translating the whole QW set-up (which is mounted on a single motorized mechanical holder) relative to the input laser beam. 
Having narrow Gaussian envelopes in the conjugate space $\bq$, these wavepackets are relatively broad Gaussians in the space of walker coordinates $\bmxy$. They are (approximate) eigenstates of the system, and therefore preserve their shape during propagation. Their center of mass $\langle \bmxy \rangle_{\Psi_g}$ obeys a dynamics which semi-classically is governed by the group velocity ${\bf v}^{(\pm)}(\bq_0)= \pm\nabla_\bq\varepsilon(\bq)|_{\bq=\bq_0}$~\cite{Cardano2015}, as shown for instance in Fig.~\ref{fig:groupvelocity}{\bf b}.
To measure experimentally the group velocity  ${\bf v}^{(\pm)}(\bq_0)$ we inject a wavepacket $\ket{\Psi_g(\bq_0,\pm)}$ in our QW, we detect its average displacement $\Delta \bmxy$ as a function of time-step $t$, and finally we perform a linear best fit on the displacements versus time (see Fig.\ \ref{fig:groupvelocity}{\bf c}). 
Figure~\ref{fig:groupvelocity}{\bf d} shows a complete mapping of the $x$ component of the upper band's velocity 
$\mathbf{v}^{(+)}$ for $\delta=\pi/2$. Correspondingly measured values of the $y$ component of $\mathbf{v}^{(+)}$ are reported in 
Fig.\ S7 in the SM. \edit{A systematic error that can affect our set-up is the possible misalignment of $g$-plates in both $x$ and $y$ directions. Our present platform permits to adjust only their position along $x$. As such, we can estimate the associated standard error by repeating the experiment after re-aligning the plates. It is not possible however to repeat the same procedure for the perpendicular direction. In this case, after measuring the effective displacements of the $T_y$ plates, that are determined by fabrication imperfections, we perform a Monte Carlo simulation of the propagation of our wavepacket and we estimate the standard deviation of the final center of mass position. The two errors are finally combined by adding their variances to obtain the error bars in Figs.\ \ref{fig:groupvelocity}{\bf c} and \ref{fig:2Danomalous}{\bf b,c}.}

\subsection{Measurement of the Chern number through the anomalous velocity}
The energy bands of the effective Hamiltonian generally possess non-zero Berry curvature. 
For a $2\times2$ Hamiltonian like ours, the latter may be written as \cite{Qi2008} 
\beq
\Omega_{xy}^{(\pm)} (\bq)= \pm \frac{1}{2} {\bf n (\bq)}\cdot \left[\frac{\partial{\bf n}}{\partial q_x} \times \frac{\partial{\bf n}}{\partial q_y} \right],
\eeq
The integral of the Berry curvature over the whole Brillouin zone (BZ) gives the Chern number:
\begin{equation}
\nu^{(\pm)}= \int_{\mathrm{BZ}} \frac{{\rm d^2}\bq}{2\pi}\,\Omega_{xy}^{(\pm)}(\bq).
\end{equation}
The Chern number $\nu^{(\pm)}$ of our QW depends on the optical retardation of our plates. 
By tuning $\delta$, we can thus switch from a trivial to a topological Chern insulator, as shown for example in Fig.~\ref{fig:2Danomalous}{\bf a}. \edit{
Our QW is a Floquet evolution, and as such its complete topological classification is not based on the Chern number only, but involves a more complex invariant introduced in Ref.\ \cite{Rudner2013}. Such classification is discussed in detail in Sec.\ S5 of the SM.}

When a constant unidirectional force is acting on the system, the Berry curvature contributes to the wave-packet displacement in a direction orthogonal to the force (as in the quantum Hall effect). 
Let us for definiteness consider a force $F_x$ acting along $x$.
Within the adiabatic approximation, the semi-classical equations of motion predict that a wavepacket $\ket{\Psi_{g}(\bq_{0},\pm)}$
will experience after a time $t$ a transverse displacement along $y$ given by \cite{
Xiao2010,Price2016}
\begin{equation}\label{eq:ydisp}
\Delta m_y = \int_0^{t} {\rm d}\tau \left[ v_y^{(\pm)}(\bq_\tau) + F_x\Omega_{xy}^{(\pm)} (\bq_\tau)\right],
\end{equation}
with $\bq_\tau=(q_{0,x}+F_x \tau ,q_{0,y})$. The contribution to the velocity coming from the Berry curvature is called anomalous velocity. 
This result is derived in the adiabatic regime, where the (dimensionless) force is much smaller than the band-gaps of the effective energy, so that inter-band transitions can be neglected. 
When we consider the overall transverse displacement of a filled band, namely when we integrate Eq.~\eqref{eq:ydisp} over the whole Brillouin zone, the group-velocity term averages to zero, while the anomalous contributions add up to the band's Chern number~\cite{Dauphin2013,Price2016} \edit{(see 
Sec.\ S4 in the SM for a detailed derivation of this result)}:
\begin{equation}
\langle \Delta m_y\rangle^{(\pm)} \approx \frac{ F_x \nu^{(\pm)}}{2 \pi} t.
\end{equation} 
\edit{As shown in Eq.\ \ref{eq:ydisp}, implementing a constant force in our set-up requires a linear shift in time of the quasi momentum component $q_x$. This degree of freedom corresponds to the $x$ coordinate in real space. Hence, we impose at each step a quasi-momentum variation by introducing a transverse spatial displacement of the light beam between each plate. Actually, rather than displacing the beam, it is equivalent (and much simpler) to displace the reference system and the setup in the opposite direction. More specifically, we shift the $g$-plate acting at time-step $t$ along the $x$ axis by an amount $\Delta x_t=t F_x \Lambda/(2\pi)$ (see 
Sec.\ S4 in the SM for further details).} 
Then, we sum up the measured displacements $\Delta m_{y}$ obtained for $11\times11$ distinct wave-packets $\ket{\Psi_{g}(\bq_{0},-)}$, which provides a homogeneous sampling of the lower band across the whole Brillouin zone and realizes a good approximation of the continuous integral yielding $\langle \Delta m_y \rangle^{(-)}$. Figure~\ref{fig:2Danomalous}{\bf b} shows the mean displacement of wavepackets prepared in the lowest energy band for a QW with $\delta=\pi/2$, corresponding to Chern number $\nu^{(-)}=1$.  The energy bandgap $\approx 1$ (see Fig.~\ref{fig:2Danomalous}{\bf a}) is sufficiently larger than the applied force $F_x=\pi/20$, thereby ensuring the validity of the adiabatic approximation.
 Experimental data (empty markers) are compared to the overall band displacement predicted by the semi-classical theory within adiabatic regime (continuous lines), namely $\langle \Delta m_y(t)\rangle^{(-)} =t\nu^{(-)} F_{x}/(2\pi)$, $\langle \Delta m_x(t)\rangle^{(-)}=0$. While $\langle  \Delta m_y \rangle^{(-)}$ follows the expected curve quite reasonably, the overall $\langle  \Delta m_x \rangle^{(-)}$ is found to be not negligible. \edit{These small differences can be understood by simulating the full dynamics of the wave packet, beyond the single band approximation (See Fig.\ S8 in the SM)}
 
To get rid of this spurious contribution, which arises mainly from residual group-velocity effects, we consider also the ``inverse protocol'' generated by the step operator $U^{-1}=W^{-1}T_x^{-1}T_y^{-1}$.
The bands of $U^{-1}$ have the same dispersion as the bands of $U$, but feature opposite Chern numbers. In this way, if filling the same band, we expect to observe identical contributions from the group velocity dispersion, while the anomalous displacement should be inverted. The step operator $U^{-1}$ can be easily implemented by swapping the $T_y$ and $W$ operators, and changing suitably their retardation (see 
Sec.\ S4 and Fig.~S2 in the SM). 
In Fig.\ \ref{fig:2Danomalous}{\bf b} we show with filled markers the difference (divided by two) of the data obtained with the protocols $U$ and $U^{-1}$.
This procedure reduces significantly the overall displacement along $x$, while in the $y$ direction we observe a very nice agreement between our data and the semi-classical predictions. The measured value of the Chern number is $\nu^{(-)}=1.19\pm0.13$, consistent with the theoretical value of 1 (errors are given at one standard deviation). A similar behaviour is also observed for larger values of the force, as shown in 
Fig.\ S9 in the SM.
In Fig.\ \ref{fig:2Danomalous}{\bf c}, we replicate the same experiment for a QW with $\delta=7\pi/8$, when the Chern numbers are zero, even though the presence of edge states witnesses non-trivial topology~\cite{Rudner2013} (see also Sec.\ S5 in the SM). In agreement with the prediction of vanishing anomalous displacement, the average wavepacket motion in both directions is observed to be negligible, yielding a Chern number $\nu^{(-)}=0.10\pm0.15$.

\section{Conclusion}
In this work we have experimentally demonstrated a conceptually new scheme for the realization of a 2D discrete-time quantum walk, that relies on encoding the walker and the coin systems into the transverse momentum of photons and in their polarization, respectively. The coin rotation and shift operators are implemented by suitably engineered liquid crystals plates, whose number scales linearly with the number of time-steps. They are arranged in a compact set-up, in which multiple degrees of freedom can be controlled dynamically, such as the plates optical retardation $\delta$ or their transverse position, allowing one to study several quantum walk architectures. If needed, different LC patterns could be written onto the plates, yielding different types of quantum dynamics. The platform 
accurately simulates the dynamics dictated by the quantum walk protocols that we tested, as witnessed by the good agreement between measured distributions and numerical results. We investigated 2D walks of both localized and extended inputs, with and without an external force. The 2D protocol we presented here simulates a Floquet Chern insulator. We probed the associated topological features by preparing wave-packets which well approximate the eigenstates of the QW Floquet Hamiltonian, and detecting their average anomalous displacement arising when a constant force is applied to the system. 

The set-up has been designed to minimize the decoherence effects caused by light diffraction and walk-off phase delays occurring when the walker follows different paths (see Sec.\ S7 in the SM). There is no fundamental limitation to scaling up our set-up to a much larger number of steps. Reflection losses at each LC plate ($\simeq 15 \%$), representing the main current limitation to the setup efficiency, could be largely reduced to the level of 1-2$\%$ by applying a standard anti-reflection coating. As such, while our experiment is carried out in a classical-wave regime, the proposed setup is perfectly suitable for single-photon quantum experiments\edit{, similarly to what already demonstrated in Ref.\ \cite{Cardano2015}}. These represent one of the most appealing future prospects for our system, particularly in view of the very large number of input and output modes that can be easily addressed.

\edit{The demonstration of a new platform for 2D quantum walks opens new avenues for the experimental study of the rich quantum dynamics in two dimensions.
In prospect, diverse directions could be investigated with our platform, such as the realization of 2D lattices with more complex topologies (for example, hexagonal), or experiments in the multi-photon regime, for instance in the context of QW applications to Boson sampling. Direct access to both walker position and quasi-momentum could be exploited to study complex dynamics in the regime of spatial disorder. By combining topology and our dynamical control of the system parameters, we could investigate dynamical quantum phase transitions in quantum walks \cite{Wang2018,Xu2018,Heyl2018}. Finally, by introducing losses for specific polarization states, this platform could be used to investigate topological features of 2D non-Hermitian systems \cite{Longhi2017,Yao2018,Gong2018,Yokomizo2019}.}


\section*{Funding}
AD'E, FC, RB and LM acknowledge financial support from the European Union Horizon 2020 program, under European Research Council (ERC) grant no. 694683 (PHOSPhOR). ADa, MM, PM and ML acknowledge Spanish MINECO (Severo Ochoa SEV-2015-0522, FisicaTeAMO FIS2016-79508-P, and SWUQM FIS2017-84114-C2-1-P), the Generalitat de Catalunya (SGR874 and CERCA), the EU (ERC AdG OSYRIS 339106, H2020-FETProAct QUIC 641122), the Fundaci\'o Privada Cellex, a Cellex-ICFO-MPQ fellowship, the "Juan de la Cierva" program (IJCI-2017-33180) and the ``Ram\'on y Cajal" program.\\

\section*{Authors Contributions}
\noindent LM conceived the 2D photon-momentum encoding strategy. AD'E, FC and LM designed the set-up. AD'E devised the 2D QW protocol. AD'E, FC, RB, with contributions from CE, performed the experiments and analyzed the data. MM, ADa and PM developed the theoretical framework and devised the scheme for the anomalous displacement detection. BP, with the help of AD'E and FC, fabricated the $g$-plates. AD'E, FC, RB, ADa, MM, PM and LM wrote the manuscript. LM and ML supervised the project. All authors discussed the results and contributed to refining the manuscript. AD'E and FC contributed equally to this work.

\bibliography{TopQW_Ale}

\begin{thebibliography}{54}%
\makeatletter
\providecommand \@ifxundefined [1]{%
 \@ifx{#1\undefined}
}%
\providecommand \@ifnum [1]{%
 \ifnum #1\expandafter \@firstoftwo
 \else \expandafter \@secondoftwo
 \fi
}%
\providecommand \@ifx [1]{%
 \ifx #1\expandafter \@firstoftwo
 \else \expandafter \@secondoftwo
 \fi
}%
\providecommand \natexlab [1]{#1}%
\providecommand \enquote  [1]{``#1''}%
\providecommand \bibnamefont  [1]{#1}%
\providecommand \bibfnamefont [1]{#1}%
\providecommand \citenamefont [1]{#1}%
\providecommand \href@noop [0]{\@secondoftwo}%
\providecommand \href [0]{\begingroup \@sanitize@url \@href}%
\providecommand \@href[1]{\@@startlink{#1}\@@href}%
\providecommand \@@href[1]{\endgroup#1\@@endlink}%
\providecommand \@sanitize@url [0]{\catcode `\\12\catcode `\$12\catcode
  `\&12\catcode `\#12\catcode `\^12\catcode `\_12\catcode `\%12\relax}%
\providecommand \@@startlink[1]{}%
\providecommand \@@endlink[0]{}%
\providecommand \url  [0]{\begingroup\@sanitize@url \@url }%
\providecommand \@url [1]{\endgroup\@href {#1}{\urlprefix }}%
\providecommand \urlprefix  [0]{URL }%
\providecommand \Eprint [0]{\href }%
\providecommand \doibase [0]{http://dx.doi.org/}%
\providecommand \selectlanguage [0]{\@gobble}%
\providecommand \bibinfo  [0]{\@secondoftwo}%
\providecommand \bibfield  [0]{\@secondoftwo}%
\providecommand \translation [1]{[#1]}%
\providecommand \BibitemOpen [0]{}%
\providecommand \bibitemStop [0]{}%
\providecommand \bibitemNoStop [0]{.\EOS\space}%
\providecommand \EOS [0]{\spacefactor3000\relax}%
\providecommand \BibitemShut  [1]{\csname bibitem#1\endcsname}%
\let\auto@bib@innerbib\@empty
\bibitem [{\citenamefont {Venegas-Andraca}(2012)}]{Venegas-Andraca2012}%
  \BibitemOpen
  \bibfield  {author} {\bibinfo {author} {\bibfnamefont {S.~E.}\ \bibnamefont
  {Venegas-Andraca}},\ }\bibfield  {title} {\bibinfo {title} {\emph {{Quantum
  walks: a comprehensive review}}},\ }\href {\doibase
  10.1007/s11128-012-0432-5} {\bibfield  {journal} {\bibinfo  {journal}
  {Quantum Inf. Process.}\ }\textbf {\bibinfo {volume} {11}},\ \bibinfo {pages}
  {1015} (\bibinfo {year} {2012})},\ \Eprint {http://arxiv.org/abs/1201.4780}
  {arXiv:1201.4780} \BibitemShut {NoStop}%
\bibitem [{\citenamefont {Shenvi}\ \emph {et~al.}(2003)\citenamefont {Shenvi},
  \citenamefont {Kempe},\ and\ \citenamefont {Whaley}}]{Shenvi2003}%
  \BibitemOpen
  \bibfield  {author} {\bibinfo {author} {\bibfnamefont {N.}~\bibnamefont
  {Shenvi}}, \bibinfo {author} {\bibfnamefont {J.}~\bibnamefont {Kempe}}, \
  and\ \bibinfo {author} {\bibfnamefont {K.~B.}\ \bibnamefont {Whaley}},\
  }\bibfield  {title} {\bibinfo {title} {\emph {{Quantum random-walk search
  algorithm}}},\ }\href {\doibase 10.1103/PhysRevA.67.052307} {\bibfield
  {journal} {\bibinfo  {journal} {Phys. Rev. A}\ }\textbf {\bibinfo {volume}
  {67}},\ \bibinfo {pages} {052307} (\bibinfo {year} {2003})},\ \Eprint
  {http://arxiv.org/abs/quant-ph/0210064} {arXiv:quant-ph/0210064} \BibitemShut
  {NoStop}%
\bibitem [{\citenamefont {Childs}(2009)}]{Childs2009}%
  \BibitemOpen
  \bibfield  {author} {\bibinfo {author} {\bibfnamefont {A.~M.}\ \bibnamefont
  {Childs}},\ }\bibfield  {title} {\bibinfo {title} {\emph {{Universal
  Computation by Quantum Walk}}},\ }\href {\doibase
  10.1103/PhysRevLett.102.180501} {\bibfield  {journal} {\bibinfo  {journal}
  {Phys. Rev. Lett.}\ }\textbf {\bibinfo {volume} {102}},\ \bibinfo {pages}
  {180501} (\bibinfo {year} {2009})},\ \Eprint {http://arxiv.org/abs/0806.1972}
  {arXiv:0806.1972} \BibitemShut {NoStop}%
\bibitem [{\citenamefont {Mohseni}\ \emph {et~al.}(2008)\citenamefont
  {Mohseni}, \citenamefont {Rebentrost}, \citenamefont {Lloyd},\ and\
  \citenamefont {Aspuru-Guzik}}]{Mohseni2008}%
  \BibitemOpen
  \bibfield  {author} {\bibinfo {author} {\bibfnamefont {M.}~\bibnamefont
  {Mohseni}}, \bibinfo {author} {\bibfnamefont {P.}~\bibnamefont {Rebentrost}},
  \bibinfo {author} {\bibfnamefont {S.}~\bibnamefont {Lloyd}}, \ and\ \bibinfo
  {author} {\bibfnamefont {A.}~\bibnamefont {Aspuru-Guzik}},\ }\bibfield
  {title} {\bibinfo {title} {\emph {{Environment-assisted quantum walks in
  photosynthetic energy transfer}}},\ }\href {\doibase 10.1063/1.3002335}
  {\bibfield  {journal} {\bibinfo  {journal} {J. Chem. Phys.}\ }\textbf
  {\bibinfo {volume} {129}},\ \bibinfo {pages} {174106} (\bibinfo {year}
  {2008})}\BibitemShut {NoStop}%
\bibitem [{\citenamefont {Kitagawa}\ \emph {et~al.}(2010)\citenamefont
  {Kitagawa}, \citenamefont {Rudner}, \citenamefont {Berg},\ and\ \citenamefont
  {Demler}}]{Kitagawa2010a}%
  \BibitemOpen
  \bibfield  {author} {\bibinfo {author} {\bibfnamefont {T.}~\bibnamefont
  {Kitagawa}}, \bibinfo {author} {\bibfnamefont {M.~S.}\ \bibnamefont
  {Rudner}}, \bibinfo {author} {\bibfnamefont {E.}~\bibnamefont {Berg}}, \ and\
  \bibinfo {author} {\bibfnamefont {E.}~\bibnamefont {Demler}},\ }\bibfield
  {title} {\bibinfo {title} {\emph {{Exploring topological phases with quantum
  walks}}},\ }\href {\doibase 10.1103/PhysRevA.82.033429} {\bibfield  {journal}
  {\bibinfo  {journal} {Physical Review A}\ }\textbf {\bibinfo {volume} {82}},\
  \bibinfo {pages} {033429} (\bibinfo {year} {2010})},\ \Eprint
  {http://arxiv.org/abs/1003.1729} {arXiv:1003.1729} \BibitemShut {NoStop}%
\bibitem [{\citenamefont {Karski}\ \emph {et~al.}(2009)\citenamefont {Karski},
  \citenamefont {Forster}, \citenamefont {Choi}, \citenamefont {Steffen},
  \citenamefont {Alt}, \citenamefont {Meschede},\ and\ \citenamefont
  {Widera}}]{Karski2009}%
  \BibitemOpen
  \bibfield  {author} {\bibinfo {author} {\bibfnamefont {M.}~\bibnamefont
  {Karski}}, \bibinfo {author} {\bibfnamefont {L.}~\bibnamefont {Forster}},
  \bibinfo {author} {\bibfnamefont {J.-M.}\ \bibnamefont {Choi}}, \bibinfo
  {author} {\bibfnamefont {A.}~\bibnamefont {Steffen}}, \bibinfo {author}
  {\bibfnamefont {W.}~\bibnamefont {Alt}}, \bibinfo {author} {\bibfnamefont
  {D.}~\bibnamefont {Meschede}}, \ and\ \bibinfo {author} {\bibfnamefont
  {A.}~\bibnamefont {Widera}},\ }\bibfield  {title} {\bibinfo {title} {\emph
  {{Quantum Walk in Position Space with Single Optically Trapped Atoms}}},\
  }\href {\doibase 10.1126/science.1174436} {\bibfield  {journal} {\bibinfo
  {journal} {Science}\ }\textbf {\bibinfo {volume} {325}},\ \bibinfo {pages}
  {174} (\bibinfo {year} {2009})}\BibitemShut {NoStop}%
\bibitem [{\citenamefont {Genske}\ \emph {et~al.}(2013)\citenamefont {Genske},
  \citenamefont {Alt}, \citenamefont {Steffen}, \citenamefont {Werner},
  \citenamefont {Werner}, \citenamefont {Meschede},\ and\ \citenamefont
  {Alberti}}]{Genske2013}%
  \BibitemOpen
  \bibfield  {author} {\bibinfo {author} {\bibfnamefont {M.}~\bibnamefont
  {Genske}}, \bibinfo {author} {\bibfnamefont {W.}~\bibnamefont {Alt}},
  \bibinfo {author} {\bibfnamefont {A.}~\bibnamefont {Steffen}}, \bibinfo
  {author} {\bibfnamefont {A.~H.}\ \bibnamefont {Werner}}, \bibinfo {author}
  {\bibfnamefont {R.~F.}\ \bibnamefont {Werner}}, \bibinfo {author}
  {\bibfnamefont {D.}~\bibnamefont {Meschede}}, \ and\ \bibinfo {author}
  {\bibfnamefont {A.}~\bibnamefont {Alberti}},\ }\bibfield  {title} {\bibinfo
  {title} {\emph {{Electric quantum walks with individual atoms}}},\ }\href
  {\doibase 10.1103/PhysRevLett.110.190601} {\bibfield  {journal} {\bibinfo
  {journal} {Phys. Rev. Lett.}\ }\textbf {\bibinfo {volume} {110}},\ \bibinfo
  {pages} {190601} (\bibinfo {year} {2013})},\ \Eprint
  {http://arxiv.org/abs/1302.2094} {arXiv:1302.2094} \BibitemShut {NoStop}%
\bibitem [{\citenamefont {Preiss}\ \emph {et~al.}(2015)\citenamefont {Preiss},
  \citenamefont {Ma}, \citenamefont {Tai}, \citenamefont {Lukin}, \citenamefont
  {Rispoli}, \citenamefont {Zupancic}, \citenamefont {Lahini}, \citenamefont
  {Islam},\ and\ \citenamefont {Greiner}}]{Preiss2015}%
  \BibitemOpen
  \bibfield  {author} {\bibinfo {author} {\bibfnamefont {P.~M.}\ \bibnamefont
  {Preiss}}, \bibinfo {author} {\bibfnamefont {R.}~\bibnamefont {Ma}}, \bibinfo
  {author} {\bibfnamefont {M.~E.}\ \bibnamefont {Tai}}, \bibinfo {author}
  {\bibfnamefont {A.}~\bibnamefont {Lukin}}, \bibinfo {author} {\bibfnamefont
  {M.}~\bibnamefont {Rispoli}}, \bibinfo {author} {\bibfnamefont
  {P.}~\bibnamefont {Zupancic}}, \bibinfo {author} {\bibfnamefont
  {Y.}~\bibnamefont {Lahini}}, \bibinfo {author} {\bibfnamefont
  {R.}~\bibnamefont {Islam}}, \ and\ \bibinfo {author} {\bibfnamefont
  {M.}~\bibnamefont {Greiner}},\ }\bibfield  {title} {\bibinfo {title} {\emph
  {{Strongly correlated quantum walks in optical lattices}}},\ }\href {\doibase
  10.1126/science.1260364} {\bibfield  {journal} {\bibinfo  {journal}
  {Science}\ }\textbf {\bibinfo {volume} {347}},\ \bibinfo {pages} {1229}
  (\bibinfo {year} {2015})},\ \Eprint {http://arxiv.org/abs/1409.3100}
  {arXiv:1409.3100} \BibitemShut {NoStop}%
\bibitem [{\citenamefont {Dadras}\ \emph {et~al.}(2018)\citenamefont {Dadras},
  \citenamefont {Gresch}, \citenamefont {Groiseau}, \citenamefont {Wimberger},\
  and\ \citenamefont {Summy}}]{Dadras2018}%
  \BibitemOpen
  \bibfield  {author} {\bibinfo {author} {\bibfnamefont {S.}~\bibnamefont
  {Dadras}}, \bibinfo {author} {\bibfnamefont {A.}~\bibnamefont {Gresch}},
  \bibinfo {author} {\bibfnamefont {C.}~\bibnamefont {Groiseau}}, \bibinfo
  {author} {\bibfnamefont {S.}~\bibnamefont {Wimberger}}, \ and\ \bibinfo
  {author} {\bibfnamefont {G.~S.}\ \bibnamefont {Summy}},\ }\bibfield  {title}
  {\bibinfo {title} {\emph {{Quantum Walk in Momentum Space with a
  Bose-Einstein Condensate}}},\ }\href {\doibase
  10.1103/PhysRevLett.121.070402} {\bibfield  {journal} {\bibinfo  {journal}
  {Phys. Rev. Lett.}\ }\textbf {\bibinfo {volume} {121}},\ \bibinfo {pages}
  {070402} (\bibinfo {year} {2018})},\ \Eprint
  {http://arxiv.org/abs/1802.08160} {arXiv:1802.08160} \BibitemShut {NoStop}%
\bibitem [{\citenamefont {Flurin}\ \emph {et~al.}(2017)\citenamefont {Flurin},
  \citenamefont {Ramasesh}, \citenamefont {Hacohen-Gourgy}, \citenamefont
  {Martin}, \citenamefont {Yao},\ and\ \citenamefont {Siddiqi}}]{Flurin2017}%
  \BibitemOpen
  \bibfield  {author} {\bibinfo {author} {\bibfnamefont {E.}~\bibnamefont
  {Flurin}}, \bibinfo {author} {\bibfnamefont {V.~V.}\ \bibnamefont
  {Ramasesh}}, \bibinfo {author} {\bibfnamefont {S.}~\bibnamefont
  {Hacohen-Gourgy}}, \bibinfo {author} {\bibfnamefont {L.~S.}\ \bibnamefont
  {Martin}}, \bibinfo {author} {\bibfnamefont {N.~Y.}\ \bibnamefont {Yao}}, \
  and\ \bibinfo {author} {\bibfnamefont {I.}~\bibnamefont {Siddiqi}},\
  }\bibfield  {title} {\bibinfo {title} {\emph {{Observing Topological
  Invariants Using Quantum Walks in Superconducting Circuits}}},\ }\href
  {\doibase 10.1103/PhysRevX.7.031023} {\bibfield  {journal} {\bibinfo
  {journal} {Phys. Rev. X}\ }\textbf {\bibinfo {volume} {7}},\ \bibinfo {pages}
  {031023} (\bibinfo {year} {2017})},\ \Eprint
  {http://arxiv.org/abs/1610.03069v1} {arXiv:1610.03069v1} \BibitemShut
  {NoStop}%
\bibitem [{\citenamefont {Broome}\ \emph {et~al.}(2010)\citenamefont {Broome},
  \citenamefont {Fedrizzi}, \citenamefont {Lanyon}, \citenamefont {Kassal},
  \citenamefont {Aspuru-Guzik},\ and\ \citenamefont {White}}]{Broome2010}%
  \BibitemOpen
  \bibfield  {author} {\bibinfo {author} {\bibfnamefont {M.~A.}\ \bibnamefont
  {Broome}}, \bibinfo {author} {\bibfnamefont {A.}~\bibnamefont {Fedrizzi}},
  \bibinfo {author} {\bibfnamefont {B.~P.}\ \bibnamefont {Lanyon}}, \bibinfo
  {author} {\bibfnamefont {I.}~\bibnamefont {Kassal}}, \bibinfo {author}
  {\bibfnamefont {A.}~\bibnamefont {Aspuru-Guzik}}, \ and\ \bibinfo {author}
  {\bibfnamefont {A.~G.}\ \bibnamefont {White}},\ }\bibfield  {title} {\bibinfo
  {title} {\emph {{Discrete Single-Photon Quantum Walks with Tunable
  Decoherence}}},\ }\href {\doibase 10.1103/PhysRevLett.104.153602} {\bibfield
  {journal} {\bibinfo  {journal} {Phys. Rev. Lett.}\ }\textbf {\bibinfo
  {volume} {104}},\ \bibinfo {pages} {153602} (\bibinfo {year} {2010})},\
  \Eprint {http://arxiv.org/abs/1002.4923} {arXiv:1002.4923} \BibitemShut
  {NoStop}%
\bibitem [{\citenamefont {Peruzzo}\ \emph {et~al.}(2010)\citenamefont
  {Peruzzo}, \citenamefont {Lobino}, \citenamefont {Matthews}, \citenamefont
  {Matsuda}, \citenamefont {Politi}, \citenamefont {Poulios}, \citenamefont
  {Zhou}, \citenamefont {Lahini}, \citenamefont {Ismail}, \citenamefont
  {Worhoff}, \citenamefont {Bromberg}, \citenamefont {Silberberg},
  \citenamefont {Thompson},\ and\ \citenamefont {OBrien}}]{Peruzzo2010}%
  \BibitemOpen
  \bibfield  {author} {\bibinfo {author} {\bibfnamefont {A.}~\bibnamefont
  {Peruzzo}}, \bibinfo {author} {\bibfnamefont {M.}~\bibnamefont {Lobino}},
  \bibinfo {author} {\bibfnamefont {J.~C.~F.}\ \bibnamefont {Matthews}},
  \bibinfo {author} {\bibfnamefont {N.}~\bibnamefont {Matsuda}}, \bibinfo
  {author} {\bibfnamefont {A.}~\bibnamefont {Politi}}, \bibinfo {author}
  {\bibfnamefont {K.}~\bibnamefont {Poulios}}, \bibinfo {author} {\bibfnamefont
  {X.-Q.}\ \bibnamefont {Zhou}}, \bibinfo {author} {\bibfnamefont
  {Y.}~\bibnamefont {Lahini}}, \bibinfo {author} {\bibfnamefont
  {N.}~\bibnamefont {Ismail}}, \bibinfo {author} {\bibfnamefont
  {K.}~\bibnamefont {Worhoff}}, \bibinfo {author} {\bibfnamefont
  {Y.}~\bibnamefont {Bromberg}}, \bibinfo {author} {\bibfnamefont
  {Y.}~\bibnamefont {Silberberg}}, \bibinfo {author} {\bibfnamefont {M.~G.}\
  \bibnamefont {Thompson}}, \ and\ \bibinfo {author} {\bibfnamefont {J.~L.}\
  \bibnamefont {OBrien}},\ }\bibfield  {title} {\bibinfo {title} {\emph
  {{Quantum Walks of Correlated Photons}}},\ }\href {\doibase
  10.1126/science.1193515} {\bibfield  {journal} {\bibinfo  {journal}
  {Science}\ }\textbf {\bibinfo {volume} {329}},\ \bibinfo {pages} {1500}
  (\bibinfo {year} {2010})},\ \Eprint {http://arxiv.org/abs/1006.4764}
  {arXiv:1006.4764} \BibitemShut {NoStop}%
\bibitem [{\citenamefont {Sansoni}\ \emph {et~al.}(2012)\citenamefont
  {Sansoni}, \citenamefont {Sciarrino}, \citenamefont {Vallone}, \citenamefont
  {Mataloni}, \citenamefont {Crespi}, \citenamefont {Ramponi},\ and\
  \citenamefont {Osellame}}]{Sansoni2012}%
  \BibitemOpen
  \bibfield  {author} {\bibinfo {author} {\bibfnamefont {L.}~\bibnamefont
  {Sansoni}}, \bibinfo {author} {\bibfnamefont {F.}~\bibnamefont {Sciarrino}},
  \bibinfo {author} {\bibfnamefont {G.}~\bibnamefont {Vallone}}, \bibinfo
  {author} {\bibfnamefont {P.}~\bibnamefont {Mataloni}}, \bibinfo {author}
  {\bibfnamefont {A.}~\bibnamefont {Crespi}}, \bibinfo {author} {\bibfnamefont
  {R.}~\bibnamefont {Ramponi}}, \ and\ \bibinfo {author} {\bibfnamefont
  {R.}~\bibnamefont {Osellame}},\ }\bibfield  {title} {\bibinfo {title} {\emph
  {{Two-Particle Bosonic-Fermionic Quantum Walk via Integrated Photonics}}},\
  }\href {\doibase 10.1103/PhysRevLett.108.010502} {\bibfield  {journal}
  {\bibinfo  {journal} {Phys. Rev. Lett.}\ }\textbf {\bibinfo {volume} {108}},\
  \bibinfo {pages} {010502} (\bibinfo {year} {2012})},\ \Eprint
  {http://arxiv.org/abs/1106.5713} {arXiv:1106.5713} \BibitemShut {NoStop}%
\bibitem [{\citenamefont {Schreiber}\ \emph {et~al.}(2012)\citenamefont
  {Schreiber}, \citenamefont {Gabris}, \citenamefont {Rohde}, \citenamefont
  {Laiho}, \citenamefont {Stefanak}, \citenamefont {Potocek}, \citenamefont
  {Hamilton}, \citenamefont {Jex},\ and\ \citenamefont
  {Silberhorn}}]{Schreiber2012}%
  \BibitemOpen
  \bibfield  {author} {\bibinfo {author} {\bibfnamefont {A.}~\bibnamefont
  {Schreiber}}, \bibinfo {author} {\bibfnamefont {A.}~\bibnamefont {Gabris}},
  \bibinfo {author} {\bibfnamefont {P.~P.}\ \bibnamefont {Rohde}}, \bibinfo
  {author} {\bibfnamefont {K.}~\bibnamefont {Laiho}}, \bibinfo {author}
  {\bibfnamefont {M.}~\bibnamefont {Stefanak}}, \bibinfo {author}
  {\bibfnamefont {V.}~\bibnamefont {Potocek}}, \bibinfo {author} {\bibfnamefont
  {C.}~\bibnamefont {Hamilton}}, \bibinfo {author} {\bibfnamefont
  {I.}~\bibnamefont {Jex}}, \ and\ \bibinfo {author} {\bibfnamefont
  {C.}~\bibnamefont {Silberhorn}},\ }\bibfield  {title} {\bibinfo {title}
  {\emph {{A 2D Quantum Walk Simulation of Two-Particle Dynamics}}},\ }\href
  {\doibase 10.1126/science.1218448} {\bibfield  {journal} {\bibinfo  {journal}
  {Science}\ }\textbf {\bibinfo {volume} {336}},\ \bibinfo {pages} {55}
  (\bibinfo {year} {2012})},\ \Eprint {http://arxiv.org/abs/1204.3555}
  {arXiv:1204.3555} \BibitemShut {NoStop}%
\bibitem [{\citenamefont {Cardano}\ \emph
  {et~al.}(2015{\natexlab{a}})\citenamefont {Cardano}, \citenamefont {Massa},
  \citenamefont {Qassim}, \citenamefont {Karimi}, \citenamefont {Slussarenko},
  \citenamefont {Paparo}, \citenamefont {de~Lisio}, \citenamefont {Sciarrino},
  \citenamefont {Santamato}, \citenamefont {Boyd},\ and\ \citenamefont
  {Marrucci}}]{Cardano2015}%
  \BibitemOpen
  \bibfield  {author} {\bibinfo {author} {\bibfnamefont {F.}~\bibnamefont
  {Cardano}}, \bibinfo {author} {\bibfnamefont {F.}~\bibnamefont {Massa}},
  \bibinfo {author} {\bibfnamefont {H.}~\bibnamefont {Qassim}}, \bibinfo
  {author} {\bibfnamefont {E.}~\bibnamefont {Karimi}}, \bibinfo {author}
  {\bibfnamefont {S.}~\bibnamefont {Slussarenko}}, \bibinfo {author}
  {\bibfnamefont {D.}~\bibnamefont {Paparo}}, \bibinfo {author} {\bibfnamefont
  {C.}~\bibnamefont {de~Lisio}}, \bibinfo {author} {\bibfnamefont
  {F.}~\bibnamefont {Sciarrino}}, \bibinfo {author} {\bibfnamefont
  {E.}~\bibnamefont {Santamato}}, \bibinfo {author} {\bibfnamefont {R.~W.}\
  \bibnamefont {Boyd}}, \ and\ \bibinfo {author} {\bibfnamefont
  {L.}~\bibnamefont {Marrucci}},\ }\bibfield  {title} {\bibinfo {title} {\emph
  {{Quantum walks and wavepacket dynamics on a lattice with twisted
  photons}}},\ }\href {\doibase 10.1126/sciadv.1500087} {\bibfield  {journal}
  {\bibinfo  {journal} {Sci. Adv.}\ }\textbf {\bibinfo {volume} {1}},\ \bibinfo
  {pages} {e1500087} (\bibinfo {year} {2015}{\natexlab{a}})},\ \Eprint
  {http://arxiv.org/abs/1407.5424v1} {arXiv:1407.5424v1} \BibitemShut {NoStop}%
\bibitem [{\citenamefont {Defienne}\ \emph {et~al.}(2016)\citenamefont
  {Defienne}, \citenamefont {Barbieri}, \citenamefont {Walmsley}, \citenamefont
  {Smith},\ and\ \citenamefont {Gigan}}]{Defienne2016}%
  \BibitemOpen
  \bibfield  {author} {\bibinfo {author} {\bibfnamefont {H.}~\bibnamefont
  {Defienne}}, \bibinfo {author} {\bibfnamefont {M.}~\bibnamefont {Barbieri}},
  \bibinfo {author} {\bibfnamefont {I.~A.}\ \bibnamefont {Walmsley}}, \bibinfo
  {author} {\bibfnamefont {B.~J.}\ \bibnamefont {Smith}}, \ and\ \bibinfo
  {author} {\bibfnamefont {S.}~\bibnamefont {Gigan}},\ }\bibfield  {title}
  {\bibinfo {title} {\emph {{Two-photon quantum walk in a multimode fiber}}},\
  }\href {\doibase 10.1126/sciadv.1501054} {\bibfield  {journal} {\bibinfo
  {journal} {Sci. Adv.}\ }\textbf {\bibinfo {volume} {2}},\ \bibinfo {pages}
  {e1501054} (\bibinfo {year} {2016})},\ \Eprint
  {http://arxiv.org/abs/1504.03178} {arXiv:1504.03178} \BibitemShut {NoStop}%
\bibitem [{\citenamefont {Schreiber}\ \emph {et~al.}(2011)\citenamefont
  {Schreiber}, \citenamefont {Cassemiro}, \citenamefont {Poto{\v{c}}ek},
  \citenamefont {G{\'{a}}bris}, \citenamefont {Jex},\ and\ \citenamefont
  {Silberhorn}}]{Schreiber2011}%
  \BibitemOpen
  \bibfield  {author} {\bibinfo {author} {\bibfnamefont {A.}~\bibnamefont
  {Schreiber}}, \bibinfo {author} {\bibfnamefont {K.~N.}\ \bibnamefont
  {Cassemiro}}, \bibinfo {author} {\bibfnamefont {V.}~\bibnamefont
  {Poto{\v{c}}ek}}, \bibinfo {author} {\bibfnamefont {A.}~\bibnamefont
  {G{\'{a}}bris}}, \bibinfo {author} {\bibfnamefont {I.}~\bibnamefont {Jex}}, \
  and\ \bibinfo {author} {\bibfnamefont {C.}~\bibnamefont {Silberhorn}},\
  }\bibfield  {title} {\bibinfo {title} {\emph {{Decoherence and Disorder in
  Quantum Walks: From Ballistic Spread to Localization}}},\ }\href {\doibase
  10.1103/PhysRevLett.106.180403} {\bibfield  {journal} {\bibinfo  {journal}
  {Phys. Rev. Lett.}\ }\textbf {\bibinfo {volume} {106}},\ \bibinfo {pages}
  {180403} (\bibinfo {year} {2011})},\ \Eprint {http://arxiv.org/abs/1101.2638}
  {arXiv:1101.2638} \BibitemShut {NoStop}%
\bibitem [{\citenamefont {Jeong}\ \emph {et~al.}(2013)\citenamefont {Jeong},
  \citenamefont {{Di Franco}}, \citenamefont {Lim}, \citenamefont {Kim},\ and\
  \citenamefont {Kim}}]{Jeong2013}%
  \BibitemOpen
  \bibfield  {author} {\bibinfo {author} {\bibfnamefont {Y.-C.}\ \bibnamefont
  {Jeong}}, \bibinfo {author} {\bibfnamefont {C.}~\bibnamefont {{Di Franco}}},
  \bibinfo {author} {\bibfnamefont {H.-T.}\ \bibnamefont {Lim}}, \bibinfo
  {author} {\bibfnamefont {M.}~\bibnamefont {Kim}}, \ and\ \bibinfo {author}
  {\bibfnamefont {Y.-H.}\ \bibnamefont {Kim}},\ }\bibfield  {title} {\bibinfo
  {title} {\emph {{Experimental realization of a delayed-choice quantum
  walk}}},\ }\href {\doibase 10.1038/ncomms3471} {\bibfield  {journal}
  {\bibinfo  {journal} {Nat. Commun.}\ }\textbf {\bibinfo {volume} {4}},\
  \bibinfo {pages} {2471} (\bibinfo {year} {2013})},\ \Eprint
  {http://arxiv.org/abs/1309.6837} {arXiv:1309.6837} \BibitemShut {NoStop}%
\bibitem [{\citenamefont {Kitagawa}\ \emph {et~al.}(2012)\citenamefont
  {Kitagawa}, \citenamefont {Broome}, \citenamefont {Fedrizzi}, \citenamefont
  {Rudner}, \citenamefont {Berg}, \citenamefont {Kassal}, \citenamefont
  {Aspuru-Guzik}, \citenamefont {Demler},\ and\ \citenamefont
  {White}}]{Kitagawa2012}%
  \BibitemOpen
  \bibfield  {author} {\bibinfo {author} {\bibfnamefont {T.}~\bibnamefont
  {Kitagawa}}, \bibinfo {author} {\bibfnamefont {M.~a.}\ \bibnamefont
  {Broome}}, \bibinfo {author} {\bibfnamefont {A.}~\bibnamefont {Fedrizzi}},
  \bibinfo {author} {\bibfnamefont {M.~S.}\ \bibnamefont {Rudner}}, \bibinfo
  {author} {\bibfnamefont {E.}~\bibnamefont {Berg}}, \bibinfo {author}
  {\bibfnamefont {I.}~\bibnamefont {Kassal}}, \bibinfo {author} {\bibfnamefont
  {A.}~\bibnamefont {Aspuru-Guzik}}, \bibinfo {author} {\bibfnamefont
  {E.}~\bibnamefont {Demler}}, \ and\ \bibinfo {author} {\bibfnamefont {A.~G.}\
  \bibnamefont {White}},\ }\bibfield  {title} {\bibinfo {title} {\emph
  {{Observation of topologically protected bound states in photonic quantum
  walks}}},\ }\href {\doibase 10.1038/ncomms1872} {\bibfield  {journal}
  {\bibinfo  {journal} {Nat. Commun.}\ }\textbf {\bibinfo {volume} {3}},\
  \bibinfo {pages} {882} (\bibinfo {year} {2012})},\ \Eprint
  {http://arxiv.org/abs/1105.5334} {arXiv:1105.5334} \BibitemShut {NoStop}%
\bibitem [{\citenamefont {Poulios}\ \emph {et~al.}(2014)\citenamefont
  {Poulios}, \citenamefont {Keil}, \citenamefont {Fry}, \citenamefont
  {Meinecke}, \citenamefont {Matthews}, \citenamefont {Politi}, \citenamefont
  {Lobino}, \citenamefont {Gr{\"{a}}fe}, \citenamefont {Heinrich},
  \citenamefont {Nolte}, \citenamefont {Szameit},\ and\ \citenamefont
  {O'Brien}}]{Poulios2014}%
  \BibitemOpen
  \bibfield  {author} {\bibinfo {author} {\bibfnamefont {K.}~\bibnamefont
  {Poulios}}, \bibinfo {author} {\bibfnamefont {R.}~\bibnamefont {Keil}},
  \bibinfo {author} {\bibfnamefont {D.}~\bibnamefont {Fry}}, \bibinfo {author}
  {\bibfnamefont {J.~D.~A.}\ \bibnamefont {Meinecke}}, \bibinfo {author}
  {\bibfnamefont {J.~C.~F.}\ \bibnamefont {Matthews}}, \bibinfo {author}
  {\bibfnamefont {A.}~\bibnamefont {Politi}}, \bibinfo {author} {\bibfnamefont
  {M.}~\bibnamefont {Lobino}}, \bibinfo {author} {\bibfnamefont
  {M.}~\bibnamefont {Gr{\"{a}}fe}}, \bibinfo {author} {\bibfnamefont
  {M.}~\bibnamefont {Heinrich}}, \bibinfo {author} {\bibfnamefont
  {S.}~\bibnamefont {Nolte}}, \bibinfo {author} {\bibfnamefont
  {A.}~\bibnamefont {Szameit}}, \ and\ \bibinfo {author} {\bibfnamefont
  {J.~L.}\ \bibnamefont {O'Brien}},\ }\bibfield  {title} {\bibinfo {title}
  {\emph {{Quantum Walks of Correlated Photon Pairs in Two-Dimensional
  Waveguide Arrays}}},\ }\href {\doibase 10.1103/PhysRevLett.112.143604}
  {\bibfield  {journal} {\bibinfo  {journal} {Phys. Rev. Lett.}\ }\textbf
  {\bibinfo {volume} {112}},\ \bibinfo {pages} {143604} (\bibinfo {year}
  {2014})},\ \Eprint {http://arxiv.org/abs/1308.2554} {arXiv:1308.2554}
  \BibitemShut {NoStop}%
\bibitem [{\citenamefont {Crespi}\ \emph {et~al.}(2013)\citenamefont {Crespi},
  \citenamefont {Osellame}, \citenamefont {Ramponi}, \citenamefont
  {Giovannetti}, \citenamefont {Fazio}, \citenamefont {Sansoni}, \citenamefont
  {{De Nicola}}, \citenamefont {Sciarrino},\ and\ \citenamefont
  {Mataloni}}]{Crespi2013a}%
  \BibitemOpen
  \bibfield  {author} {\bibinfo {author} {\bibfnamefont {A.}~\bibnamefont
  {Crespi}}, \bibinfo {author} {\bibfnamefont {R.}~\bibnamefont {Osellame}},
  \bibinfo {author} {\bibfnamefont {R.}~\bibnamefont {Ramponi}}, \bibinfo
  {author} {\bibfnamefont {V.}~\bibnamefont {Giovannetti}}, \bibinfo {author}
  {\bibfnamefont {R.}~\bibnamefont {Fazio}}, \bibinfo {author} {\bibfnamefont
  {L.}~\bibnamefont {Sansoni}}, \bibinfo {author} {\bibfnamefont
  {F.}~\bibnamefont {{De Nicola}}}, \bibinfo {author} {\bibfnamefont
  {F.}~\bibnamefont {Sciarrino}}, \ and\ \bibinfo {author} {\bibfnamefont
  {P.}~\bibnamefont {Mataloni}},\ }\bibfield  {title} {\bibinfo {title} {\emph
  {{Anderson localization of entangled photons in an integrated quantum
  walk}}},\ }\href {\doibase 10.1038/nphoton.2013.26} {\bibfield  {journal}
  {\bibinfo  {journal} {Nat. Photon.}\ }\textbf {\bibinfo {volume} {7}},\
  \bibinfo {pages} {322} (\bibinfo {year} {2013})},\ \Eprint
  {http://arxiv.org/abs/1304.1012v1} {arXiv:1304.1012v1} \BibitemShut {NoStop}%
\bibitem [{\citenamefont {Harris}\ \emph {et~al.}(2017)\citenamefont {Harris},
  \citenamefont {Steinbrecher}, \citenamefont {Prabhu}, \citenamefont {Lahini},
  \citenamefont {Mower}, \citenamefont {Bunandar}, \citenamefont {Chen},
  \citenamefont {Wong}, \citenamefont {Baehr-Jones}, \citenamefont {Hochberg},
  \citenamefont {Lloyd},\ and\ \citenamefont {Englund}}]{Harris2017}%
  \BibitemOpen
  \bibfield  {author} {\bibinfo {author} {\bibfnamefont {N.~C.}\ \bibnamefont
  {Harris}}, \bibinfo {author} {\bibfnamefont {G.~R.}\ \bibnamefont
  {Steinbrecher}}, \bibinfo {author} {\bibfnamefont {M.}~\bibnamefont
  {Prabhu}}, \bibinfo {author} {\bibfnamefont {Y.}~\bibnamefont {Lahini}},
  \bibinfo {author} {\bibfnamefont {J.}~\bibnamefont {Mower}}, \bibinfo
  {author} {\bibfnamefont {D.}~\bibnamefont {Bunandar}}, \bibinfo {author}
  {\bibfnamefont {C.}~\bibnamefont {Chen}}, \bibinfo {author} {\bibfnamefont
  {F.~N.~C.}\ \bibnamefont {Wong}}, \bibinfo {author} {\bibfnamefont
  {T.}~\bibnamefont {Baehr-Jones}}, \bibinfo {author} {\bibfnamefont
  {M.}~\bibnamefont {Hochberg}}, \bibinfo {author} {\bibfnamefont
  {S.}~\bibnamefont {Lloyd}}, \ and\ \bibinfo {author} {\bibfnamefont
  {D.}~\bibnamefont {Englund}},\ }\bibfield  {title} {\bibinfo {title} {\emph
  {{Quantum transport simulations in a programmable nanophotonic processor}}},\
  }\href {\doibase 10.1038/nphoton.2017.95} {\bibfield  {journal} {\bibinfo
  {journal} {Nat. Photon.}\ }\textbf {\bibinfo {volume} {11}},\ \bibinfo
  {pages} {447} (\bibinfo {year} {2017})},\ \Eprint
  {http://arxiv.org/abs/1507.03406} {arXiv:1507.03406} \BibitemShut {NoStop}%
\bibitem [{\citenamefont {Zeuner}\ \emph {et~al.}(2015)\citenamefont {Zeuner},
  \citenamefont {Rechtsman}, \citenamefont {Plotnik}, \citenamefont {Lumer},
  \citenamefont {Nolte}, \citenamefont {Rudner}, \citenamefont {Segev},\ and\
  \citenamefont {Szameit}}]{Zeuner2015}%
  \BibitemOpen
  \bibfield  {author} {\bibinfo {author} {\bibfnamefont {J.~M.}\ \bibnamefont
  {Zeuner}}, \bibinfo {author} {\bibfnamefont {M.~C.}\ \bibnamefont
  {Rechtsman}}, \bibinfo {author} {\bibfnamefont {Y.}~\bibnamefont {Plotnik}},
  \bibinfo {author} {\bibfnamefont {Y.}~\bibnamefont {Lumer}}, \bibinfo
  {author} {\bibfnamefont {S.}~\bibnamefont {Nolte}}, \bibinfo {author}
  {\bibfnamefont {M.~S.}\ \bibnamefont {Rudner}}, \bibinfo {author}
  {\bibfnamefont {M.}~\bibnamefont {Segev}}, \ and\ \bibinfo {author}
  {\bibfnamefont {A.}~\bibnamefont {Szameit}},\ }\bibfield  {title} {\bibinfo
  {title} {\emph {{Observation of a Topological Transition in the Bulk of a
  Non-Hermitian System}}},\ }\href {\doibase 10.1103/PhysRevLett.115.040402}
  {\bibfield  {journal} {\bibinfo  {journal} {Phys. Rev. Lett.}\ }\textbf
  {\bibinfo {volume} {115}},\ \bibinfo {pages} {040402} (\bibinfo {year}
  {2015})},\ \Eprint {http://arxiv.org/abs/1408.2191} {arXiv:1408.2191}
  \BibitemShut {NoStop}%
\bibitem [{\citenamefont {Cardano}\ \emph
  {et~al.}(2015{\natexlab{b}})\citenamefont {Cardano}, \citenamefont {Maffei},
  \citenamefont {Massa}, \citenamefont {Piccirillo}, \citenamefont {de~Lisio},
  \citenamefont {{De Filippis}}, \citenamefont {Cataudella}, \citenamefont
  {Santamato},\ and\ \citenamefont {Marrucci}}]{Cardano2016}%
  \BibitemOpen
  \bibfield  {author} {\bibinfo {author} {\bibfnamefont {F.}~\bibnamefont
  {Cardano}}, \bibinfo {author} {\bibfnamefont {M.}~\bibnamefont {Maffei}},
  \bibinfo {author} {\bibfnamefont {F.}~\bibnamefont {Massa}}, \bibinfo
  {author} {\bibfnamefont {B.}~\bibnamefont {Piccirillo}}, \bibinfo {author}
  {\bibfnamefont {C.}~\bibnamefont {de~Lisio}}, \bibinfo {author}
  {\bibfnamefont {G.}~\bibnamefont {{De Filippis}}}, \bibinfo {author}
  {\bibfnamefont {V.}~\bibnamefont {Cataudella}}, \bibinfo {author}
  {\bibfnamefont {E.}~\bibnamefont {Santamato}}, \ and\ \bibinfo {author}
  {\bibfnamefont {L.}~\bibnamefont {Marrucci}},\ }\bibfield  {title} {\bibinfo
  {title} {\emph {{Dynamical moments reveal a topological quantum transition in
  a photonic quantum walk}}},\ }\href {\doibase 10.1038/ncomms11439} {\bibfield
   {journal} {\bibinfo  {journal} {Nat. Commun.}\ }\textbf {\bibinfo {volume}
  {7}},\ \bibinfo {pages} {11439} (\bibinfo {year} {2015}{\natexlab{b}})},\
  \Eprint {http://arxiv.org/abs/1507.01785} {arXiv:1507.01785} \BibitemShut
  {NoStop}%
\bibitem [{\citenamefont {Cardano}\ \emph {et~al.}(2017)\citenamefont
  {Cardano}, \citenamefont {D'Errico}, \citenamefont {Dauphin}, \citenamefont
  {Maffei}, \citenamefont {Piccirillo}, \citenamefont {de~Lisio}, \citenamefont
  {{De Filippis}}, \citenamefont {Cataudella}, \citenamefont {Santamato},
  \citenamefont {Marrucci}, \citenamefont {Lewenstein},\ and\ \citenamefont
  {Massignan}}]{Cardano2017}%
  \BibitemOpen
  \bibfield  {author} {\bibinfo {author} {\bibfnamefont {F.}~\bibnamefont
  {Cardano}}, \bibinfo {author} {\bibfnamefont {A.}~\bibnamefont {D'Errico}},
  \bibinfo {author} {\bibfnamefont {A.}~\bibnamefont {Dauphin}}, \bibinfo
  {author} {\bibfnamefont {M.}~\bibnamefont {Maffei}}, \bibinfo {author}
  {\bibfnamefont {B.}~\bibnamefont {Piccirillo}}, \bibinfo {author}
  {\bibfnamefont {C.}~\bibnamefont {de~Lisio}}, \bibinfo {author}
  {\bibfnamefont {G.}~\bibnamefont {{De Filippis}}}, \bibinfo {author}
  {\bibfnamefont {V.}~\bibnamefont {Cataudella}}, \bibinfo {author}
  {\bibfnamefont {E.}~\bibnamefont {Santamato}}, \bibinfo {author}
  {\bibfnamefont {L.}~\bibnamefont {Marrucci}}, \bibinfo {author}
  {\bibfnamefont {M.}~\bibnamefont {Lewenstein}}, \ and\ \bibinfo {author}
  {\bibfnamefont {P.}~\bibnamefont {Massignan}},\ }\bibfield  {title} {\bibinfo
  {title} {\emph {{Detection of Zak phases and topological invariants in a
  chiral quantum walk of twisted photons}}},\ }\href {\doibase
  10.1038/ncomms15516} {\bibfield  {journal} {\bibinfo  {journal} {Nat.
  Commun.}\ }\textbf {\bibinfo {volume} {8}},\ \bibinfo {pages} {15516}
  (\bibinfo {year} {2017})},\ \Eprint {http://arxiv.org/abs/1610.06322}
  {arXiv:1610.06322} \BibitemShut {NoStop}%
\bibitem [{\citenamefont {Xiao}\ \emph {et~al.}(2017)\citenamefont {Xiao},
  \citenamefont {Zhan}, \citenamefont {Bian}, \citenamefont {Wang},
  \citenamefont {Zhang}, \citenamefont {Wang}, \citenamefont {Li},
  \citenamefont {Mochizuki}, \citenamefont {Kim}, \citenamefont {Kawakami},
  \citenamefont {Yi}, \citenamefont {Obuse}, \citenamefont {Sanders},\ and\
  \citenamefont {Xue}}]{Xiao2017}%
  \BibitemOpen
  \bibfield  {author} {\bibinfo {author} {\bibfnamefont {L.}~\bibnamefont
  {Xiao}}, \bibinfo {author} {\bibfnamefont {X.}~\bibnamefont {Zhan}}, \bibinfo
  {author} {\bibfnamefont {Z.~H.}\ \bibnamefont {Bian}}, \bibinfo {author}
  {\bibfnamefont {K.~K.}\ \bibnamefont {Wang}}, \bibinfo {author}
  {\bibfnamefont {X.}~\bibnamefont {Zhang}}, \bibinfo {author} {\bibfnamefont
  {X.~P.}\ \bibnamefont {Wang}}, \bibinfo {author} {\bibfnamefont
  {J.}~\bibnamefont {Li}}, \bibinfo {author} {\bibfnamefont {K.}~\bibnamefont
  {Mochizuki}}, \bibinfo {author} {\bibfnamefont {D.}~\bibnamefont {Kim}},
  \bibinfo {author} {\bibfnamefont {N.}~\bibnamefont {Kawakami}}, \bibinfo
  {author} {\bibfnamefont {W.}~\bibnamefont {Yi}}, \bibinfo {author}
  {\bibfnamefont {H.}~\bibnamefont {Obuse}}, \bibinfo {author} {\bibfnamefont
  {B.~C.}\ \bibnamefont {Sanders}}, \ and\ \bibinfo {author} {\bibfnamefont
  {P.}~\bibnamefont {Xue}},\ }\bibfield  {title} {\bibinfo {title} {\emph
  {{Observation of topological edge states in parity--time-symmetric quantum
  walks}}},\ }\href {\doibase 10.1038/nphys4204} {\bibfield  {journal}
  {\bibinfo  {journal} {Nat. Phys.}\ }\textbf {\bibinfo {volume} {13}},\
  \bibinfo {pages} {1117} (\bibinfo {year} {2017})}\BibitemShut {NoStop}%
\bibitem [{\citenamefont {Zhan}\ \emph {et~al.}(2017)\citenamefont {Zhan},
  \citenamefont {Xiao}, \citenamefont {Bian}, \citenamefont {Wang},
  \citenamefont {Qiu}, \citenamefont {Sanders}, \citenamefont {Yi},\ and\
  \citenamefont {Xue}}]{Zhan2017}%
  \BibitemOpen
  \bibfield  {author} {\bibinfo {author} {\bibfnamefont {X.}~\bibnamefont
  {Zhan}}, \bibinfo {author} {\bibfnamefont {L.}~\bibnamefont {Xiao}}, \bibinfo
  {author} {\bibfnamefont {Z.}~\bibnamefont {Bian}}, \bibinfo {author}
  {\bibfnamefont {K.}~\bibnamefont {Wang}}, \bibinfo {author} {\bibfnamefont
  {X.}~\bibnamefont {Qiu}}, \bibinfo {author} {\bibfnamefont {B.~C.}\
  \bibnamefont {Sanders}}, \bibinfo {author} {\bibfnamefont {W.}~\bibnamefont
  {Yi}}, \ and\ \bibinfo {author} {\bibfnamefont {P.}~\bibnamefont {Xue}},\
  }\bibfield  {title} {\bibinfo {title} {\emph {{Detecting Topological
  Invariants in Nonunitary Discrete-Time Quantum Walks}}},\ }\href {\doibase
  10.1103/PhysRevLett.119.130501} {\bibfield  {journal} {\bibinfo  {journal}
  {Phys. Rev. Lett.}\ }\textbf {\bibinfo {volume} {119}},\ \bibinfo {pages}
  {130501} (\bibinfo {year} {2017})},\ \Eprint
  {http://arxiv.org/abs/1709.10287} {arXiv:1709.10287} \BibitemShut {NoStop}%
\bibitem [{\citenamefont {Khanikaev}\ and\ \citenamefont
  {Shvets}(2017)}]{Khanikaev2017}%
  \BibitemOpen
  \bibfield  {author} {\bibinfo {author} {\bibfnamefont {A.~B.}\ \bibnamefont
  {Khanikaev}}\ and\ \bibinfo {author} {\bibfnamefont {G.}~\bibnamefont
  {Shvets}},\ }\bibfield  {title} {\bibinfo {title} {\emph {{Two-dimensional
  topological photonics}}},\ }\href {\doibase 10.1038/s41566-017-0048-5}
  {\bibfield  {journal} {\bibinfo  {journal} {Nat. Photon.}\ }\textbf {\bibinfo
  {volume} {11}},\ \bibinfo {pages} {763} (\bibinfo {year} {2017})}\BibitemShut
  {NoStop}%
\bibitem [{\citenamefont {Ozawa}\ \emph {et~al.}(2019)\citenamefont {Ozawa},
  \citenamefont {Price}, \citenamefont {Amo}, \citenamefont {Goldman},
  \citenamefont {Hafezi}, \citenamefont {Lu}, \citenamefont {Rechtsman},
  \citenamefont {Schuster}, \citenamefont {Simon}, \citenamefont {Zilberberg},\
  and\ \citenamefont {Carusotto}}]{Ozawa2018}%
  \BibitemOpen
  \bibfield  {author} {\bibinfo {author} {\bibfnamefont {T.}~\bibnamefont
  {Ozawa}}, \bibinfo {author} {\bibfnamefont {H.~M.}\ \bibnamefont {Price}},
  \bibinfo {author} {\bibfnamefont {A.}~\bibnamefont {Amo}}, \bibinfo {author}
  {\bibfnamefont {N.}~\bibnamefont {Goldman}}, \bibinfo {author} {\bibfnamefont
  {M.}~\bibnamefont {Hafezi}}, \bibinfo {author} {\bibfnamefont
  {L.}~\bibnamefont {Lu}}, \bibinfo {author} {\bibfnamefont {M.~C.}\
  \bibnamefont {Rechtsman}}, \bibinfo {author} {\bibfnamefont {D.}~\bibnamefont
  {Schuster}}, \bibinfo {author} {\bibfnamefont {J.}~\bibnamefont {Simon}},
  \bibinfo {author} {\bibfnamefont {O.}~\bibnamefont {Zilberberg}}, \ and\
  \bibinfo {author} {\bibfnamefont {I.}~\bibnamefont {Carusotto}},\ }\bibfield
  {title} {\bibinfo {title} {\emph {{Topological photonics}}},\ }\href
  {\doibase 10.1103/RevModPhys.91.015006} {\bibfield  {journal} {\bibinfo
  {journal} {Reviews of Modern Physics}\ }\textbf {\bibinfo {volume} {91}},\
  \bibinfo {pages} {015006} (\bibinfo {year} {2019})},\ \Eprint
  {http://arxiv.org/abs/1802.04173} {arXiv:1802.04173} \BibitemShut {NoStop}%
\bibitem [{\citenamefont {Aidelsburger}\ \emph {et~al.}(2018)\citenamefont
  {Aidelsburger}, \citenamefont {Nascimbene},\ and\ \citenamefont
  {Goldman}}]{Aidelsburger_2018}%
  \BibitemOpen
  \bibfield  {author} {\bibinfo {author} {\bibfnamefont {M.}~\bibnamefont
  {Aidelsburger}}, \bibinfo {author} {\bibfnamefont {S.}~\bibnamefont
  {Nascimbene}}, \ and\ \bibinfo {author} {\bibfnamefont {N.}~\bibnamefont
  {Goldman}},\ }\bibfield  {title} {\bibinfo {title} {\emph {Artificial gauge
  fields in materials and engineered systems}},\ }\href {\doibase
  10.1016/j.crhy.2018.03.002} {\bibfield  {journal} {\bibinfo  {journal} {C. R.
  Phys.}\ }\textbf {\bibinfo {volume} {19}},\ \bibinfo {pages} {394} (\bibinfo
  {year} {2018})}\BibitemShut {NoStop}%
\bibitem [{\citenamefont {Chen}\ \emph {et~al.}(2018)\citenamefont {Chen},
  \citenamefont {Ding}, \citenamefont {Qin}, \citenamefont {He}, \citenamefont
  {Luo}, \citenamefont {Chen}, \citenamefont {Liu}, \citenamefont {Wang},
  \citenamefont {Zhang}, \citenamefont {Li}, \citenamefont {You}, \citenamefont
  {Wang}, \citenamefont {Wang}, \citenamefont {Sanders}, \citenamefont {Lu},\
  and\ \citenamefont {Pan}}]{Chen2018}%
  \BibitemOpen
  \bibfield  {author} {\bibinfo {author} {\bibfnamefont {C.}~\bibnamefont
  {Chen}}, \bibinfo {author} {\bibfnamefont {X.}~\bibnamefont {Ding}}, \bibinfo
  {author} {\bibfnamefont {J.}~\bibnamefont {Qin}}, \bibinfo {author}
  {\bibfnamefont {Y.}~\bibnamefont {He}}, \bibinfo {author} {\bibfnamefont
  {Y.-H.}\ \bibnamefont {Luo}}, \bibinfo {author} {\bibfnamefont {M.-C.}\
  \bibnamefont {Chen}}, \bibinfo {author} {\bibfnamefont {C.}~\bibnamefont
  {Liu}}, \bibinfo {author} {\bibfnamefont {X.-L.}\ \bibnamefont {Wang}},
  \bibinfo {author} {\bibfnamefont {W.-J.}\ \bibnamefont {Zhang}}, \bibinfo
  {author} {\bibfnamefont {H.}~\bibnamefont {Li}}, \bibinfo {author}
  {\bibfnamefont {L.-X.}\ \bibnamefont {You}}, \bibinfo {author} {\bibfnamefont
  {Z.}~\bibnamefont {Wang}}, \bibinfo {author} {\bibfnamefont {D.-W.}\
  \bibnamefont {Wang}}, \bibinfo {author} {\bibfnamefont {B.~C.}\ \bibnamefont
  {Sanders}}, \bibinfo {author} {\bibfnamefont {C.-Y.}\ \bibnamefont {Lu}}, \
  and\ \bibinfo {author} {\bibfnamefont {J.-W.}\ \bibnamefont {Pan}},\
  }\bibfield  {title} {\bibinfo {title} {\emph {{Observation of Topologically
  Protected Edge States in a Photonic Two-Dimensional Quantum Walk}}},\ }\href
  {\doibase 10.1103/PhysRevLett.121.100502} {\bibfield  {journal} {\bibinfo
  {journal} {Phys. Rev. Lett.}\ }\textbf {\bibinfo {volume} {121}},\ \bibinfo
  {pages} {100502} (\bibinfo {year} {2018})}\BibitemShut {NoStop}%
\bibitem [{\citenamefont {Chalabi}\ \emph {et~al.}(2019)\citenamefont
  {Chalabi}, \citenamefont {Barik}, \citenamefont {Mittal}, \citenamefont
  {Murphy}, \citenamefont {Hafezi},\ and\ \citenamefont {Waks}}]{Chalabi2019}%
  \BibitemOpen
  \bibfield  {author} {\bibinfo {author} {\bibfnamefont {H.}~\bibnamefont
  {Chalabi}}, \bibinfo {author} {\bibfnamefont {S.}~\bibnamefont {Barik}},
  \bibinfo {author} {\bibfnamefont {S.}~\bibnamefont {Mittal}}, \bibinfo
  {author} {\bibfnamefont {T.~E.}\ \bibnamefont {Murphy}}, \bibinfo {author}
  {\bibfnamefont {M.}~\bibnamefont {Hafezi}}, \ and\ \bibinfo {author}
  {\bibfnamefont {E.}~\bibnamefont {Waks}},\ }\bibfield  {title} {\bibinfo
  {title} {\emph {{Synthetic Gauge Field for Two-Dimensional Time-Multiplexed
  Quantum Random Walks}}},\ }\href {\doibase 10.1103/PhysRevLett.123.150503}
  {\bibfield  {journal} {\bibinfo  {journal} {Physical Review Letters}\
  }\textbf {\bibinfo {volume} {123}},\ \bibinfo {pages} {150503} (\bibinfo
  {year} {2019})},\ \Eprint {http://arxiv.org/abs/1902.06331}
  {arXiv:1902.06331} \BibitemShut {NoStop}%
\bibitem [{\citenamefont {Wang}\ \emph {et~al.}(2018)\citenamefont {Wang},
  \citenamefont {Chen},\ and\ \citenamefont {Zhang}}]{Wang2018b}%
  \BibitemOpen
  \bibfield  {author} {\bibinfo {author} {\bibfnamefont {B.}~\bibnamefont
  {Wang}}, \bibinfo {author} {\bibfnamefont {T.}~\bibnamefont {Chen}}, \ and\
  \bibinfo {author} {\bibfnamefont {X.}~\bibnamefont {Zhang}},\ }\bibfield
  {title} {\bibinfo {title} {\emph {{Experimental Observation of Topologically
  Protected Bound States with Vanishing Chern Numbers in a Two-Dimensional
  Quantum Walk}}},\ }\href {\doibase 10.1103/PhysRevLett.121.100501} {\bibfield
   {journal} {\bibinfo  {journal} {Phys. Rev. Lett.}\ }\textbf {\bibinfo
  {volume} {121}},\ \bibinfo {pages} {100501} (\bibinfo {year}
  {2018})}\BibitemShut {NoStop}%
\bibitem [{\citenamefont {Tang}\ \emph
  {et~al.}(2018{\natexlab{a}})\citenamefont {Tang}, \citenamefont {Lin},
  \citenamefont {Feng}, \citenamefont {Chen}, \citenamefont {Gao},
  \citenamefont {Sun}, \citenamefont {Wang}, \citenamefont {Lai}, \citenamefont
  {Xu}, \citenamefont {Wang}, \citenamefont {Qiao}, \citenamefont {Yang},\ and\
  \citenamefont {Jin}}]{Tang2018}%
  \BibitemOpen
  \bibfield  {author} {\bibinfo {author} {\bibfnamefont {H.}~\bibnamefont
  {Tang}}, \bibinfo {author} {\bibfnamefont {X.-F.}\ \bibnamefont {Lin}},
  \bibinfo {author} {\bibfnamefont {Z.}~\bibnamefont {Feng}}, \bibinfo {author}
  {\bibfnamefont {J.-Y.}\ \bibnamefont {Chen}}, \bibinfo {author}
  {\bibfnamefont {J.}~\bibnamefont {Gao}}, \bibinfo {author} {\bibfnamefont
  {K.}~\bibnamefont {Sun}}, \bibinfo {author} {\bibfnamefont {C.-Y.}\
  \bibnamefont {Wang}}, \bibinfo {author} {\bibfnamefont {P.-C.}\ \bibnamefont
  {Lai}}, \bibinfo {author} {\bibfnamefont {X.-Y.}\ \bibnamefont {Xu}},
  \bibinfo {author} {\bibfnamefont {Y.}~\bibnamefont {Wang}}, \bibinfo {author}
  {\bibfnamefont {L.-F.}\ \bibnamefont {Qiao}}, \bibinfo {author}
  {\bibfnamefont {A.-L.}\ \bibnamefont {Yang}}, \ and\ \bibinfo {author}
  {\bibfnamefont {X.-M.}\ \bibnamefont {Jin}},\ }\bibfield  {title} {\bibinfo
  {title} {\emph {{Experimental two-dimensional quantum walk on a photonic
  chip}}},\ }\href {\doibase 10.1126/sciadv.aat3174} {\bibfield  {journal}
  {\bibinfo  {journal} {Sci. Adv.}\ }\textbf {\bibinfo {volume} {4}},\ \bibinfo
  {pages} {eaat3174} (\bibinfo {year} {2018}{\natexlab{a}})},\ \Eprint
  {http://arxiv.org/abs/1704.08242} {arXiv:1704.08242} \BibitemShut {NoStop}%
\bibitem [{\citenamefont {Tang}\ \emph
  {et~al.}(2018{\natexlab{b}})\citenamefont {Tang}, \citenamefont {{Di
  Franco}}, \citenamefont {Shi}, \citenamefont {He}, \citenamefont {Feng},
  \citenamefont {Gao}, \citenamefont {Sun}, \citenamefont {Li}, \citenamefont
  {Jiao}, \citenamefont {Wang}, \citenamefont {Kim},\ and\ \citenamefont
  {Jin}}]{Tang2018b}%
  \BibitemOpen
  \bibfield  {author} {\bibinfo {author} {\bibfnamefont {H.}~\bibnamefont
  {Tang}}, \bibinfo {author} {\bibfnamefont {C.}~\bibnamefont {{Di Franco}}},
  \bibinfo {author} {\bibfnamefont {Z.-Y.}\ \bibnamefont {Shi}}, \bibinfo
  {author} {\bibfnamefont {T.-S.}\ \bibnamefont {He}}, \bibinfo {author}
  {\bibfnamefont {Z.}~\bibnamefont {Feng}}, \bibinfo {author} {\bibfnamefont
  {J.}~\bibnamefont {Gao}}, \bibinfo {author} {\bibfnamefont {K.}~\bibnamefont
  {Sun}}, \bibinfo {author} {\bibfnamefont {Z.-M.}\ \bibnamefont {Li}},
  \bibinfo {author} {\bibfnamefont {Z.-Q.}\ \bibnamefont {Jiao}}, \bibinfo
  {author} {\bibfnamefont {T.-Y.}\ \bibnamefont {Wang}}, \bibinfo {author}
  {\bibfnamefont {M.~S.}\ \bibnamefont {Kim}}, \ and\ \bibinfo {author}
  {\bibfnamefont {X.-M.}\ \bibnamefont {Jin}},\ }\bibfield  {title} {\bibinfo
  {title} {\emph {{Experimental quantum fast hitting on hexagonal graphs}}},\
  }\href {\doibase 10.1038/s41566-018-0282-5} {\bibfield  {journal} {\bibinfo
  {journal} {Nat. Photon.}\ }\textbf {\bibinfo {volume} {32}},\ \bibinfo
  {pages} {534} (\bibinfo {year} {2018}{\natexlab{b}})},\ \Eprint
  {http://arxiv.org/abs/1807.06625} {arXiv:1807.06625} \BibitemShut {NoStop}%
\bibitem [{\citenamefont {Marrucci}\ \emph {et~al.}(2006)\citenamefont
  {Marrucci}, \citenamefont {Manzo},\ and\ \citenamefont
  {Paparo}}]{Marrucci2006}%
  \BibitemOpen
  \bibfield  {author} {\bibinfo {author} {\bibfnamefont {L.}~\bibnamefont
  {Marrucci}}, \bibinfo {author} {\bibfnamefont {C.}~\bibnamefont {Manzo}}, \
  and\ \bibinfo {author} {\bibfnamefont {D.}~\bibnamefont {Paparo}},\
  }\bibfield  {title} {\bibinfo {title} {\emph {Optical Spin-to-Orbital Angular
  Momentum Conversion in Inhomogeneous Anisotropic Media}},\ }\href {\doibase
  10.1103/PhysRevLett.96.163905} {\bibfield  {journal} {\bibinfo  {journal}
  {Phys. Rev. Lett.}\ }\textbf {\bibinfo {volume} {96}},\ \bibinfo {pages}
  {163905} (\bibinfo {year} {2006})}\BibitemShut {NoStop}%
\bibitem [{\citenamefont {Rubano}\ \emph {et~al.}(2019)\citenamefont {Rubano},
  \citenamefont {Cardano}, \citenamefont {Piccirillo},\ and\ \citenamefont
  {Marrucci}}]{Rubano2019}%
  \BibitemOpen
  \bibfield  {author} {\bibinfo {author} {\bibfnamefont {A.}~\bibnamefont
  {Rubano}}, \bibinfo {author} {\bibfnamefont {F.}~\bibnamefont {Cardano}},
  \bibinfo {author} {\bibfnamefont {B.}~\bibnamefont {Piccirillo}}, \ and\
  \bibinfo {author} {\bibfnamefont {L.}~\bibnamefont {Marrucci}},\ }\bibfield
  {title} {\bibinfo {title} {\emph {{Q-plate technology: a progress review
  [Invited]}}},\ }\href {\doibase 10.1364/JOSAB.36.000D70} {\bibfield
  {journal} {\bibinfo  {journal} {JOSA B}\ }\textbf {\bibinfo {volume} {36}},\
  \bibinfo {pages} {D70} (\bibinfo {year} {2019})}\BibitemShut {NoStop}%
\bibitem [{\citenamefont {{Di Franco}}\ \emph
  {et~al.}(2011{\natexlab{a}})\citenamefont {{Di Franco}}, \citenamefont {{Mc
  Gettrick}},\ and\ \citenamefont {Busch}}]{DiFranco2011}%
  \BibitemOpen
  \bibfield  {author} {\bibinfo {author} {\bibfnamefont {C.}~\bibnamefont {{Di
  Franco}}}, \bibinfo {author} {\bibfnamefont {M.}~\bibnamefont {{Mc
  Gettrick}}}, \ and\ \bibinfo {author} {\bibfnamefont {T.}~\bibnamefont
  {Busch}},\ }\bibfield  {title} {\bibinfo {title} {\emph {{Mimicking the
  Probability Distribution of a Two-Dimensional Grover Walk with a Single-Qubit
  Coin}}},\ }\href {\doibase 10.1103/PhysRevLett.106.080502} {\bibfield
  {journal} {\bibinfo  {journal} {Phys. Rev. Lett.}\ }\textbf {\bibinfo
  {volume} {106}},\ \bibinfo {pages} {080502} (\bibinfo {year}
  {2011}{\natexlab{a}})},\ \Eprint {http://arxiv.org/abs/1010.2470}
  {arXiv:1010.2470} \BibitemShut {NoStop}%
\bibitem [{\citenamefont {{Di Franco}}\ \emph
  {et~al.}(2011{\natexlab{b}})\citenamefont {{Di Franco}}, \citenamefont {{Mc
  Gettrick}}, \citenamefont {Machida},\ and\ \citenamefont
  {Busch}}]{DiFranco2011a}%
  \BibitemOpen
  \bibfield  {author} {\bibinfo {author} {\bibfnamefont {C.}~\bibnamefont {{Di
  Franco}}}, \bibinfo {author} {\bibfnamefont {M.}~\bibnamefont {{Mc
  Gettrick}}}, \bibinfo {author} {\bibfnamefont {T.}~\bibnamefont {Machida}}, \
  and\ \bibinfo {author} {\bibfnamefont {T.}~\bibnamefont {Busch}},\ }\bibfield
   {title} {\bibinfo {title} {\emph {{Alternate two-dimensional quantum walk
  with a single-qubit coin}}},\ }\href {\doibase 10.1103/PhysRevA.84.042337}
  {\bibfield  {journal} {\bibinfo  {journal} {Phys. Rev. A}\ }\textbf {\bibinfo
  {volume} {84}},\ \bibinfo {pages} {042337} (\bibinfo {year}
  {2011}{\natexlab{b}})},\ \Eprint {http://arxiv.org/abs/1107.4400}
  {arXiv:1107.4400} \BibitemShut {NoStop}%
\bibitem [{\citenamefont {Aharonov}\ \emph {et~al.}(1993)\citenamefont
  {Aharonov}, \citenamefont {Davidovich},\ and\ \citenamefont
  {Zagury}}]{Aharanov1993}%
  \BibitemOpen
  \bibfield  {author} {\bibinfo {author} {\bibfnamefont {Y.}~\bibnamefont
  {Aharonov}}, \bibinfo {author} {\bibfnamefont {L.}~\bibnamefont
  {Davidovich}}, \ and\ \bibinfo {author} {\bibfnamefont {N.}~\bibnamefont
  {Zagury}},\ }\bibfield  {title} {\bibinfo {title} {\emph {Quantum random
  walks}},\ }\href {\doibase 10.1103/PhysRevA.48.1687} {\bibfield  {journal}
  {\bibinfo  {journal} {Phys. Rev. A}\ }\textbf {\bibinfo {volume} {48}},\
  \bibinfo {pages} {1687} (\bibinfo {year} {1993})}\BibitemShut {NoStop}%
\bibitem [{\citenamefont {Piccirillo}\ \emph {et~al.}(2010)\citenamefont
  {Piccirillo}, \citenamefont {D'Ambrosio}, \citenamefont {Slussarenko},
  \citenamefont {Marrucci},\ and\ \citenamefont {Santamato}}]{Piccirillo2010}%
  \BibitemOpen
  \bibfield  {author} {\bibinfo {author} {\bibfnamefont {B.}~\bibnamefont
  {Piccirillo}}, \bibinfo {author} {\bibfnamefont {V.}~\bibnamefont
  {D'Ambrosio}}, \bibinfo {author} {\bibfnamefont {S.}~\bibnamefont
  {Slussarenko}}, \bibinfo {author} {\bibfnamefont {L.}~\bibnamefont
  {Marrucci}}, \ and\ \bibinfo {author} {\bibfnamefont {E.}~\bibnamefont
  {Santamato}},\ }\bibfield  {title} {\bibinfo {title} {\emph {{Photon
  spin-to-orbital angular momentum conversion via an electrically tunable
  {\$}q{\$}-plate}}},\ }\href {\doibase 10.1063/1.3527083} {\bibfield
  {journal} {\bibinfo  {journal} {Appl. Phys. Lett.}\ }\textbf {\bibinfo
  {volume} {97}},\ \bibinfo {pages} {4085} (\bibinfo {year} {2010})},\ \Eprint
  {http://arxiv.org/abs/1010.4473} {arXiv:1010.4473} \BibitemShut {NoStop}%
\bibitem [{\citenamefont {Qi}\ \emph {et~al.}(2008)\citenamefont {Qi},
  \citenamefont {Hughes},\ and\ \citenamefont {Zhang}}]{Qi2008}%
  \BibitemOpen
  \bibfield  {author} {\bibinfo {author} {\bibfnamefont {X.-L.}\ \bibnamefont
  {Qi}}, \bibinfo {author} {\bibfnamefont {T.~L.}\ \bibnamefont {Hughes}}, \
  and\ \bibinfo {author} {\bibfnamefont {S.-C.}\ \bibnamefont {Zhang}},\
  }\bibfield  {title} {\bibinfo {title} {\emph {Topological field theory of
  time-reversal invariant insulators}},\ }\href {\doibase
  10.1103/PhysRevB.78.195424} {\bibfield  {journal} {\bibinfo  {journal} {Phys.
  Rev. B}\ }\textbf {\bibinfo {volume} {78}},\ \bibinfo {pages} {195424}
  (\bibinfo {year} {2008})}\BibitemShut {NoStop}%
\bibitem [{\citenamefont {Rudner}\ \emph {et~al.}(2013)\citenamefont {Rudner},
  \citenamefont {Lindner}, \citenamefont {Berg},\ and\ \citenamefont
  {Levin}}]{Rudner2013}%
  \BibitemOpen
  \bibfield  {author} {\bibinfo {author} {\bibfnamefont {M.~S.}\ \bibnamefont
  {Rudner}}, \bibinfo {author} {\bibfnamefont {N.~H.}\ \bibnamefont {Lindner}},
  \bibinfo {author} {\bibfnamefont {E.}~\bibnamefont {Berg}}, \ and\ \bibinfo
  {author} {\bibfnamefont {M.}~\bibnamefont {Levin}},\ }\bibfield  {title}
  {\bibinfo {title} {\emph {Anomalous Edge States and the Bulk-Edge
  Correspondence for Periodically Driven Two-Dimensional Systems}},\ }\href
  {\doibase 10.1103/PhysRevX.3.031005} {\bibfield  {journal} {\bibinfo
  {journal} {Phys. Rev. X}\ }\textbf {\bibinfo {volume} {3}},\ \bibinfo {pages}
  {031005} (\bibinfo {year} {2013})}\BibitemShut {NoStop}%
\bibitem [{\citenamefont {Xiao}\ \emph {et~al.}(2010)\citenamefont {Xiao},
  \citenamefont {Chang},\ and\ \citenamefont {Niu}}]{Xiao2010}%
  \BibitemOpen
  \bibfield  {author} {\bibinfo {author} {\bibfnamefont {D.}~\bibnamefont
  {Xiao}}, \bibinfo {author} {\bibfnamefont {M.-C.}\ \bibnamefont {Chang}}, \
  and\ \bibinfo {author} {\bibfnamefont {Q.}~\bibnamefont {Niu}},\ }\bibfield
  {title} {\bibinfo {title} {\emph {Berry phase effects on electronic
  properties}},\ }\href {\doibase 10.1103/RevModPhys.82.1959} {\bibfield
  {journal} {\bibinfo  {journal} {Rev. Mod. Phys.}\ }\textbf {\bibinfo {volume}
  {82}},\ \bibinfo {pages} {1959} (\bibinfo {year} {2010})}\BibitemShut
  {NoStop}%
\bibitem [{\citenamefont {Price}\ \emph {et~al.}(2016)\citenamefont {Price},
  \citenamefont {Zilberberg}, \citenamefont {Ozawa}, \citenamefont
  {Carusotto},\ and\ \citenamefont {Goldman}}]{Price2016}%
  \BibitemOpen
  \bibfield  {author} {\bibinfo {author} {\bibfnamefont {H.~M.}\ \bibnamefont
  {Price}}, \bibinfo {author} {\bibfnamefont {O.}~\bibnamefont {Zilberberg}},
  \bibinfo {author} {\bibfnamefont {T.}~\bibnamefont {Ozawa}}, \bibinfo
  {author} {\bibfnamefont {I.}~\bibnamefont {Carusotto}}, \ and\ \bibinfo
  {author} {\bibfnamefont {N.}~\bibnamefont {Goldman}},\ }\bibfield  {title}
  {\bibinfo {title} {\emph {Measurement of Chern numbers through center-of-mass
  responses}},\ }\href {\doibase 10.1103/PhysRevB.93.245113} {\bibfield
  {journal} {\bibinfo  {journal} {Phys. Rev. B}\ }\textbf {\bibinfo {volume}
  {93}},\ \bibinfo {pages} {245113} (\bibinfo {year} {2016})}\BibitemShut
  {NoStop}%
\bibitem [{\citenamefont {Dauphin}\ and\ \citenamefont
  {Goldman}(2013)}]{Dauphin2013}%
  \BibitemOpen
  \bibfield  {author} {\bibinfo {author} {\bibfnamefont {A.}~\bibnamefont
  {Dauphin}}\ and\ \bibinfo {author} {\bibfnamefont {N.}~\bibnamefont
  {Goldman}},\ }\bibfield  {title} {\bibinfo {title} {\emph {Extracting the
  Chern Number from the Dynamics of a Fermi Gas: Implementing a Quantum Hall
  Bar for Cold Atoms}},\ }\href {\doibase 10.1103/PhysRevLett.111.135302}
  {\bibfield  {journal} {\bibinfo  {journal} {Phys. Rev. Lett.}\ }\textbf
  {\bibinfo {volume} {111}},\ \bibinfo {pages} {135302} (\bibinfo {year}
  {2013})}\BibitemShut {NoStop}%
\bibitem [{\citenamefont {Wang}\ \emph {et~al.}(2019)\citenamefont {Wang},
  \citenamefont {Qiu}, \citenamefont {Xiao}, \citenamefont {Zhan},
  \citenamefont {Bian}, \citenamefont {Yi},\ and\ \citenamefont
  {Xue}}]{Wang2018}%
  \BibitemOpen
  \bibfield  {author} {\bibinfo {author} {\bibfnamefont {K.}~\bibnamefont
  {Wang}}, \bibinfo {author} {\bibfnamefont {X.}~\bibnamefont {Qiu}}, \bibinfo
  {author} {\bibfnamefont {L.}~\bibnamefont {Xiao}}, \bibinfo {author}
  {\bibfnamefont {X.}~\bibnamefont {Zhan}}, \bibinfo {author} {\bibfnamefont
  {Z.}~\bibnamefont {Bian}}, \bibinfo {author} {\bibfnamefont {W.}~\bibnamefont
  {Yi}}, \ and\ \bibinfo {author} {\bibfnamefont {P.}~\bibnamefont {Xue}},\
  }\bibfield  {title} {\bibinfo {title} {\emph {{Simulating Dynamic Quantum
  Phase Transitions in Photonic Quantum Walks}}},\ }\href {\doibase
  10.1103/PhysRevLett.122.020501} {\bibfield  {journal} {\bibinfo  {journal}
  {Phys. Rev. Lett.}\ }\textbf {\bibinfo {volume} {122}},\ \bibinfo {pages}
  {020501} (\bibinfo {year} {2019})},\ \Eprint
  {http://arxiv.org/abs/1806.10871} {arXiv:1806.10871} \BibitemShut {NoStop}%
\bibitem [{\citenamefont {Xu}\ \emph {et~al.}(2018)\citenamefont {Xu},
  \citenamefont {Wang}, \citenamefont {Heyl}, \citenamefont {Budich},
  \citenamefont {Pan}, \citenamefont {Chen}, \citenamefont {Jan}, \citenamefont
  {Sun}, \citenamefont {Xu}, \citenamefont {Han}, \citenamefont {Li},\ and\
  \citenamefont {Guo}}]{Xu2018}%
  \BibitemOpen
  \bibfield  {author} {\bibinfo {author} {\bibfnamefont {X.-Y.}\ \bibnamefont
  {Xu}}, \bibinfo {author} {\bibfnamefont {Q.-Q.}\ \bibnamefont {Wang}},
  \bibinfo {author} {\bibfnamefont {M.}~\bibnamefont {Heyl}}, \bibinfo {author}
  {\bibfnamefont {J.~C.}\ \bibnamefont {Budich}}, \bibinfo {author}
  {\bibfnamefont {W.-W.}\ \bibnamefont {Pan}}, \bibinfo {author} {\bibfnamefont
  {Z.}~\bibnamefont {Chen}}, \bibinfo {author} {\bibfnamefont {M.}~\bibnamefont
  {Jan}}, \bibinfo {author} {\bibfnamefont {K.}~\bibnamefont {Sun}}, \bibinfo
  {author} {\bibfnamefont {J.-S.}\ \bibnamefont {Xu}}, \bibinfo {author}
  {\bibfnamefont {Y.-J.}\ \bibnamefont {Han}}, \bibinfo {author} {\bibfnamefont
  {C.-F.}\ \bibnamefont {Li}}, \ and\ \bibinfo {author} {\bibfnamefont {G.-C.}\
  \bibnamefont {Guo}},\ }\bibfield  {title} {\bibinfo {title} {\emph
  {{Measuring a Dynamical Topological Order Parameter in Quantum Walks}}},\
  }\href {http://arxiv.org/abs/1808.03930} {\bibfield  {journal} {\bibinfo
  {journal} {arXiv:1808.03930}\ } (\bibinfo {year} {2018})}\BibitemShut
  {NoStop}%
\bibitem [{\citenamefont {Heyl}(2018)}]{Heyl2018}%
  \BibitemOpen
  \bibfield  {author} {\bibinfo {author} {\bibfnamefont {M.}~\bibnamefont
  {Heyl}},\ }\bibfield  {title} {\bibinfo {title} {\emph {{Dynamical quantum
  phase transitions: a review}}},\ }\href {\doibase 10.1088/1361-6633/aaaf9a}
  {\bibfield  {journal} {\bibinfo  {journal} {Rep. Prog. Phys.}\ }\textbf
  {\bibinfo {volume} {81}},\ \bibinfo {pages} {054001} (\bibinfo {year}
  {2018})},\ \Eprint {http://arxiv.org/abs/1701.08851} {arXiv:1701.08851}
  \BibitemShut {NoStop}%
\bibitem [{\citenamefont {Longhi}(2017)}]{Longhi2017}%
  \BibitemOpen
  \bibfield  {author} {\bibinfo {author} {\bibfnamefont {S.}~\bibnamefont
  {Longhi}},\ }\bibfield  {title} {\bibinfo {title} {\emph {Parity-time
  symmetry meets photonics: A new twist in non-Hermitian optics}},\ }\href
  {\doibase 10.1209/0295-5075/120/64001} {\bibfield  {journal} {\bibinfo
  {journal} {{EPL} (Europhysics Letters)}\ }\textbf {\bibinfo {volume} {120}},\
  \bibinfo {pages} {64001} (\bibinfo {year} {2017})}\BibitemShut {NoStop}%
\bibitem [{\citenamefont {Yao}\ \emph {et~al.}(2018)\citenamefont {Yao},
  \citenamefont {Song},\ and\ \citenamefont {Wang}}]{Yao2018}%
  \BibitemOpen
  \bibfield  {author} {\bibinfo {author} {\bibfnamefont {S.}~\bibnamefont
  {Yao}}, \bibinfo {author} {\bibfnamefont {F.}~\bibnamefont {Song}}, \ and\
  \bibinfo {author} {\bibfnamefont {Z.}~\bibnamefont {Wang}},\ }\bibfield
  {title} {\bibinfo {title} {\emph {{Non-Hermitian Chern Bands}}},\ }\href
  {\doibase 10.1103/PhysRevLett.121.136802} {\bibfield  {journal} {\bibinfo
  {journal} {Physical Review Letters}\ }\textbf {\bibinfo {volume} {121}},\
  \bibinfo {pages} {136802} (\bibinfo {year} {2018})},\ \Eprint
  {http://arxiv.org/abs/arXiv:1804.04672v1} {arXiv:arXiv:1804.04672v1}
  \BibitemShut {NoStop}%
\bibitem [{\citenamefont {Gong}\ \emph {et~al.}(2018)\citenamefont {Gong},
  \citenamefont {Ashida}, \citenamefont {Kawabata}, \citenamefont {Takasan},
  \citenamefont {Higashikawa},\ and\ \citenamefont {Ueda}}]{Gong2018}%
  \BibitemOpen
  \bibfield  {author} {\bibinfo {author} {\bibfnamefont {Z.}~\bibnamefont
  {Gong}}, \bibinfo {author} {\bibfnamefont {Y.}~\bibnamefont {Ashida}},
  \bibinfo {author} {\bibfnamefont {K.}~\bibnamefont {Kawabata}}, \bibinfo
  {author} {\bibfnamefont {K.}~\bibnamefont {Takasan}}, \bibinfo {author}
  {\bibfnamefont {S.}~\bibnamefont {Higashikawa}}, \ and\ \bibinfo {author}
  {\bibfnamefont {M.}~\bibnamefont {Ueda}},\ }\bibfield  {title} {\bibinfo
  {title} {\emph {{Topological Phases of Non-Hermitian Systems}}},\ }\href
  {\doibase 10.1103/PhysRevX.8.031079} {\bibfield  {journal} {\bibinfo
  {journal} {Physical Review X}\ }\textbf {\bibinfo {volume} {8}},\ \bibinfo
  {pages} {031079} (\bibinfo {year} {2018})},\ \Eprint
  {http://arxiv.org/abs/1802.07964} {arXiv:1802.07964} \BibitemShut {NoStop}%
\bibitem [{\citenamefont {Yokomizo}\ and\ \citenamefont
  {Murakami}(2019)}]{Yokomizo2019}%
  \BibitemOpen
  \bibfield  {author} {\bibinfo {author} {\bibfnamefont {K.}~\bibnamefont
  {Yokomizo}}\ and\ \bibinfo {author} {\bibfnamefont {S.}~\bibnamefont
  {Murakami}},\ }\bibfield  {title} {\bibinfo {title} {\emph {{Non-Bloch Band
  Theory of Non-Hermitian Systems}}},\ }\href {\doibase
  10.1103/PhysRevLett.123.066404} {\bibfield  {journal} {\bibinfo  {journal}
  {Physical Review Letters}\ }\textbf {\bibinfo {volume} {123}},\ \bibinfo
  {pages} {066404} (\bibinfo {year} {2019})},\ \Eprint
  {http://arxiv.org/abs/1902.10958} {arXiv:1902.10958} \BibitemShut {NoStop}%
\bibitem [{\citenamefont {Nathan}\ and\ \citenamefont
  {Rudner}(2015)}]{Nathan2015}%
  \BibitemOpen
  \bibfield  {author} {\bibinfo {author} {\bibfnamefont {F.}~\bibnamefont
  {Nathan}}\ and\ \bibinfo {author} {\bibfnamefont {M.~S.}\ \bibnamefont
  {Rudner}},\ }\bibfield  {title} {\bibinfo {title} {\emph {Topological
  singularities and the general classification of Floquet--Bloch systems}},\
  }\href {http://stacks.iop.org/1367-2630/17/i=12/a=125014} {\bibfield
  {journal} {\bibinfo  {journal} {New J. Phys.}\ }\textbf {\bibinfo {volume}
  {17}},\ \bibinfo {pages} {125014} (\bibinfo {year} {2015})}\BibitemShut
  {NoStop}%
\end{thebibliography}%

\clearpage

\onecolumngrid
\clearpage
\renewcommand{\theequation}{S\arabic{equation}}
\renewcommand{\thesection}{S\arabic{section}}
\renewcommand{\thefigure}{S\arabic{figure}}
\renewcommand{\theHfigure}{Supplement.\thefigure}
\renewcommand{\figurename}{Supplementary Figure}

\setcounter{equation}{0}
\setcounter{section}{0}
\setcounter{subsection}{0}
\setcounter{figure}{0}
\setcounter{page}{1}

\begin{center}
\textbf{Supplementary Material}
\end{center}

\vspace{1 EM}

\section{Photonic spatial modes 
for the encoding of the walker degree of freedom}\label{supp:spatialmodes}
\subsection{Gaussian modes encoding single walker positions}
The state $\ket{\bmxy,\phi}$ of a walker is encoded in our set-up in a light beam described by Eq.
1 in the main manuscript
\beq
\label{eq:gaussianbeamtiltedSuppMat}
\ket{\bmxy,\phi}=A(x,y,z) e^{i[\Delta k_\perp (m_x x+m_y y) +k_z z]}\otimes \ket{\phi},
\eeq
where $\bmxy = (m_x,m_y)$ are the integer coordinates giving the discrete position of the walker, and the light's polarization $\ket{\phi}$ encodes the state of the coin of the walker.
The spatial profile of the beam is determined by the Gaussian envelope function
\beq
\label{eq: gaussianenvelope}
A(x,y,z)= \frac{w_0}{w(z)}\text{e}^{-\frac{x^2+y^2}{w(z)^2}}\text{e}^{i k\frac{x^2+y^2}{2R(z)}} \text{e}^{-i \xi(z)},
\eeq
where $k=2\pi/\lambda$ is the wavenumber, and the beam radius $w(z)$, the beam curvature $R(z)$ and the Gouy phase $\xi(z)$ are defined as follows:
\begin{align}
w(z)=&w_0\sqrt{1+(z/z_0)^2},\\
R(z)&=z[1+(z_0/z)^2],\\
\xi(z)&=\arctan(z/z_0).
\end{align}
Here $w_0=w(z=0)$ is the beam radius at the waist position $z=0$, and the parameter $z_0=\pi w_0^2/\lambda$ is known as Rayleigh range.
In our experiment, we set $w_0=5$ mm, which yields $z_0 \approx 120$ m, so that across the whole QW setup (about 30 cm long) the beam radius is approximately constant ($w(z)\approx w_0$), and both the Gouy phase $\xi(z)$ and the inverse beam curvature $1/R(z)$ are entirely negligible.

If we place a converging lens at the end of the quantum walk, in the focal plane the field distribution is proportional to the distribution of the transverse wavevector, that is:
\beq
\label{eq:gaussian_FT} 
A(X,Y)\propto \int_{\Omega} A(x, y, d)e^{i[\Delta k_\perp(m_xx+m_yy)+k_zd]} \text{e}^{i (Xx+Yy)k/f}{\rm d}x{\rm d}y,
\eeq
where $\Omega$ is the transverse plane, $d$ the distance of the lens from the beam waist, $f$ the focal length of the lens, and $\bR=(X,Y)$ the spatial coordinates in the focal plane of the lens. It is well known that, independently of the distance $d$, in the focal plane the field intensity is proportional to the Fourier transform of the field impinging on the lens, that is
\begin{align}
\label{eq:int_fourier}
\abs{A(X,Y )}^2\propto \abs{g(k_x,k_y)}^2
\end{align} 
where $g(k_x,k_y)$ is the Fourier transform of $A(x,y)$, provided that one sets $X=k_xf/k$ and $Y=k_yf/k$. In the case of a Gaussian beam, we have 
\begin{align}
\label{eq:int_Ft}
\abs{A(X,Y )}^2\propto \text{e}^{\bigr( -2 \frac{(m_x\Delta k_\perp-kX/f)^2+(m_y\Delta k_\perp-kY/f)^2}{\tilde{w}_k^2}\bigr)},
\end{align} 
with $\tilde{w}_k=2/w_0$. Thus $\tilde{w}_k$ is a measure of the radius of the spots that appear in the focal plane, provided that one \edit{converts the transverse momentum of the photons into a position on the camera. The spatial position  $\bR$ on the camera is related the transverse momentum of light by the relation 
\beq\label{eq:koncamera}
\bR= \frac{f\lambda \,\bk_\perp}{2\pi}.
\eeq
}

\subsection{Extended wavepacket walker states and their optical implementation}
\begin{figure*}[t!]
\centering
\includegraphics[width=0.5\linewidth]{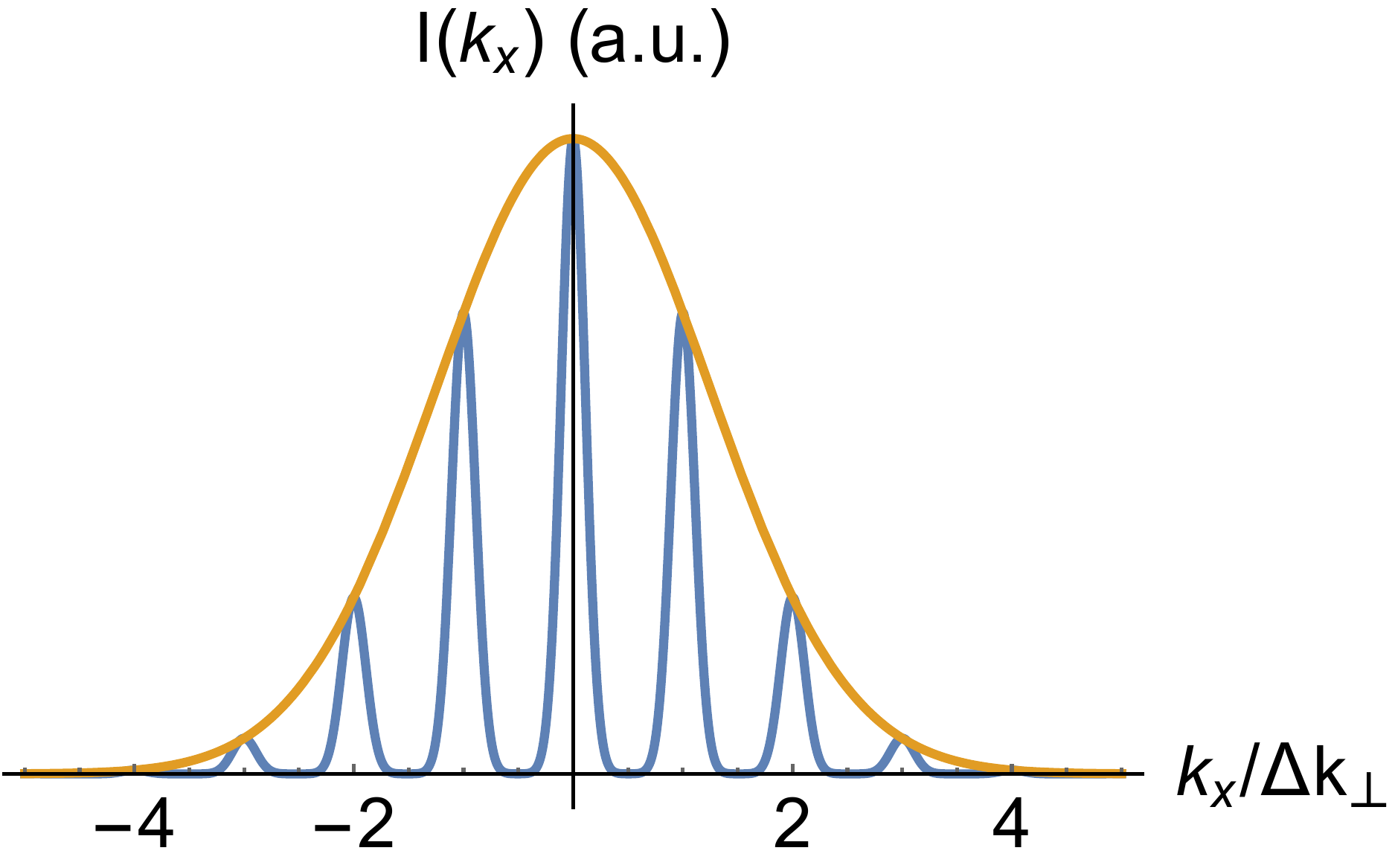}
\caption{{\bf Gaussian wavepackets.} Comparison between the intensity distribution $I(k_x)$ (we set $k_y$=0) of the wavepacket states $\ket{G}$ (blue curve), with $\sigma_G=2.5$, and of a single Gaussian mode $\ket{\Psi_g}$ (orange curve), whose beam radius in Fourier space is $\tilde \sigma_G=\sigma_G\Delta k_\perp$. In this plot, individual modes $\ket{\bmxy}$ contained in $\ket{G}$ are characterized by their actual beam radius $w_k=0.23 \Delta k_\perp$.
}
\label{fig:locvsgauss}
\end{figure*}
As discussed above, attention must be paid to the beam radius of modes $\ket{\bmxy}$. Indeed, once the lattice spacing $\Delta k_\perp$ is fixed, $w_0$ must be selected so that the overlap between adjacent modes is negligible. In our set-up, $\Lambda=5$ mm and our choice of setting the beam waist to $w_0= 5$ mm leads to an overlap between adjacent modes of around 0.8$\%$. In general, initial states other than localized ones can be prepared, such as for instance Gaussian wavepackets $\ket{G}=\mathcal N\sum_{\bmxy}\ket{\bmxy}e^{-(m_x^2+m_y^2)/\sigma_G^2}$, where $\mathcal N$ is a normalization factor and $\sigma_G$ is the width of the overall Gaussian envelope (in dimensionless units). In 
Fig.\ \ref{fig:locvsgauss}, we plot a 1D cut of the corresponding beam-intensity wavevector profile versus $k_x$, at $k_y=0$ (blue curve), showing several peaks modulated by a Gaussian envelope, each peak corresponding to a mode $\ket{\bmxy}$ included in the wavepacket. The QW dynamics of such an input state is equivalent to that of a single Gaussian beam $\ket{\Psi_g}$, whose beam radius is $w_g=2/(\sigma_G\Delta k_\perp)$.
The preparation of such a state is much simpler, since it requires a modulation of the beam radius only, which is simply achieved with a confocal pair of lenses.  In 
Fig.\ \ref{fig:locvsgauss} we provide a comparison between the intensity distribution associated with $\ket{G}$ (blue curve) and $\ket{\Psi_g}$ (orange curve).

In our experiments, we are particularly interested at wavepackets $\mathcal N\sum_{\bmxy}\ket{\bmxy}e^{i\bq_0\cdot\bmxy}e^{-(m_x^2+m_y^2)/\sigma_G^2}$, that include the phase factor $e^{i\bq_0\cdot \bmxy}$. Indeed, in quasi-momentum space these feature a Gaussian distribution with $\tilde \sigma_G=2/\sigma_G$, peaked around a specific quasi-momentum $\bq_0$. Their expression reads
\begin{align}\label{eq: gaussian_wp}
\ket{\Psi_g(\bq_0)}=\mathcal{N'}\int_{\mathrm{BZ}}\frac{d^2\bf{q}}{4\pi^2}e^{-\frac{(\bf{q}-\bf{q}_0)^2}{\tilde \sigma_G^2}}\ket{\bq},
\end{align}
where $\mathcal{N'}$ is a normalization factor, and $\mathrm{BZ}=[-\pi,\pi]^2$ is the Brillouin zone. We want these beams to be sharply peaked, that is $\tilde \sigma_G\ll1$, so that they approximate as much as possibile the individual state $\ket{\bq_0}$. Being the simulated quasi-momentum encoded in the physical transverse position $\br_\perp$, these wavepackets are realized by standard Gaussian beams, whose central position is set to $\br_\perp=-\Lambda\bq_0/(2\pi)$, and which are characterized by a beam radius that is much smaller than the spatial period $\Lambda$.

In the focal plane of the lens, these beams display a continuous distribution, as shown for instance in Fig.\ 
3{\bf b} in the main manuscript. Being 
 sharply peaked in the space of the walker quasi-momentum, we expect them to cover multiple lattice sites in the space of walker position. If one is interested in obtaining the associated walker probability distribution, our standard procedure described in Sec.\ \sref{supp:dataexctraction} can be applied. However, in our experiments, we are interested in detecting the wavepacket center of mass, which can be determined by analyzing directly the overall intensity pattern recorded by the camera.
 
As shown in the main text, we use these beams to prepare photonic states:
\begin{align}\label{eq: gaussian_wp_complete}
\ket{\Psi_g(\bq_0,\pm)}=\ket{\Psi_g(\bq_0)}\otimes\ket{\phi_\pm(\bf{q}_0)}
\end{align}
where the coin part corresponds to the eigenstates $\ket{\phi_\pm(\bq_0)}$ of the effective Hamiltonian. These states are extremely useful to probe the QW dispersion and the associated topological features (see Figs.\ 
3, 4 in the main manuscript).

\begin{figure*}[h!]
\centering
\includegraphics[width=\linewidth]{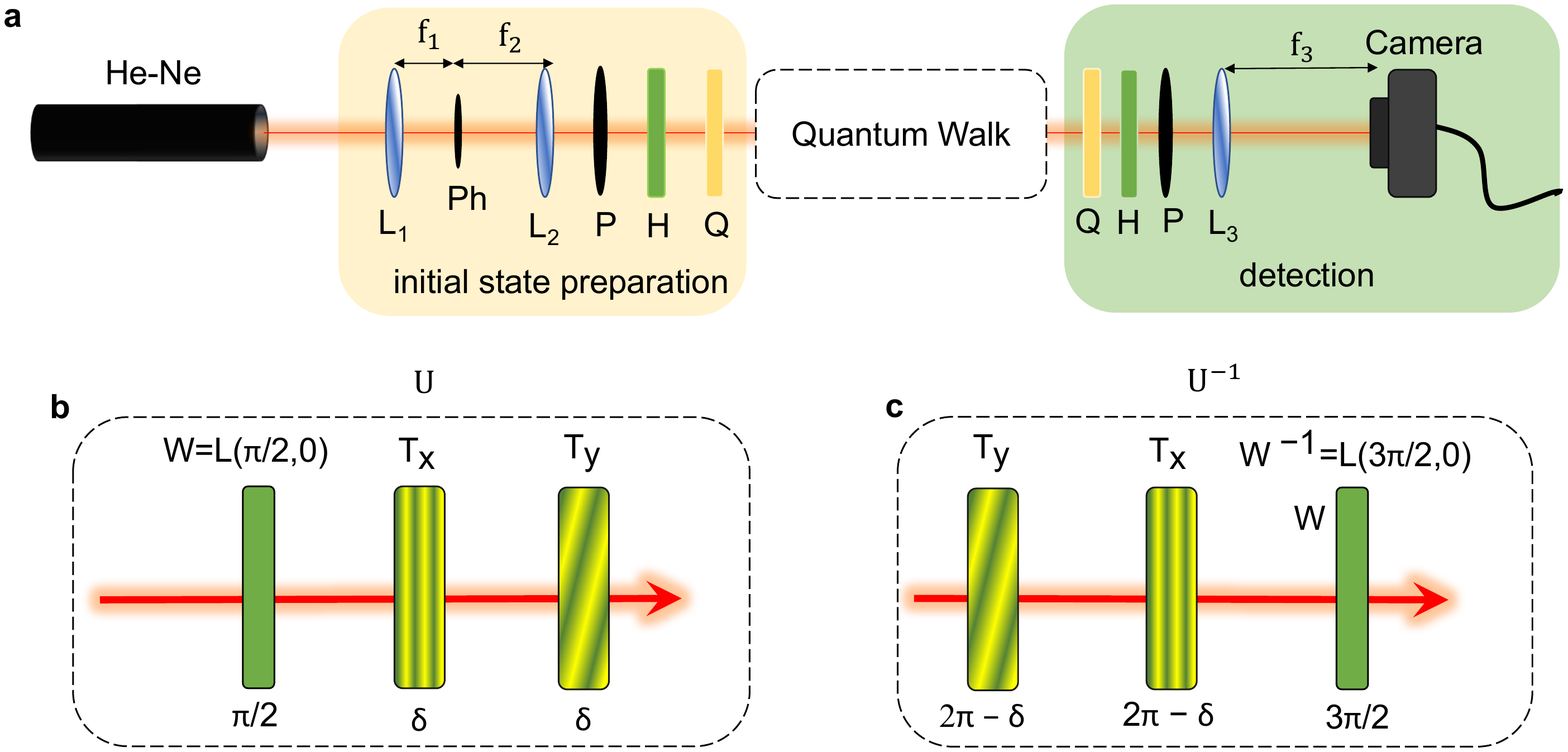}
\caption{{\bf Details of the experimental set-up and protocols.}
{\bf a.} Sketch of the experimental set-up. The waist of a laser beam emitted from a He-Ne laser is modified with a pair of convex lenses $L_1$ and $L_2$, with focal lengths $f_1$ and $f_2$.
 A pinhole (Ph), placed in the common focus of the two lenses, is used as spatial filter to erase higher-order spatial frequencies. The polarization of the input state is selected by means of a polarizer (P), a half-wave plate (H) and a quarter-wave plate (Q). These wave plates are mounted on motorized rotation stages allowing for automatic selection of the coin states. After the QW an additional set of waveplates can be used to analyze the probability distributions of specific polarization components. The probability distribution is visualized by focusing the laser beam  on a camera with a lens $L_3$ of focal length $f_3=50$ cm.
{\bf b.} A single step of our QW with protocol U is obtained by cascading three optical plates: a spin rotation $W$, followed by a coin-conditioned translation along $x$ and another along $y$, with optical retardations $\delta$ as indicated below every device. {\bf c.} Set-up yielding a single step of the inverse protocol $U^{-1}$.
}
\label{fig:suppsetup}
\end{figure*}

\section{Details of the experimental platform}\label{supp:setup}
A complete scheme of the set-up implementing our QW dynamics is reported in 
Fig.\ \ref{fig:suppsetup}{\bf a}. A laser beam is produced by a Helium:Neon source with wavelength $\lambda=632.8$ nm, and propagates along the $\hat{z}$ direction.
 We use a system of two lenses and a pinhole ($L_1$, Ph, $L_2$) to set the beam waist to $w_0=(5.0\pm0.2)$ mm or to $w_0=(0.62\pm0.02)$ mm, depending on our necessity to start the walk with either ``localized states'' or ``extended wave-packets'', respectively (see Sec.\ \ref{supp:spatialmodes} for further details on the spatial features of the light beam). In the last stage of the preparation, a polarizer and two waveplates (P,H,Q) are used to prepare a given polarization.

The beam undergoes the proposed QW dynamics by passing through a sequence of wave-plates and $g$-plates. In panels {\bf b} and {\bf c} we display the combination of plates realizing the protocols $U$ and $U^{-1}$, respectively. All operators are physically implemented by thin optical plates, which allows us to mount them in a very compact mechanical holder realized by a 3D printing technique. The distance between consecutive steps is currently $\simeq 2$ cm, yet it could be significantly reduced by optimizing the thickness of the glass and of the plastic mounts.
Within each plate, the active layer containing liquid crystals (LCs) is 6 $\mu$m thick.
The spatial period of $g$-plates pattern is $\Lambda=5$ mm, yielding a transverse momentum displacement of $\Delta k_\perp=2\pi/\Lambda=1.26 $ (mm)$^{-1}$, that provides the spacing between neighbouring sites in our squared lattice. In order to have each mode $\ket{\bmxy}$ entirely localized on the associated site $(m_x,m_y)$, without ``cross-talk'', the single-mode beam radius $w_0$ must be properly selected. In Fourier space, where the lattice of walker positions is defined, these beams are characterized by a radius $w_k=\sqrt 2/w_0$. If one chooses $w_0 \simeq \Lambda$, one gets that the ratio between the beam radius (in Fourier space) and the lattice spacing $w_k/\Delta k_\perp \simeq1/\pi=0.32$ is sufficiently small and the overlap between adjacent modes is negligible (see Fig.\ 
1{\bf b} in the main manuscript).

At the exit of the walk, two wave-plates and one polarizer (Q,H,P) are used to analyze individual polarization components. Finally a lens ($L_3$) focuses the field on a camera that records the distribution of light intensity, operating an all-optical Fourier transform. In the focal plane, light is spread over several spots, according to the walker distribution over the lattice. A single image contains the overall probability distribution, and the latter can be monitored in real time. Since the Rayleigh range of the input beam is much longer than the total distance of the walk (see Sec.\ \ref{supp:spatialmodes}), the latter takes place in the near field and the beam remains collimated.

\section{Extraction of the probability distributions from the intensity patterns}\label{supp:dataexctraction}
When injecting modes with beam radius $w_0\simeq 5$ mm, at the end of the walk we record the light intensity in the focal plane of the camera, that is distributed over many spots corresponding to the walker lattice sites. The probability distribution of the associated quantum walk can be extracted by measuring the amount of light in each region. In principle, the lattice site positions on the camera could be determined by (i) individuating the axes origin $(0,0)$ (setting all $g$-plates at $\delta=0$, so that a single spot appears on the camera), and (ii) determining the expected positions of the other sites in terms of the spacing $\Delta k_\perp$. However, imperfections of all plates can cause small deviations between actual spot positions and the expected ones. For instance, one contribution can be ascribed to undesired modulations in the $g$-plates patterns, that can be modelled by the local optic axis orientation $\alpha(x,y)=\alpha_0+(\Delta k_\perp/2)x+\epsilon(x,y)$, where $\epsilon(x,y)$ is a small
random error. Another source of errors can be a small tilt in the polarization gratings, so that the coordinate $x$ in $\alpha(x)$ should be replaced by $x'=\cos(\theta)x+\sin(\theta)y$ (with $\theta$ small, and different for each grating).

\begin{figure*}[t!]
\centering
\includegraphics[width=\linewidth]{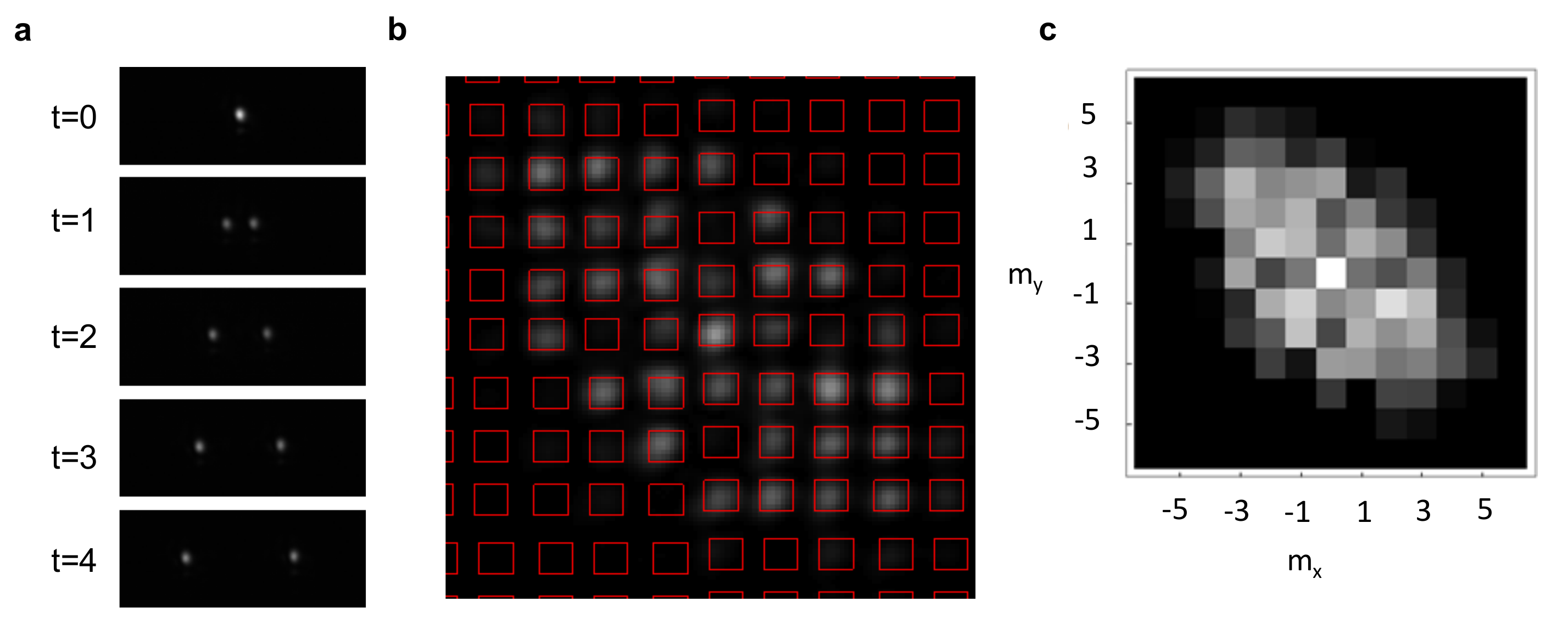}
\caption{{\bf Extracting probability distributions from the recorded intensity patterns.}
\textbf{a}, Sample intensity patterns obtained with the simple protocol $U=S_x(\pi)\cdot H_{wp}$. An input linearly polarized state is split into two spots with opposite circular polarizations. At each step these spots are shifted by a quantity corresponding to the equivalent of a lattice spacing. This process can be used to identify the coordinates of the lattice sites on the camera, getting rid of some of the experimental imperfections explained in the text. Panels \textbf{b} and \textbf{c} show the procedure used to extract the probability distributions from the intensity patterns. The red squares in \textbf{b} represent the regions over which we obtain the  total intensities (= powers) associated to specific lattice sites (since the single spots occupy a small number of pixels there is no substantial difference in using square or circular integration regions). Normalizing to $1$ the sum of all these intensities we obtain the probability distribution shown in panel \textbf{c}. 
}
\label{fig:calibration}
\end{figure*}

To improve the calibration procedure, we follow therefore a different approach, which is illustrated in 
Fig.\ \ref{fig:calibration}.
Let us first consider the 1D set-up defined by the single step operator $U_{x}=T_x(\delta=\pi)\cdot H_{\rm wp}$, where $H_{\rm wp}$ is a half-waveplate (that can be described by the operator $\sigma_x$). 
The particle dynamics, shown in 
Fig.\ \ref{fig:calibration}\textbf{a}, is indeed very simple: at each step the positions of $L/R$ polarized components are shifted respectively by $\pm1$. If we start with a linearly polarized input beam, in the following steps we will see two spots (with opposite circular polarizations), which will be located, at the time $t$, at the effective positions $m_x=t$ and $m_x=-t$, respectively. In this way we reconstruct the coordinates of each site by performing Gaussian fits for the two spots. By repeating the same analysis with the protocol $U_y=T_yW$ we measure the $y$ coordinate of each site. In the actual set-up, we can realize both protocols $U_x$ and $U_y$ by turning off ($\delta=0$) plates $T_x$ or $T_y$. After the site coordinates have been determined, we draw squared regions around each point (see 
Fig.~\ref{fig:calibration}\textbf{b}). Light detected within one of these regions is automatically associated to the corresponding lattice sites. By integrating the light intensity measured within each square, and by dividing each of these values by their total sum, we obtain a properly normalized probability distribution for the walker position (see 
Fig.~\ref{fig:calibration}\textbf{c}). 

\section{Operators in quasi-momentum space and displacement of a wavepacket in the presence of a constant force}
\label{supp:anomalous}

We analyze the building blocks of the quantum walk in the reciprocal quasi-momentum space. For the W operator defined in Eq.~
5 of the main text, the expression remains the same, as it does not depend on the position:

\begin{align}
W &= e^{i \frac{\pi}{4}\sigma_x}=
\frac{1}{\sqrt{2}} \left( \begin{array}{cc} 1 & i \\ i & 1 \end{array} \right),
\end{align}
where the basis of the polarization space has been chosen to be $\left\lbrace\ket{L},\ket{R}\right\rbrace$.

The operator  $T_{x}$ \edit{can be obtained by the operator described in Eq.\ 2 in the main text}. By inserting the explicit expression of the angle $\alpha(x)=x\pi/\Lambda +\alpha_0$, one obtains the expression of $T_{x}$ in momentum space:
\begin{align}
T_x(q_{x}) &= e^{i \frac{\delta}{2}(\cos(q_x) \sigma_x -\sin(q_x) \sigma_y )}= \nonumber\\
&=\left(
\begin{array}{cc}
 \cos (\delta/2) & i e^{i q_x} \sin (\delta/2) \\
 i e^{-i q_x }  \sin (\delta/2) & \cos (\delta/2) \\
\end{array}
\right),
\end{align}
and similarly for $T_y$:
\begin{align}
T_y(q_{y}) &= e^{i \frac{\delta}{2}(\cos(q_y) \sigma_x -\sin(q_y) \sigma_y )}= \nonumber \\
&=\left(
\begin{array}{cc}
 \cos (\delta/2) & i e^{i q_y} \sin (\delta/2) \\
 i e^{-i q_y}  \sin (\delta/2) & \cos (\delta/2) \\
\end{array}
\right),
\end{align}
where we used the mapping $q_{x}= \frac{- 2\pi \, x}{\Lambda}, q_{y}= \frac{- 2\pi \, y}{\Lambda}$ and we set $\alpha_0=0$. This allows one to use the standard convention for the normalized plane waves:
\beq
\bra{{\bf m}}{\bf q}\rangle =\frac{e^{i {\bf q}\cdot{\bf m}}}{2\pi},
\eeq
where ${\bf q}$ is the quasi-momentum of the walker and ${\bf m}$ is its position on the 2D lattice, according to the convention chosen in the paper.\\ 

\edit{As discussed in the main text, in quasi-momentum space the single step operator is 
\begin{align}
U(q_x,q_y,\delta)=T_y(q_y, \delta) T_x(q_x,\delta) W=\exp\left(-i H_{\mathrm{eff}}(q_x,q_y,\delta)\right), 
\end{align}
where $H_{\mathrm{eff}} ({\bf q})=\varepsilon({\bf q}) {\bf n}({\bf q}) \cdot \sigma$ is the effective Hamiltonian, $\varepsilon({\bf q})$ is the quasi-energy and ${\bf n}({\bf q})$ is a unit vector representing the system eigenstates on the Bloch sphere. The quasi-energy $\varepsilon({\bf q})$ is given by:
\begin{align}\label{eq:qenergy}
\cos\varepsilon= \frac{1}{\sqrt{2}}
\left(A^2-A B(\cos(q_x)+\cos(q_y))-B^2\cos(q_x-q_y) \right)
\end{align} 
while the components of ${\bf n}({\bf q})$ are
\begin{align}\label{eq:eigenstates}
n_x&=\frac{-A^2- A B\left(\cos(q_x)+\cos(q_y)\right)+B^2\cos(q_x-q_y)}{\sqrt{2}\sin\varepsilon}\nonumber\\
n_y&=\frac{-A B\left(\sin(q_x)+\sin(q_y)\right)+B^2\sin(q_x-q_y)}{\sqrt{2}\sin\varepsilon}\nonumber\\
n_z&=\frac{-A B\left(\sin(q_x)+\sin(q_y)\right)-B^2\sin(q_x-q_y)}{\sqrt{2}\sin\varepsilon}.
\end{align}
In these equations we defined the quantities $A=\cos(\delta/2)$ and $B=\sin(\delta/2)$. By looking at Eqs.\ \ref{eq:qenergy}-\ref{eq:eigenstates} it is clear that the effective Hamiltonian cannot be expressed as the sum of two terms acting along orthogonal directions, i.e., it is not separable.}

The operator implementing the potential of the constant dimensionless force, $F_x \hat{m}_x$, can be regarded as a translation of the walker's quasi-momentum component $q_x$ of a quantity $F_x$ at each step. This operation is nondiagonal in momentum space, so the step operator in momentum space is described by $\bra{\bq'}\tilde{U}(t)\ket{\bq} = \tilde{U}(\bq',\bq,t)$, where $\tilde{U}(\bq',\bq,t)$ is a $2\times2$ matrix operating in coin space only. In turn, the latter is given by $\tilde{U}({\bq',\bq},t) = \delta(q'_x-q_x+F_xt)\delta(q'_y-q_y)U(q_x+F_x t,q_y)$, where $\delta(\cdot)$ denotes a Dirac delta function and $U = e^{-iH_{\mathrm{eff}}}$ is the step operator in coin space for a given quasi-momentum $\bq=(q_x,q_y)$ for a vanishing force. We also have
\begin{align}
U(q_x+F_x t,q_y)=T_y(q_{y}) e^{i t\frac{F_x}{2}\sigma_z}T_x(q_{x})e^{-it\frac{F_x}{2}\sigma_z} W.
\end{align}
\edit{The operator $e^{it\frac{F_x}{2}\sigma_z}T_x(\delta,\alpha_0)e^{-it\frac{F_x}{2}\sigma_z}=T_x(\delta,\alpha_0-tF_x/2) = L(\delta, \pi x/\Lambda+\alpha_0-tF_x/2)$ is obtained by shifting a $g$-plate by $\Delta x=tF_x\Lambda/(2\pi)$ along the $x$ axis, which corresponds to the transformation $x\rightarrow x-\Delta x$. The same reference-system transformation realizes also the operation $q_x \rightarrow q_x+F_xt$ in quasi-momentum space.}\\


In the adiabatic limit within the single band approximation, the semi-classical equations of motion of a wave-packet initially peaked around an energy eigenstate $ e^{i {\bf q}_{0}\cdot{\bf m}_{0}}\ket{\phi_{\pm}({\bf q}_{0})} $ read~\cite{
Price2016,Dauphin2013}
\begin{align}
\dot{m}_{i}&=\partial_{q_{i}}\varepsilon_{\pm}({\bf q})-\dot{q}_{j} \Omega^{(\pm)}_{ij}({\bf q}), \\ \nonumber
\dot{q}_{i}&=F_{i}.
\end{align}
Here $\{i,j\} \in \{ x,y \}$, $\pm$ denote the upper/lower band, $\varepsilon_{\pm}(\mathbf{q})=\pm\varepsilon(\mathbf{q})$ is the quasi-energy dispersion and $\Omega^{(\pm)}_{ji}(\mathbf{q})=-\Omega^{(\pm)}_{ij}(\mathbf{q})$ is the Berry curvature

\beq\label{eq:IQH_curv}
 \Omega^{(\pm)}_{xy} (\textbf{q})= i \left[\partial_{q_{x}} \bra{\phi_\pm({\bf q})}\partial_{q_{y}}\phi_\pm({\bf q})\rangle - \partial_{q_{y}} \bra{\phi_\pm({\bf q})}\partial_{q_{x}}\phi_\pm({\bf q})\rangle \right],
\eeq
where the $\ket{\phi_\pm}$ are the eigenvectors of the Bloch effective Hamiltonian $H_{\rm eff}$
\beq
H_{\rm eff}({\bf q}) \ket{\phi_\pm({\bf q})}= \varepsilon_{\pm} ({\bf q})\ket{\phi_\pm({\bf q})}.
\eeq

In our two-band system, the Berry curvature can also be written as~\cite{Qi2008}
\beq
\Omega_{xy}^{(\pm)} (\bq)= \pm \frac{1}{2} {\bf n (\bq)}\cdot \left[\frac{\partial{\bf n}}{\partial q_x} \times \frac{\partial{\bf n}}{\partial q_y} \right],
\eeq
${\bf n}({\bf q})$ being the unitary vector giving the Floquet Hamiltonian $H_{\mathrm{eff}} ({\bf q})=\varepsilon({\bf q}) {\bf n}({\bf q}) \cdot \sigma$.
Therefore, for a force in the $x$-direction, the semi-classical equations of motion for a wavepacket center-of-mass read
\begin{align}\label{eq:velox2D}
\dot{m}_{x}^{(\pm)}&= \partial_{q_{x}}\varepsilon_{\pm}({\bf q})\\
\dot{m}_{y}^{(\pm)}&= \partial_{q_{y}}\varepsilon_{\pm}({\bf q})- F_{x} \Omega^{(\pm)}_{yx}({\bf q})=\partial_{q_{y}}\varepsilon_{\pm}({\bf q})+ F_{x} \Omega^{(\pm)}_{xy}({\bf q}).
\end{align}
We now sum the displacement of the wavepackets located on a grid $q_{x,y}=-\pi+2\pi i/N$, where $i=1...N$. In the limit of $N\rightarrow\infty$, the mean displacement of the sum of the wavepackets corresponds to the average displacement of a filled band, i.e.,
\begin{align} \label{eq:transv_mot}
\langle \Delta m_x (t)\rangle^{(\pm)}&=0\\
\langle \Delta m_y (t)\rangle^{(\pm)} &= \frac{F_{x} \nu^{(\pm)}}{2\pi} t,
\end{align}
where the Chern number of the $\pm$-th band is defined as:
\beq\label{eq:Chern}
\nu^{(\pm)}= \frac{1}{2\pi} \int_{\mathrm{BZ}} {\rm d^2}\bq\,  \Omega^{(\pm)}_{xy} (\textbf{q}),
\eeq
One finds that the Chern number of the lower band for $\delta=\pi/2$ is $\nu^{(-)}=1$, so that the displacement will be positive and proportional to time in the transverse direction for a positive force. By numerical tests, we have confirmed that a finite grid with $11\times11$ points in the BZ is sufficient in our system to obtain a good approximation of the continuous integral over the whole BZ.

As discussed in the main text, to provide a more accurate read-out of the Chern numbers of our QW, we also measured the displacements for the inverse protocol $U^{-1}$. This is obtained by properly tuning the plate retardations. 
Since $L(\delta_1)L(\delta_2)=L(\delta_1+\delta_2)$ and $L(2\pi)$ is the identity operator (up to a global phase factor), the inverse operator is simply $L^{-1}(\delta)=L(2\pi-\delta)$.
 Recalling that $W=L(\pi/2,0)$ [see Eq.~2 in the main text] and that $U=T_y(\delta)T_x(\delta)W$, it is straightforward to show that $U^{-1}=L(3/2\pi,0)T_x(2\pi-\delta)T_y(2\pi-\delta)$.

\section{Topological characterization}\label{supp:chern}

In static models, the bulk-edge correspondence guarantees that no edge modes may be present when all bands have trivial topological invariants. However, our QW protocol is described by an effective Floquet Hamiltonian. 
Depending on the values of $\delta$, our effective Hamiltonian may or may not be deformed continuously into its static counterpart~\cite{Cardano2017,Rudner2013,Nathan2015}. In particular, this can lead to regimes where the topological invariant of the static system, the Chern number, does not describe completely the topology of the system and does not predict the presence of protected topological edge states.

In the present work, for example, we measured the anomalous displacement of the system for two values of the parameter $\delta$, namely $\delta=\pi/2$ and $\delta=7\pi/8$, associated respectively with Chern numbers $\nu^{(\mp)}=\pm1$ and $\nu^{(\mp)}=0$. As we will see in a few lines, the latter case displays protected edge modes, even though all its bands have trivial topological invariants.

\begin{figure*}[h!]
\includegraphics[width=\linewidth]{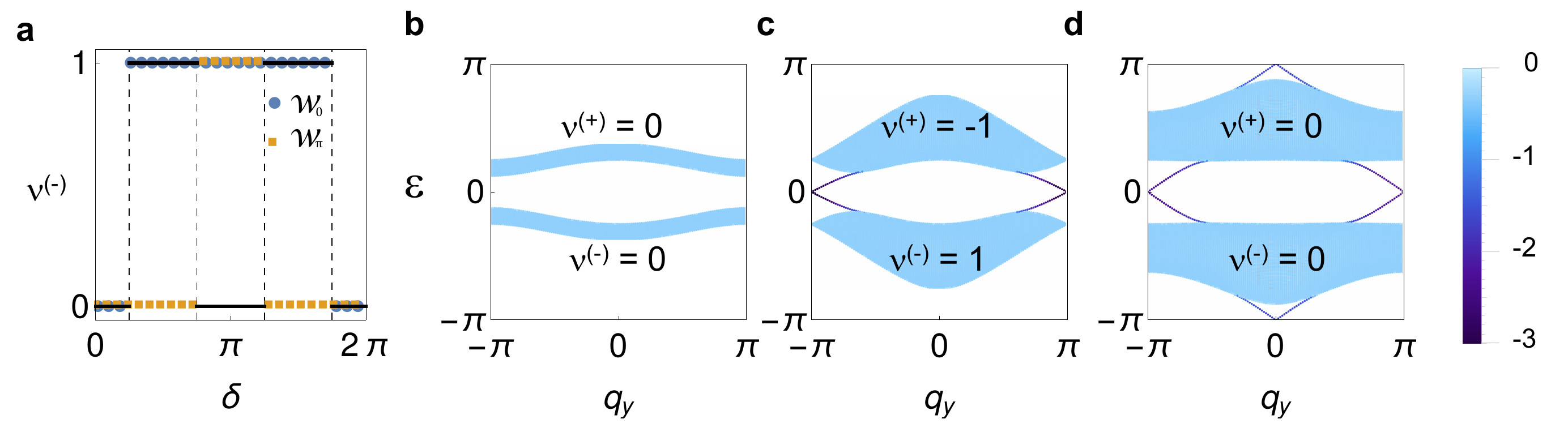}
\caption{{\bf Complete topological characterization of the QW protocol and bulk-edge correspondence.} {\bf a.} Phase diagram of our QW protocol showing the Chern number $\nu$ and the topological invariants $\mathcal{W}_{0}$, $\mathcal{W}_{\pi}$ defined in Ref.~\cite{Rudner2013}.
{\bf b-c-d.} Quasi-energy spectra computed on a cylinder open along $x$ for $\delta=\pi/8$ ({\bf b}),  $\delta=\pi/2$ ({\bf c}) and  $\delta=7\pi/8$ ({\bf d}). 
The color scale depicts the function $\lambda=\log_{10} (1-\langle |\hat{x}|\rangle_{\psi}/ N)$, which indicates 
the degree of localization of each state $\psi$. The two edges of the cylinder are located at $m_x=-N$  and $m_x=N$, so that lighter points denote bulk states, while darker points denote states that are closer to the edges.}
\label{fig:topo_class}
\end{figure*}

The bulk-edge correspondence of such systems was studied in detail by Rudner \emph{et al.} in Ref.~\cite{Rudner2013}. In the specific case of our model, characterized by two bands which are symmetric around zero quasi-energy, edge states may appear independently within the gap centered at quasi-energy 0, or within the gap at quasi-energy $\pm\pi$.
For definiteness, figure~\ref{fig:topo_class}{\bf a} shows the Chern number of the lowest band $\nu^{(-)}$, together with the topological invariants $\mathcal{W}_0$ and $\mathcal{W}_\pi$, which count, respectively, the number of pairs of edge modes in the 0-energy and $\pi-$energy gaps. Ref.~\cite{Rudner2013} provides their analytical expression, which is rather involved, but nonetheless straightforward to compute.

To see the emergence of edge states, Figs.~\ref{fig:topo_class}{\bf b,c,d} show the spectrum of our model on a cylinder which is open (closed) along the direction $x$ $(y)$, for three values of the optical retardation $\delta=\pi/8,\,\pi/2,$ and $7\pi/8$. 
In the vicinity of $\delta=0$ both bands have trivial Chern numbers, and no edge states are visible (see Fig.~\ref{fig:topo_class}{\bf b}).
As $\delta$ is increased further, a first gap closing happens at quasi-energy 0 when $\delta=\pi/4$. As the gap re-opens, a pair of protected edge modes appear around zero-energy, and the Chern numbers switch from 0 to $\pm1$ (see Fig.~\ref{fig:topo_class}{\bf c}).
The next gap closing happens at quasi-energy $\pm\pi$ when $\delta=3\pi/4$. Upon re-opening of this gap, another pair of protected edge modes appears inside it, and the Chern numbers switch back from $\pm1$ to 0 (see Fig.~\ref{fig:topo_class}{\bf d}). 

The work by Rudner {\it et al.} \cite{Rudner2013} in particular showed that the Chern number is related to the number of 0 and $\pi$ edge modes by $\nu^{(-)} = \mathcal{W}_0-\mathcal{W}_\pi$.
 We computed numerically these invariants for our model, and we correspondently recovered the bulk-edge correspondence, as can be seen in Fig.~\ref{fig:topo_class}{\bf a}.

\section{Supplementary data}\label{supp:suppdata}
\begin{figure*}[h!]
\centering
\includegraphics[width=\linewidth]{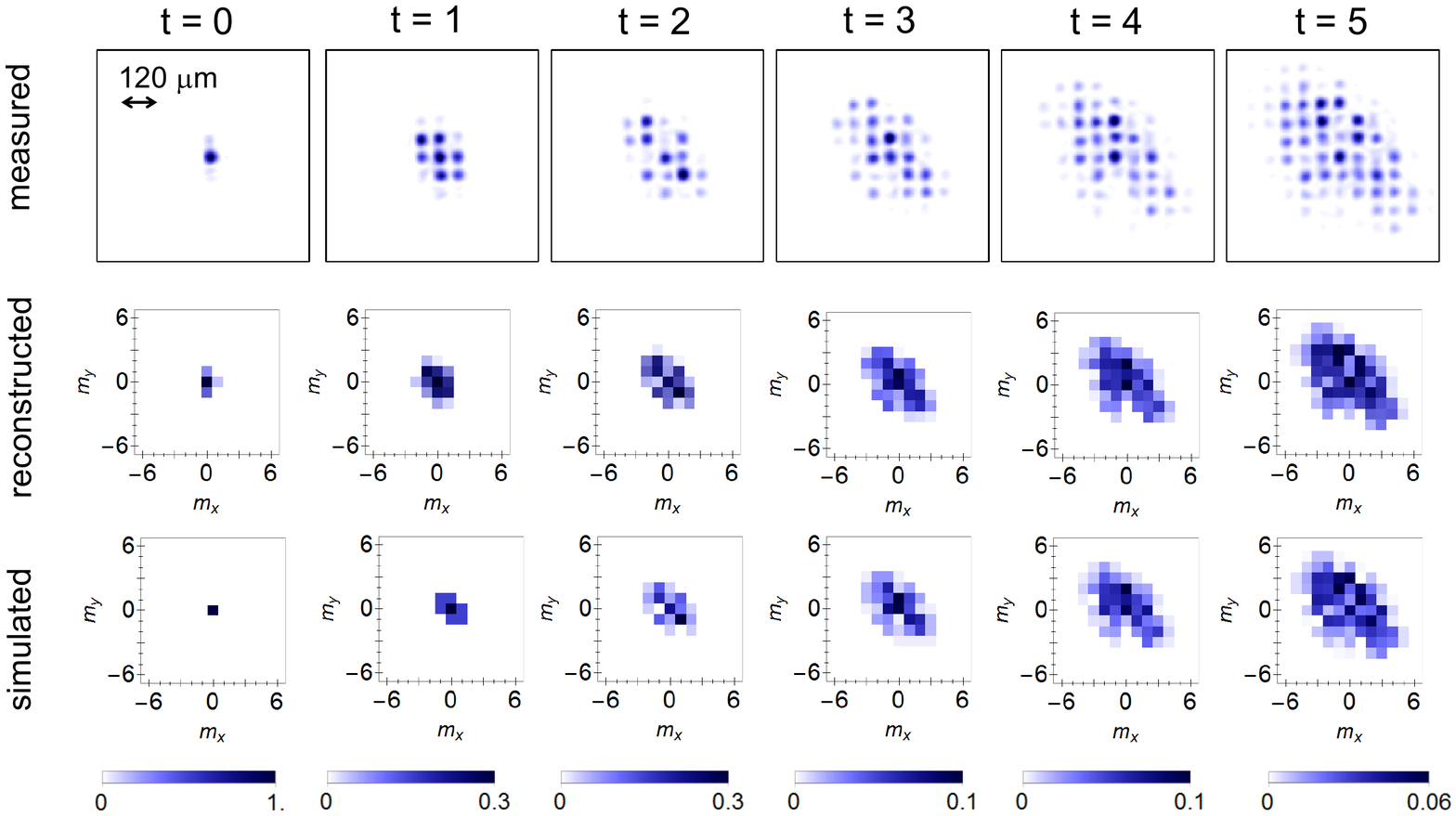}
\caption{{\bf 2D QW for input polarization $\ket{A}=(\ket{L}-i\ket{R})/\sqrt 2$.}
Spatial probability distributions for a quantum walk with initial condition $\ket{0,0,A}$ and optical retardation
 $\delta=\pi/2$. From left to right, we display results after 0 to 5 evolution steps. 
Datapoints are averages of four independent measures.
}
\label{fig: A_2d}
\end{figure*}

\begin{figure*}[h!]
\centering
\includegraphics[width=\linewidth]{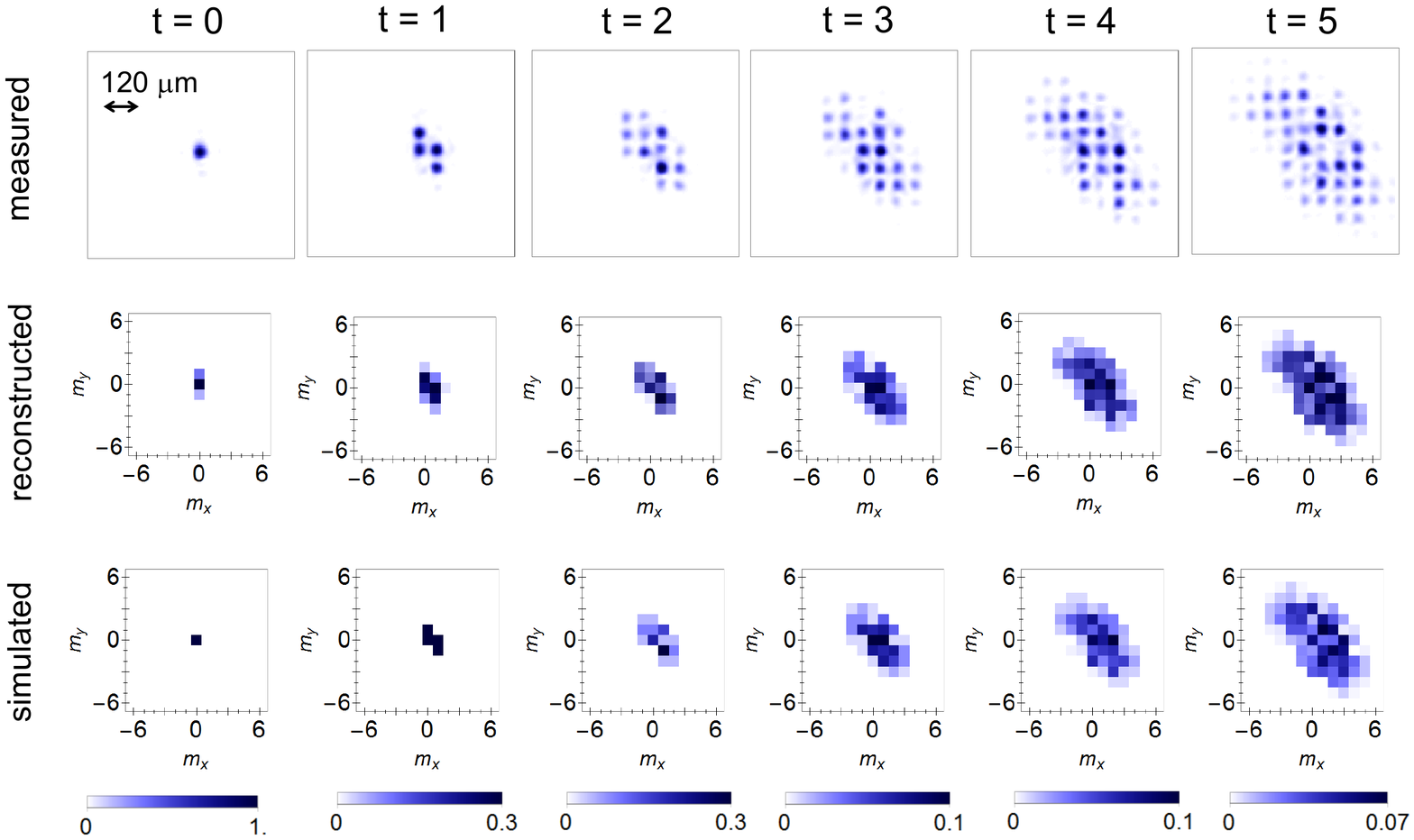}
\caption{{\bf 2D QW for input polarization $\ket{L}$.}
Spatial probability distributions for a quantum walk with initial condition $\ket{0,0,L}$ and optical retardation
 $\delta=\pi/2$. From left to right, we display results after 0 to 5 evolution steps. 
Datapoints are averages of four independent measures.
}
\label{fig: L_2d}
\end{figure*}

\begin{figure*}[h!]
\centering
\includegraphics[width=0.5\linewidth]{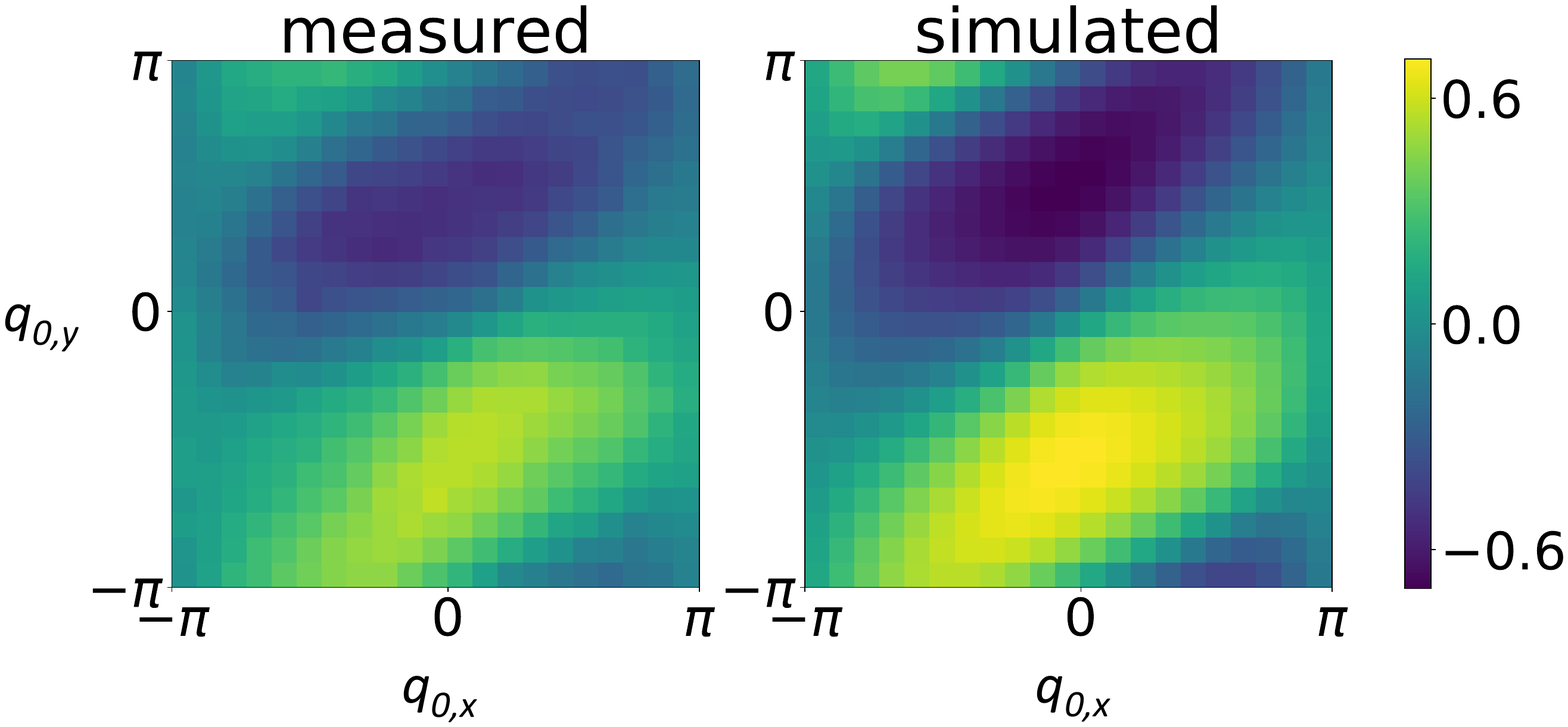}
\caption{{\bf Group velocity detection.}
Measurement of the $y$ component of the group velocity $\mathbf{v}^{(+)}$, for the upper band of a QW with $\delta=\pi/2$. 
Each datapoint is obtained from a linear fit of the center of mass displacement of a Gaussian wavepacket.
}
\label{fig:vgy}
\end{figure*}

\begin{figure*}[h!]
\centering
\includegraphics[width=0.7\linewidth]{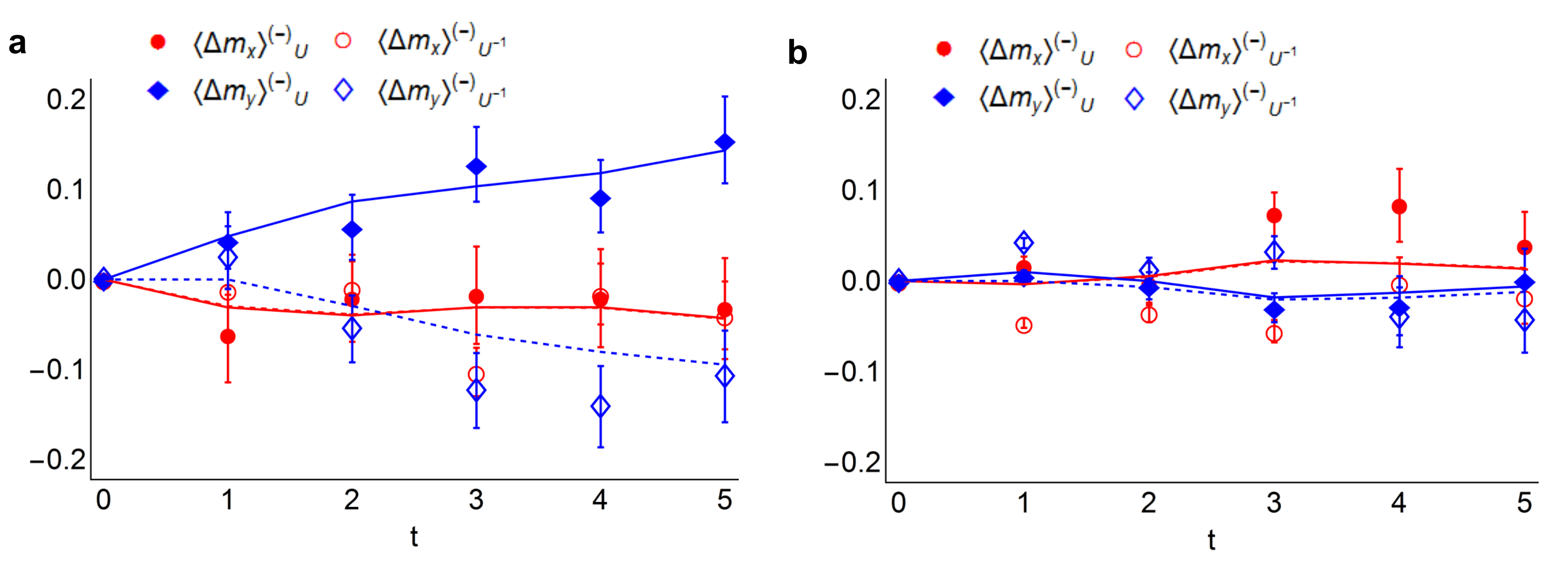}
\caption{{\bf Anomalous and individual displacements for protocols $U$ and $U^{-1}$}
We report experimental data compared with theoretical simulations of the wavepacket evolution of the anomalous $\me{\Delta m_y}^{(-)}$ and longitudinal $\me{\Delta m_x}^{(-)}$ displacements, obtained for the protocols $U$  and its inverse $U^{-1}$. {\bf a} Results for $\delta=\pi$. {\bf b} Results for $\delta=7\pi/8$. In both panels the blue lines correspond to simulations for displacements along $y$ (continuous line for $U$ and dashed for $U^{-1}$) while the red lines are simulations for displacement along $x$ (continuous line for $U$ and dashed for $U^{-1}$). Subtraction of these data allows to retrieve the results in Fig. 4 of the main text. 
}
\label{fig:forced wavepacket simulations}
\end{figure*}

\begin{figure*}[h!]
\centering
\includegraphics[width=0.5\linewidth]{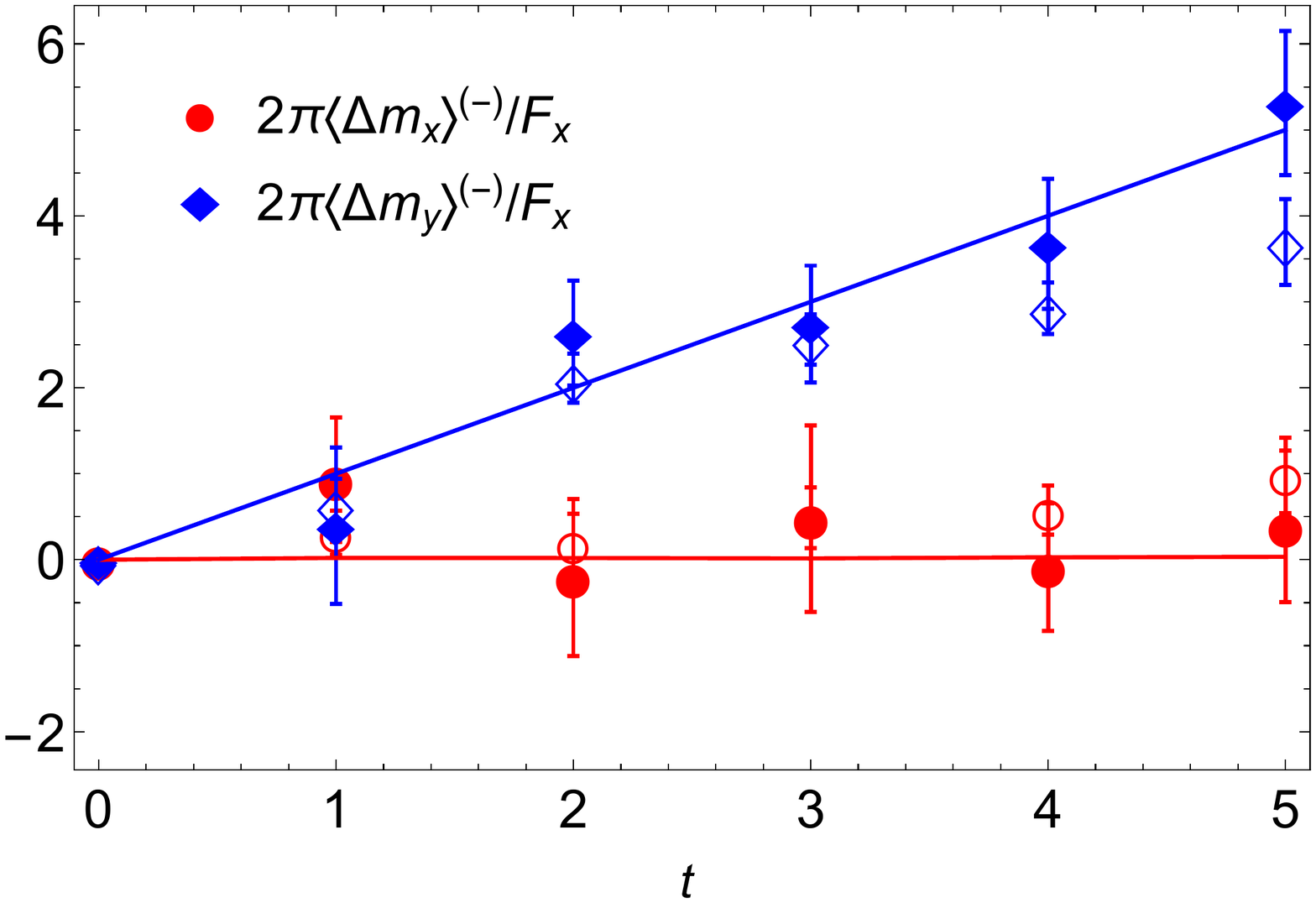}
\caption{{\bf Wavepacket displacements for larger values of the force.}
Band-averaged wavepacket displacements in the $x$ and $y$ directions, for $F=\pi/10$ (filled markers) and $F=\pi/5$ (empty markers), obtained combining measurements from the direct and inverse protocols, with $\delta=\pi/2$, for the lower band. Datapoints are experimental data, the continuous lines represent semi-classical predictions. 
}
\label{fig:forces}
\end{figure*}

In 
Figs.\ \ref{fig: A_2d}-\ref{fig:forces}, we show supplementary data supporting our results described in the main text.

\section{Possible deviations from the ideal QW evolution}\label{supp:deviations}
\begin{figure*}[t!]
\centering
\includegraphics[width=\linewidth]{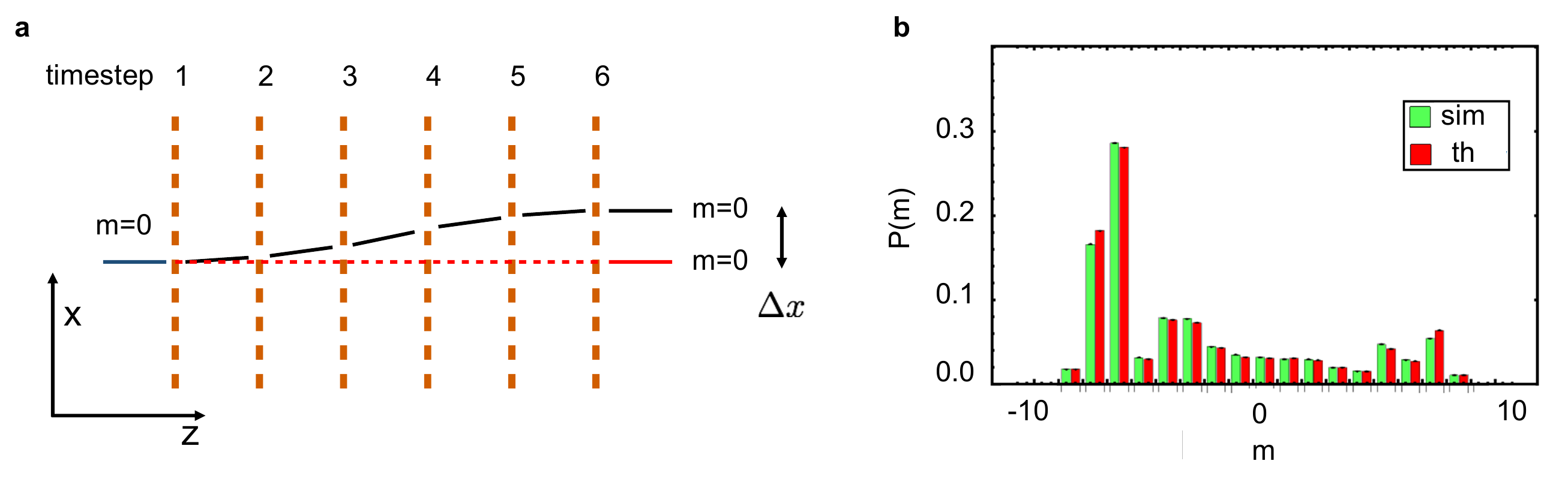}
\caption{{\bf Deviations from the perfect simulation of a QW process.} {\bf a}, At the input of the 1D QW we have a single beam with $\bk_\perp=0$, localized at the lattice site $m=0$. At the exit of a 6-steps QW, we consider two contributions to the final wavefunction at site $m=0$. One is resulting from the part of the input state that has been left unchanged (red). The second has gained $\Delta k$ transverse momentum at the first three steps, and then has acquired opposite momentum at steps 4 to 6 (black). At the exit of the walk, also this component corresponds to the lattice site $m=0$. However, the associated beams have some differences, which represent deviations from the ideal QW; being related to the same lattice site, they should be identical. First, they exit the walk laterally displaced by $\Delta x$, and the lack of overlap may reduce the interference, similarly to a decoherence effect. Second, the upper beam suffered a longer optical path, hence it accumulated a relative phase with respect to the central one. Finally, at each $g$-plate, the effective value of $\alpha_0$ changes at each step for the deflected beam. {\bf b}, Numerical simulation of a 1D walk with protocol $U=T_xW$, for $\delta=\pi/2$ and for an input state $\ket{0,R}$ (green columns), compared to a theoretical predictions of the ideal QW dynamics (red columns). After 10 steps differences are minimal.
}
\label{fig:deviations}
\end{figure*}
During propagation through the QW set-up, effects related to free space propagation of modes $\ket{\bmxy}$ can act as perturbations to the ideal QW dynamics. In this section we describe the main phenomena that can take place in our system, and investigate their effect with the help of a numerical analysis. For simplicity, we will refer to a 1D QW, where modes $\ket{\bmxy}$ are characterized by a single integer $m_x=m$.

There are essentially three \qo{undesired} effects that may arise when considering the tilt in propagation direction \textsl{inside} the quantum walk:
\begin{enumerate}
	\item At the end of the quantum walk, to each value of $\mathbf{k}_{\perp}$ will correspond a superposition of waves that have followed different paths in the wavevector space, as illustrated in 
Fig.\ \ref{fig:deviations}{\bf a}. These trajectories are actually associated with different optical paths. Associated relative phases are absent in the ideal QW dynamics and can modify the interference of the wavefunction components.  To simulate this, the final amplitude of each mode $\ket{\bmxy}$ can be calculated as the sum of all components related to these optical paths, multiplied by their relative phase. When propagating between two consecutive timesteps, mode $\ket{0}$ and a different mode $\ket{s}$ accumulate a phase delay $\Delta \phi_1 \approx (2\pi \lambda d s^2)/(\Lambda^2)$, where $d$ is the distance between consecutive steps.
	\item The two beams considered in 
Fig. \ref{fig:deviations}{\bf a}, that exit the walk in the same mode with $m=0$, have an imperfect overlap that makes them partly distinguishable, since they are propagating along axes that are parallel but laterally displaced. These modes have a finite extension, and the absence of perfect spatial overlap results in a reduction of the interference visibility, similarly to a decoherence effect. Referring to the case presented above, between consecutive timesteps the two modes accumulate a lateral shift $|\Delta x|=d\lambda s/\Lambda$.
	\item A tilted beam hits two consecutive $g$-plates at points that have a relative shift. Eq.\ 
	4 in the main text, describing the action of a $g$-plate, is derived by considering a Gaussian beam that hits the plate with its central position at $(x,y)=(0,0)$. If the beam center is displaced by $\Delta x$, Eq.\ (
4) in the main text remains valid after replacing $\alpha_0$ with $\alpha_0'=\alpha_0+\Delta x\, \pi/\Lambda$, which represents the effective LC orientation at the beam central position. By looking at Eq.\ (
4) in the main text, we can observe that this effect results in additional phases accumulated by modes $\ket{m}$ during propagation, which have to be added to the phases associated with the different path lengths (see the previous point 1). 
	\end{enumerate}
All these effects are estimated to be negligible for our setup. To make a quantitative check, in 
Fig. \ref{fig:deviations}{\bf b}, we provide a comparison between an ideal QW evolution and the simulation of the real beam propagation through our set-up, by taking into account effects described in points (1-3) and using the real system parameters. After 10 steps of a 1D walk, we observe no significant deviations. This guarantees that also 5 steps of a 2D would not suffer any deviations. Indeed, such effects strongly depends on the typical order of the modes that are excited during the walk, which is of course limited by the highest possible order. In the first case, when starting from a localized input, the highest-order mode that can be excited has $|m|=\pm10$. In the 2D case, highest-order excited modes have $|\bmxy|=\sqrt{m_x^2+m_y^2}=5\sqrt{2}\approx 7$.

In prospect, when increasing the number of steps, we expect these systematic deviations from ideality to become progressively more relevant. On the other hand, if needed, these issues could be tackled by (i) changing the system parameters, in particular increasing $\Lambda$ and reducing the step distance $d$, or (ii) by adopting a loop architecture combined with an imaging system. Indeed, by imaging the output of each step to the input of the following one, all the effects discussed above can be eliminated.%

\end{document}